\documentclass[twocolumn,tighten]{aastex62}

\providecommand\bibsty{yahapj_twoauthor}
\providecommand\tablesize{\scriptsize}

\usepackage{soul} 
\usepackage{amsmath}
\usepackage{xspace}
\usepackage{xifthen}
\usepackage{eso-pic}
\usepackage{multirow}
\usepackage{makecell}
\usepackage{comment}
\usepackage[htt]{hyphenat}
\usepackage{apjfonts}
\usepackage{listings}
\usepackage{subfigure}
\newcommand{\NEW}[1]{\ignorespaces}
\newcommand{\FIXME}[1]{}
\newcommand{\CHECK}[1]{}
\newcommand{\COMMENT}[3]{}
\newcommand{\highlight}[1]{}
\newcommand{\response}[1]{}

\mathchardef\mhyphen="2D

\newcommand{\roughly}{\ensuremath{ {\sim}\,} }

\newlength{\dhatheight}

\newcommand{\code}[1]{\texttt{#1}\xspace}

\newcommand{\var}[1]{\ensuremath{\texttt{\MakeUppercase{#1}}}\xspace}

\newcommand{\unit}[1]{\ensuremath{\mathrm{\,#1}}\xspace}

\newcommand{\degree}{\ensuremath{{}^{\circ}}\xspace}
\newcommand{\mas}{\unit{mas}}
\newcommand{\amin}{\unit{arcmin}}
\newcommand{\asec}{\unit{arcsec}}
\newcommand{\pix}{\unit{pix}}

\newcommand{\nm}{\unit{nm}}

\newcommand{\jy}{\unit{Jy}}

\newcommand{\magn}{\unit{mag}}
\newcommand{\mmag}{\unit{mmag}}
\providecommand{\deg}{}
\renewcommand{\deg}{\unit{deg}}

\newcommand{\secref}[1]{Section~\ref{sec:#1}}
\newcommand{\appref}[1]{Appendix~\ref{app:#1}}
\newcommand{\tabref}[1]{Table~\ref{tab:#1}}

\newcommand{\figref}[1]{Figure~\ref{fig:#1}}

\newcommand{\eqnref}[1]{Equation~\eqref{eqn:#1}}

\newcommand{\bandvar}[2][]{%
  \ifthenelse{\isempty{#1}}{\var{#2}}{\var{#2\_#1}}%
}

\newcommand{\magpsf}[1][]{\bandvar[#1]{mag\_psf}}
\newcommand{\magapereight}[1][]{\bandvar[#1]{mag\_aper\_8}}

\newcommand{\magsof}[1][]{\bandvar[#1]{mag\_sof}}

\newcommand{\nitermodel}[1][]{\bandvar[#1]{niter\_model}}

\newcommand{\sofmash}[1][]{\bandvar[#1]{extended\_class\_mash\_sof}}
\newcommand{\mofmash}[1][]{\bandvar[#1]{extended\_class\_mash\_mof}}
\newcommand{\sofcmt}[1][]{\bandvar[#1]{sof\_cm\_t}}
\newcommand{\sofmag}[1][]{\bandvar[#1]{sof\_cm\_mag\_i}}
\newcommand{\sofpsfmag}[1][]{\bandvar[#1]{sof\_psf\_mag\_i}}
\newcommand{\flagsfootprint}{\var{flags\_footprint}}
\newcommand{\flagsforeground}{\var{flags\_foreground}}
\newcommand{\flagsbadregions}{\var{flags\_badregions}}
\newcommand{\flagsgold}{\var{flags\_gold}}

\newcommand{\numimage}{\var{num\_image}}
\newcommand{\fracdet}{\var{fracdet}}

\newcommand{\nexposures}{38850\xspace}

\newcommand{\nexposuressn}{6877\xspace}

\newcommand{\areanimagesgrizy}{5186} % from the DR1 paper

\newcommand{\footprintareaappyone}{1786 } % square degrees, from Y1 paper
\newcommand{\footprintareaapp}{4946 } % square degrees
\newcommand{\foregroundareaapp}{551 } % square degrees

\newcommand{\astroabs}{158} %{145} mean {54} corrected

\newcommand{\astrorel}{28}

\newcommand{\photgaia}{2.2} %from Eli directly (Nacho has file locally), older plots from https://cdcvs.fnal.gov/redmine/projects/descalibration/wiki/FGCM_Y4A1_Zeropoints#Comparisons-to-Gaia-DR2

\newcommand{\maglimmangyonei}{23.29}

\newcommand{\maglimmofyoneg}{23.7}
\newcommand{\maglimmofyoner}{23.5}
\newcommand{\maglimmofyonei}{22.9}
\newcommand{\maglimmofyonez}{22.2}

\newcommand{\maglimmangi}{23.34}

\newcommand{\maglimmanggapp}{24.3}
\newcommand{\maglimmangrapp}{24.0}
\newcommand{\maglimmangiapp}{23.3}
\newcommand{\maglimmangzapp}{22.6}
\newcommand{\maglimmangyapp}{21.4 }

\newcommand{\maglimmangdri}{23.33}

\newcommand{\maglimmangsvi}{23.68}

\newcommand{\maglimsofgapp}{23.8}
\newcommand{\maglimsofrapp}{23.6}
\newcommand{\maglimsofiapp}{23.0}
\newcommand{\maglimsofzapp}{22.4 }
\newcommand{\maglimsofgup}{0.2}
\newcommand{\maglimsofrup}{0.2}
\newcommand{\maglimsofiup}{0.2}
\newcommand{\maglimsofzup}{0.2}
\newcommand{\maglimsofglo}{0.3}
\newcommand{\maglimsofrlo}{0.3}
\newcommand{\maglimsofilo}{0.2}
\newcommand{\maglimsofzlo}{0.2}

\newcommand{\magcompleteg}{23.4}
\newcommand{\magcompleter}{23.0}
\newcommand{\magcompletei}{22.6}
\newcommand{\magcompletez}{22.2}

\newcommand{\medfwhmg}{1.14} % 1.119
\newcommand{\medfwhmr}{0.98} % 0.958
\newcommand{\medfwhmi}{0.89} % 0.880
\newcommand{\medfwhmz}{0.85} % 0.836
\newcommand{\medfwhmy}{0.95} % 0.904

\newcommand{\medcoaddskybrightg}{420}
\newcommand{\medcoaddskybrightr}{1113} 
\newcommand{\medcoaddskybrighti}{3386} 
\newcommand{\medcoaddskybrightz}{7600} 
\newcommand{\medcoaddskybrighty}{2807} 

\newcommand{\medcoaddskyuncg}{26.0}
\newcommand{\medcoaddskyuncr}{25.6}
\newcommand{\medcoaddskyunci}{25.0}
\newcommand{\medcoaddskyuncz}{24.3}
\newcommand{\medcoaddskyuncy}{23.1}

\newcommand{\medfwhmdeepg}{1.1}
\newcommand{\medfwhmdeepr}{0.96}
\newcommand{\medfwhmdeepi}{0.86}
\newcommand{\medfwhmdeepz}{0.73}
\newcommand{\medfwhmdeepy}{1.22}
\newcommand{\maglimdeepg}{26.03}
\newcommand{\maglimdeepr}{25.63}
\newcommand{\maglimdeepi}{25.06}
\newcommand{\maglimdeepz}{24.31}

\newcommand{\maglimdeepgapp}{26.0}
\newcommand{\maglimdeeprapp}{25.6}
\newcommand{\maglimdeepiapp}{25.0}
\newcommand{\maglimdeepzapp}{24.3}
\newcommand{\maglimdeepyapp}{22.5}
\newcommand{\footprintareadeep}{5.88}

\newcommand{\starefficiency}{95} 
\newcommand{\starcontamination}{8}
\newcommand{\galefficiency}{98.5}
\newcommand{\galcontamination}{1}
\newcommand{\tilings}{4 }

\newcommand{\objdensity}{15.5}
\newcommand{\galdensity}{10.5}

\newcommand{\photoz}{photo-$z$\xspace}

\newcommand{\mof}{MOF\xspace}
\newcommand{\MOF}{\mof}
\newcommand{\sof}{SOF\xspace}
\newcommand{\SOF}{\sof}
\newcommand{\metacal}{\code{METACALIBRATION}}

\newcommand{\SExtractor}{\code{SourceExtractor}}
\newcommand{\sextractor}{\SExtractor}
\newcommand{\PSFEx}{\code{PSFEx}}

\newcommand{\SWARP}{\code{SWarp}}
\newcommand{\swarp}{\SWARP}
\newcommand{\SCAMP}{\code{SCAMP}}
\newcommand{\scamp}{\SCAMP}
\newcommand{\ngmix}{\code{ngmix}}

\newcommand{\balrog}{\code{Balrog}}
\newcommand{\BIGMACS}{\code{big-macs-calibrate}}

\newcommand{\redmagic}{\code{redMaGiC}}

\newcommand{\HEALPix}{\code{HEALPix}}
\newcommand{\healpix}{\HEALPix}

\newcommand{\nside}{\code{nside}}

\newcommand{\mangle}{\code{mangle}}

\newcommand{\easyaccess}{\code{easyaccess}}
\newcommand{\bpz}{\code{BPZ}}
\newcommand{\dnf}{\code{DNF}}
\newcommand{\ann}{\code{ANNz2}}
\newcommand{\sompz}{\code{SOMpz}}
\newcommand{\lephare}{\code{LePhare}}

\newcommand{\SNR}{\ensuremath{\mathrm{S/N}}\xspace}

\newcommand{\gaia}{\textit{Gaia}\xspace}

\newcommand{\svgold}{\code{SVA1\,GOLD}}
\newcommand{\yonegold}{\code{Y1\,GOLD}}

\newcommand{\gold}{\code{Y3\,GOLD}}
\newcommand{\coaddtruth}{\code{COADD\_TRUTH}}

\newcommand{\drurl}{\url{https://des.ncsa.illinois.edu/releases}}

\graphicspath{{./}{figures/}}

\received{XXX, 2021}
\revised{XXX, 2021}
\accepted{XXX, 2021}
\submitjournal{ApJS}

\reportnum{DES-2019-0451}
\reportnum{FERMILAB-PUB-20-569-AE}

\shorttitle{The \gold data set}
\shortauthors{The Dark Energy Survey Collaboration}

\begin{document}

\title{Dark Energy Survey Year 3 Results:
Photometric Data Set for Cosmology}

\author{I.~Sevilla-Noarbe}
\affiliation{Centro de Investigaciones Energ\'eticas, Medioambientales y Tecnol\'ogicas (CIEMAT), E-28040 Madrid, Spain}

\author{K.~Bechtol}
\affiliation{Physics Department, 2320 Chamberlin Hall, University of Wisconsin-Madison, 1150 University Avenue Madison, WI  53706-1390}

\author{M.~Carrasco~Kind}
\affiliation{Department of Astronomy, University of Illinois at Urbana-Champaign, 1002 W. Green Street, Urbana, IL 61801, USA}
\affiliation{National Center for Supercomputing Applications, 1205 West Clark St., Urbana, IL 61801, USA}

\author{A. Carnero Rosell}
\affiliation{Centro de Investigaciones Energ\'eticas, Medioambientales y Tecnol\'ogicas (CIEMAT), E-28040 Madrid, Spain}
\affiliation{Instituto de Astrof\'isica de Canarias, E-38205 La Laguna, Tenerife, Spain}
\affiliation{Universidad de La Laguna, Dpto. Astrof\'isica, E-38206 La Laguna, Tenerife, Spain}

\author{M.~R.~Becker}
\affiliation{HEP Division, Argonne National Laboratory, Lemont, IL 60439, USA}

\author{A.~Drlica-Wagner}
\affiliation{Department of Astronomy and Astrophysics, University of Chicago, Chicago, IL 60637, USA}
\affiliation{Fermi National Accelerator Laboratory, P. O. Box 500, Batavia, IL 60510, USA}
\affiliation{Kavli Institute for Cosmological Physics, University of Chicago, Chicago, IL 60637, USA}

\author{R.~A.~Gruendl}
\affiliation{Department of Astronomy, University of Illinois at Urbana-Champaign, 1002 W. Green Street, Urbana, IL 61801, USA}
\affiliation{National Center for Supercomputing Applications, 1205 West Clark St., Urbana, IL 61801, USA}

\author{E.~S.~Rykoff}
\affiliation{Kavli Institute for Particle Astrophysics \& Cosmology, P. O. Box 2450, Stanford University, Stanford, CA 94305, USA}
\affiliation{SLAC National Accelerator Laboratory, Menlo Park, CA 94025, USA}

\author{E.~Sheldon}
\affiliation{Brookhaven National Laboratory, Bldg 510, Upton, NY 11973, USA}

\author{B.~Yanny}
\affiliation{Fermi National Accelerator Laboratory, P. O. Box 500, Batavia, IL 60510, USA}

% ...

\collaboration{(DES Collaboration)}

\correspondingauthor{Ignacio Sevilla-Noarbe}
\email{ignacio.sevilla@ciemat.es}
\correspondingauthor{Keith Bechtol}
\email{kbechtol@wisc.edu}
\correspondingauthor{Matias Carrasco Kind}
\email{mcarras2@illinois.edu}

\begin{abstract}
We describe the Dark Energy Survey (DES) photometric data set assembled from the first three years of science operations to support DES Year 3 cosmology analyses, and provide usage notes aimed at the broad astrophysics community. \gold improves on previous releases from DES, \yonegold and Data Release 1 (DES DR1), presenting an expanded and curated data set that incorporates algorithmic developments in image detrending and processing, photometric calibration, and object classification. \gold comprises nearly $5000 \deg^2$ of $grizY$ imaging in the south Galactic cap, including nearly 390 million objects, with depth reaching $\SNR \sim10$ for extended objects up to $i_{AB}\sim  \maglimsofiapp$, and top-of-the-atmosphere photometric uniformity $< 3 \mmag$. 
Compared to DR1, photometric residuals with respect to \gaia are reduced by 50\%, and per-object chromatic corrections are introduced.
\gold augments DES DR1 with simultaneous fits to multi-epoch photometry for more robust galaxy color measurements and corresponding photometric redshift estimates.
\gold features improved morphological star-galaxy classification with efficiency $>98\%$ and purity $>99\%$ for galaxies with $19 < i_{AB} < 22.5$.
Additionally, it includes per-object quality information, and accompanying maps of the footprint coverage, masked regions, imaging depth, survey conditions, and astrophysical foregrounds that are used to select the cosmology analysis samples.
This paper will be complemented by online resources. 
\end{abstract}


\section{Introduction} 
\label{sec:intro}

Optical and near-infrared imaging surveys have become one of the most widely used tools to study new physics at the cosmic frontier, including dark energy, dark matter, neutrino properties, and inflation.
The current generation of imaging surveys, such as Pan-STARRS1 \citep[PS1;][]{panstarrs_2016_surveys}, Hyper Suprime-Cam Subaru Strategic Program \citep[HSC-SSP;][]{hscdr2}, Kilo-Degree Survey \citep[KiDS;][]{kids_2012_dr4}, DESI Legacy Imaging Surveys \citep{desi_2019_legacy_imaging_surveys}, and the Dark Energy Survey \citep[DES;][]{DES:2005,DES:2016ktf} have collectively provided deep multi-band imaging over nearly the entire high-Galactic-latitude sky, and cataloged more than a billion galaxies and thousands of supernovae spanning 10 billion years of cosmic history. 
Together with spectroscopic surveys (e.g. \citet{eboss_2020_cosmology}, \citet{desi_description}), these imaging surveys yield measurements of the expansion rate and large-scale structure in the late-time universe \citep[e.g.,][]{desy1cosmo,hsc_dr1shear_ps,kidsviking} that are complementary to the high-precision measurements of the early Universe \citep{planck18}. 
Wide-area imaging surveys provide access to the largest number of galaxies for statistical analyses, and the opportunity to combine several complementary probes of the cosmic expansion history and growth of structure into the same study \citep[e.g.,][]{des18multi,kids_2020_multi}.
Ground-based imaging surveys of the next decade, including the Vera C. Rubin Observatory Legacy Survey of Space and Time \citep[LSST;][]{ivezic_2019_lsst}, aim to catalog $>10^{10}$ galaxies and $>10^5$ Type Ia supernovae to further test the Cold Dark Matter with a Cosmological Constant ($\Lambda$CDM) Universe paradigm and its extensions.

The DES Collaboration has found significant benefits to developing, validating, and curating a shared reference data set to be used as the basis for most cosmological analyses \citep{y1gold}. The creation of this value-added `Gold' catalog involves close collaboration between the data pipeline team and science working groups to define and validate a set of high-quality data products that are broadly useful for science analysis.
We use this iterative process to prioritize algorithmic development and introduction of new data products as needed to support accurate cosmology. 

The DES data set is assembled from an imaging survey using the Blanco 4m telescope at the Cerro Tololo Inter-American Observatory (CTIO) in Chile to observe $\roughly 5000 \deg^2$ of the southern sky in five broadband filters, $grizY$, ranging from $\roughly400\nm$ to $\roughly1060\nm$ in wavelength, with the Dark Energy Camera \citep[DECam;][]{Flaugher:2015}. DES completed observations in January 2019, after 6 years of operations, with 10 overlapping dithered exposures at predefined positions in the sky on each filter. The primary goal of DES is to study the origin of cosmic acceleration and the nature of dark energy, using a variety of cosmological probes enabled by this rich data set.

Many DES Year 1 (Y1) cosmology results \citep{desy1cosmo} used the \yonegold catalog described in \citet{y1gold}. The emphasis of that work was to detail the data pipelines, calibration and curation of the coadded catalog. The Y1 data set was publicly released in October 2018\footnote{\url{https://des.ncsa.illinois.edu/releases/y1a1}}, including the aforementioned \yonegold catalog spanning an area of $\sim 1800 \deg^2$, together with ancillary maps of the survey properties \citep{Leistedt:2016},  shear catalogs \citep*{zuntz}, photometric redshift catalogs \citep*{y1photoz}, the \redmagic \citep*{Rozo:2016} catalogs used in DES Y1 results, and value-added catalogs \citep{y1sgsep, tarsitano}.

The coadded catalog from the first three years of data (Y3) was publicly released as DES Data Release 1 \citep[DR1;][]{dr1}.\footnote{\url{https://des.ncsa.illinois.edu/releases/dr1}} DR1 is the first DES catalog that spans the whole footprint ($\sim 5000 \deg^2$).
DR1 was produced as part of an annual data processing campaign with the DES Data Management pipeline \citep[DESDM;][]{desdm}, with photometric calibration described in \citet{Y3FGCM}.

Here, we present the core data set used in Y3 cosmology analyses. \gold builds upon the DR1 coadded catalog described in \citet{dr1}, with additional enhancements described in \citet{y1gold}, and introduces several new products and algorithmic developments. A summary of previous DES data releases appears in \tabref{releases}. Key attributes of the \gold data set are listed in \tabref{summary}. 

\begin{deluxetable*}{c c c c c c c}
\tablewidth{0pt}
\tabletypesize{\tablesize}
\tablecaption{Dark Energy Survey data releases\label{tab:releases} 
}
\tablehead{
\colhead{Release} & \colhead{Area} & \colhead{Depth} & \colhead{Nb. objects} & \colhead{Photometry uniformity} & \colhead{Supplemental data} & \colhead{Reference}
\\
\colhead{} & \colhead{(sq.deg.)} & \colhead{($i$ band)} & \colhead{} & \colhead{(mmag)} & \colhead{} & \colhead{}
}
\startdata
\svgold & $\sim 250$ & \maglimmangsvi & 25M & $< 15$ & Photo-zs & \url{https://des.ncsa.illinois.edu/sva1} \\
\yonegold & \footprintareaappyone & \maglimmangyonei & 137M & $< 15$ & Photo-zs, MOF, maps& \citet{y1gold} \\
DR1 & \areanimagesgrizy & \maglimmangdri & 399M & $< 7$ & None & \citet{dr1} \\
\gold & \footprintareaapp & \maglimmangi & 388M & $< 3$ & \makecell{Photo-zs, SOF/MOF, maps, \\improved classification} & This work \\
%DR2 & $\sim 5000$ & $\sim 23.7$ & \sim 700M & & & exp. 2021 \\
%Y6 Gold & $\sim 5000$ & $\sim 23.7$ & \sim 700M & & & exp. 2022  \\
\enddata
\tablecomments{All releases are made public at \url{https://des.ncsa.illinois.edu/}. Quoted depth corresponds to ${\rm S/N} = 10$ in $2\asec$ diameter apertures. \sof and \mof are multi-epoch pipelines described in \secref{mofsof}.}
\end{deluxetable*}

\begin{deluxetable*}{l c c c c c c}
\tablewidth{0pt}
\tabletypesize{\tablesize}
\tablecaption{Key numbers and data quality summary for the DES Wide Survey (\gold; this work) and Deep Fields \citep*[\coaddtruth; reproduced from][]{y3-deepfields}.
For parameters representing a distribution, the median or mean values are quoted as specified in the main text. All magnitudes are in the AB system. 
\label{tab:summary}}
\tablehead{
\colhead{Parameter} & \multicolumn{5}{c}{Band} \\
 & $g$ & $r$ & $i$ & $z$ & $Y$
}
\startdata
\multicolumn{6}{c}{Wide Survey (this work)} \\
\hline
Median PSF FWHM ($\asec$) & $\medfwhmg$ & $\medfwhmr$ & $\medfwhmi$ & $\medfwhmz$ & $\medfwhmy$ \\
Median Sky Brightness (electrons/pixel) & $\medcoaddskybrightg$ & $\medcoaddskybrightr$ & \medcoaddskybrighti & $\medcoaddskybrightz$ & $\medcoaddskybrighty$\tablenotemark{a} \\
Median Sky Brightness uncertainty ($\magn/\asec^2$) & $\medcoaddskyuncg$ & $\medcoaddskyuncr$ & $\medcoaddskyunci$ & $\medcoaddskyuncz$ & $\medcoaddskyuncy$ \\
Sky Coverage ($grizY$ intersection, deg$^{2}$) & \multicolumn{5}{c}{$\footprintareaapp$} \\ 
Coadd Astrometric Precision (total distance, \mas) & \multicolumn{5}{c}{$\astrorel$~(internal); $\astroabs$~(vs. \gaia DR2)} \\ 
SOF Photometric Uniformity vs. Gaia (mmag)\tablenotemark{b} & \multicolumn{3}{c}{$\photgaia$} & \nodata & \nodata \\ 
Median Coadd Magnitude Limit, $2 \asec$ diameter ($\SNR = 10$) & $\maglimmanggapp$ & $\maglimmangrapp$ & $\maglimmangiapp$ & $\maglimmangzapp$ & $\maglimmangyapp$ \\ 
Coadd $90\%$ Completeness Limit for extended objects (mag) & $\magcompleteg$ & $\magcompleter$ & $\magcompletei$ & $\magcompletez$ & \nodata \\ 
Multi-Epoch Galaxy Magnitude Limit ($\SNR = 10$, SOF)\tablenotemark{c} & $\maglimsofgapp^{+\maglimsofgup}_{-\maglimsofglo}$ & $\maglimsofrapp^{+\maglimsofrup}_{-\maglimsofrlo}$ & $\maglimsofiapp^{+\maglimsofiup}_{-\maglimsofilo}$ & $\maglimsofzapp^{+\maglimsofzup}_{-\maglimsofzlo}$ & \ldots \\ 
Coadd Galaxy Selection ($\sofmash \geq 2$, $\magsof[i] \leq 22.5$) & \multicolumn{5}{c}{Efficiency $>\galefficiency\%$; Contamination $<\galcontamination\%$} \\
Coadd Stellar Selection ($\sofmash \leq 1$, $\magsof[i] \leq 22.5$) & \multicolumn{5}{c}{Efficiency $>\starefficiency\%$; Contamination $<\starcontamination\%$}  \\
Object density ($\amin^{-2}$) \tablenotemark{d}& \multicolumn{5}{c}{Overall: \objdensity; Galaxies: \galdensity} \\
\hline
\multicolumn{6}{c}{Deep Fields \citep*{y3-deepfields}} \\
\hline
Median PSF FWHM (\asec) & $\medfwhmdeepg$ & $\medfwhmdeepr$ & $\medfwhmdeepi$ & $\medfwhmdeepz$ & $\medfwhmdeepy$\tablenotemark{e} \\
Median Coadd Magnitude Limit, $2 \asec$ diameter, $\SNR = 10$) & $\maglimdeepgapp$ & $\maglimdeeprapp$ & $\maglimdeepiapp$ & $\maglimdeepzapp$ & $\maglimdeepyapp$\tablenotemark{e}  \\
Sky Coverage ($ugrizJHKs$ intersection, $\deg^{2}$) & \multicolumn{5}{c}{$\footprintareadeep$}  \\
\enddata
\tablenotetext{a}{$Y$-band exposures are half the exposure time of the other bands, only after Y4 were 90 s exposures taken.}
\tablenotetext{b}{Photometric uniformity measured vs. Gaia's $G$ band, which encompasses DECam's $gri$, see footnote on \secref{photometry_performance}.}
\tablenotetext{c}{Median values with $16\%$ and $84\%$ percentile errors from the magnitude limit distribution.}
\tablenotetext{d}{Object density determined for all objects in \gold footprint outside foreground and bad regions, and the subset of those classified as high-confidence galaxies (\sofmash = 3).} \tablenotetext{e}{NB not every deep field has Y-band measurements, see \citet*{y3-deepfields} for details.}
\end{deluxetable*}

\begin{figure*}
	\includegraphics[width=\textwidth]{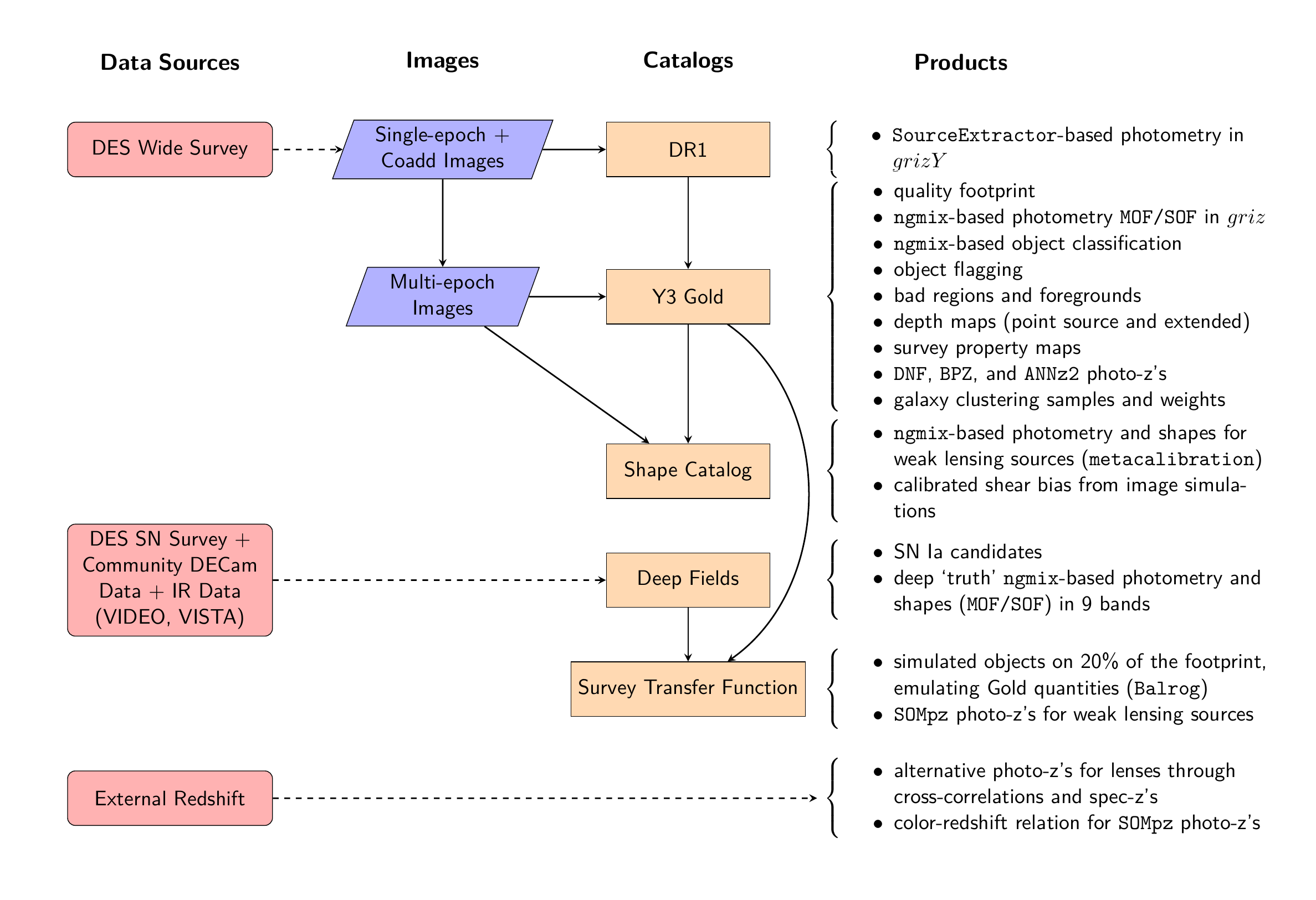}
    \caption{The Y3 DES core data sets and their relationships. Arrows indicate information flow from data sources (observations, dashed lines) or processed images and catalogs (continuous lines) to another catalog or data product. The ``Products'' column indicates the outputs associated with the catalogs immediately to their left.
    }
\label{fig:datasets}
\end{figure*}

The \gold data set and associated documentation are a core element of DES Y3 cosmology, and are complemented and enhanced by several additional data products described in companion papers to this one, such as refined photometric redshift estimates, shear catalogs, cosmological simulations and mock DES data sets.
\figref{datasets} shows relationships between the various DES Y3 data products.
In this work, mainly devoted to \gold, we will highlight these relationships as appropriate in the text.  

\begin{figure*}
	\includegraphics[width=\textwidth]{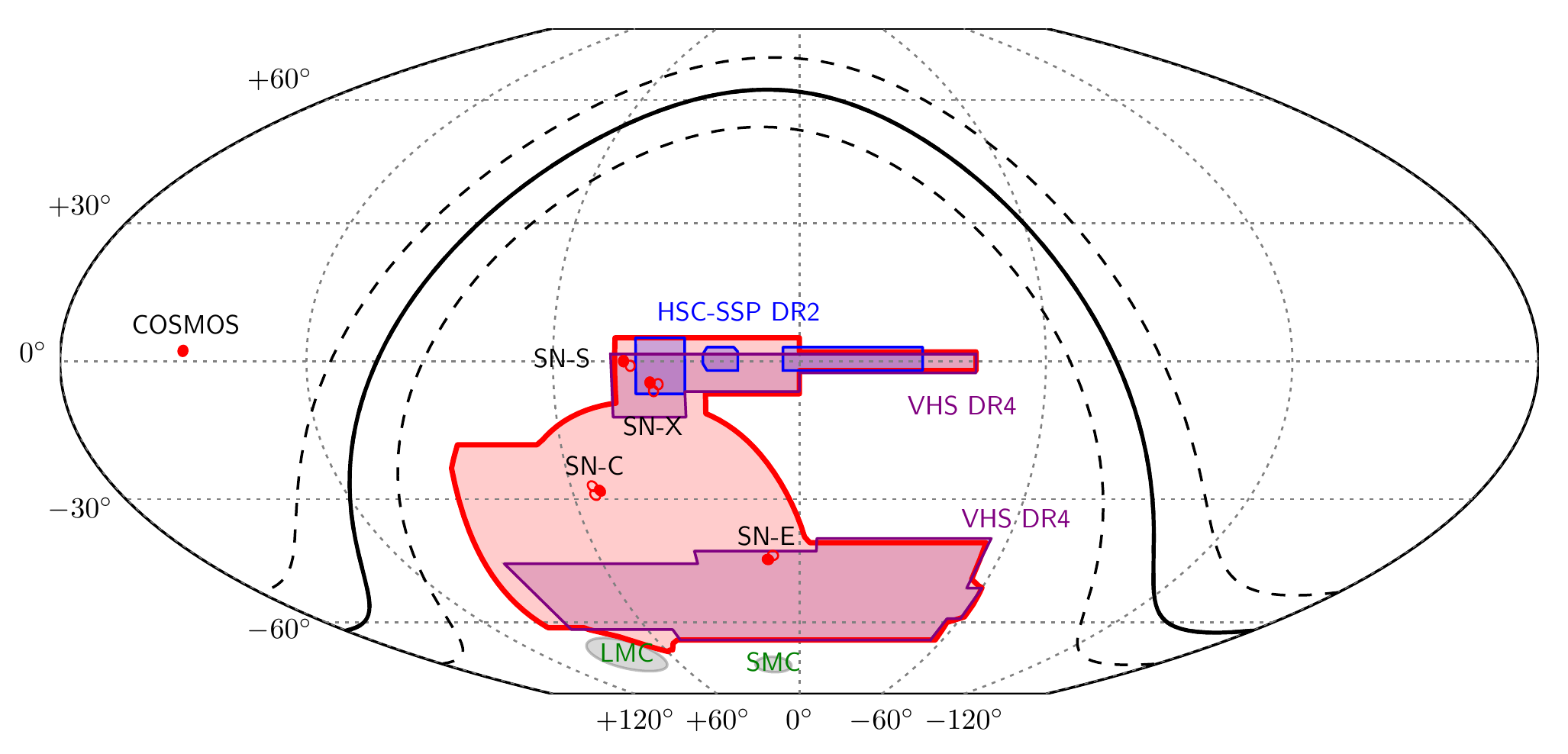}
    \caption{The Dark Energy Survey footprint in equatorial coordinates, including the Wide Survey, the Supernova Survey fields \citep[labeled SN;][]{snsurvey}, the COSMOS field, as well as relevant HSC-SSP DR2 \citep{hscdr2} and VHS DR4 \citep{vhs} data sets used in this work (only the approximate overlapping regions are indicated for clarity). The fields used for DES Deep Field processing are filled red (\secref{deepfields}). The DES footprint avoids the Galactic plane (solid black line with dashed lines at $b = \pm10 \degree$) and central regions of the Large and Small Magellanic Clouds (LMC, SMC).}
\label{fig:footmap}
\end{figure*}

In \secref{threeyears} we review DES science operations and major upgrades appearing in \gold.
We then detail the data processing for this particular release in \secref{dataprocessing}, going into some additional detail for astrometric and photometric calibration and performance in \secref{astrophoto}. We characterize the depth of the survey in \secref{depth} and describe several value-added quantities in \secref{characterization}. \secref{maps} contains a description of the maps that accompany the release.  
\secref{usage} presents usage notes for \gold to facilitate exploration by the wider community, and we conclude in \secref{conclusions}.

All magnitude quantities in this work are expressed in the AB scale unless otherwise specified.

\section{The first three years of DES data}
\label{sec:threeyears}

In this section, we review the Wide and Supernova Survey components of DES, and detail differences between \gold and previous releases. The data included in \gold spans 345 distinct nights of observations with at least one observation passing quality tests from 2013 August 15 to 2016 February 12.

\subsection{Survey overview}
\label{sec:strategy}

DES used two survey modes \citep{neilsen:2019} to meet the specific requirements of multiple cosmological probes:

\begin{itemize}
\item The \textbf{Wide Survey} is optimized for gravitational weak lensing, galaxy clustering, and galaxy cluster cosmological probes.
The Wide Survey spans $\sim5000 \deg^2$ imaged with 10 dithered exposures at each position in each of five broad photometric bands $grizY$ (90 second exposures, except for $Y$, which employed some 45 second exposures).
During the first three years of DES, most of the Wide Survey footprint was covered with \tilings overlapping images in each band.
The Wide Survey is the basis for the \gold data set. 
\item The \textbf{Supernova Survey} involves repeated observations of 10 DECam fields, amounting to a total of $27 \deg^2$, imaged in $griz$ with an approximately weekly cadence \citep{Kessler:2015,snsurvey}.
Difference imaging analysis of the Supernovae Survey fields enables the discovery the discovery of thousands of Type Ia supernovae (SN Ia) and precision photometric lightcurves are computed following \citet{Brout:2019}. Cosmology results based on the analysis of a subset of spectroscopically confirmed SN Ia in the redshift range of $0.2 < z < 0.85$ from the first three years of data taking, combined with other sets, have been presented in \citet{y3sncosmo}. The SN exposures are coadded to produce the Y3 Deep Field data set. \textbf{Deep Field} processing of some of the Supernovae Survey fields, together with DECam imaging of the COSMOS\footnote{\url{http://cosmos.astro.caltech.edu}} field, enables high S/N measurements of galaxies approximately $1.5-2$ \magn fainter than the Wide Survey \citep*{y3-deepfields}.
A subset of these data have been combined with deep near-infrared imaging to produce a reference object catalog used for various applications in DES Y3 cosmology analyses. The Supernovae Survey data and the Y3 Deep field data are not part of the \gold data release.
\end{itemize}

Exposures were acquired during the allocated nights for DES at the Blanco Telescope and transferred to the National Center for Supercomputing Applications (NCSA) \citep{sispi} for further processing (\secref{dataprocessing}).
A total of \nexposures exposures were acquired, across all bands, and included in the Y3 Wide Survey processing \citep{desdm}. The supernova cosmology program used \nexposuressn exposures \citep{snsurvey} from the Y3 period. 

The DES footprint, including the Wide and Supernovae Surveys, as well as relevant external data sets mentioned in this paper, are shown in \figref{footmap}.

\subsection{\gold data set and differences relative to previous DES releases}

\begin{figure*}
	\includegraphics[width=0.45\textwidth,trim=0cm 0cm 0cm 0cm,clip]{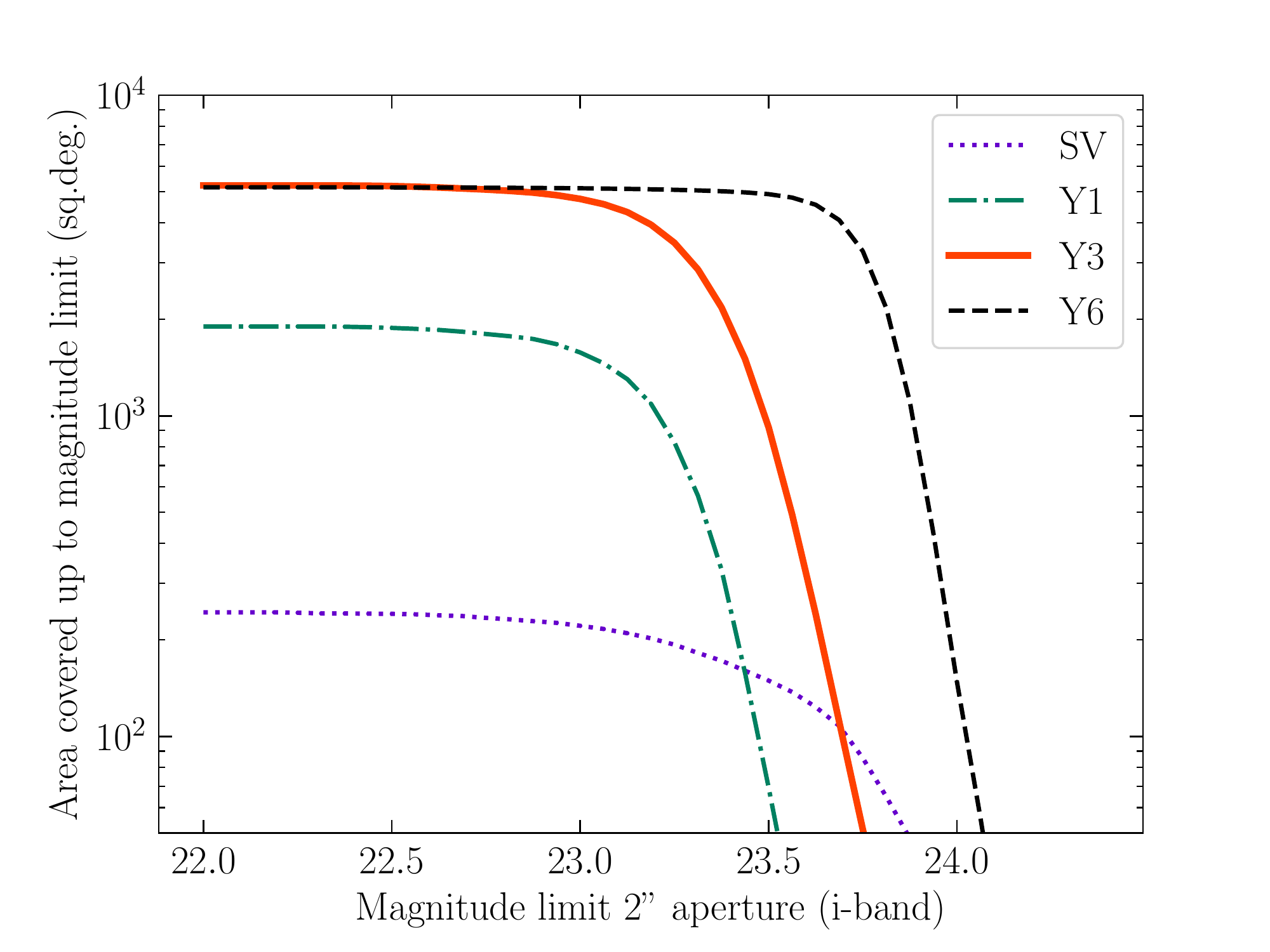}
	\includegraphics[width=0.50\textwidth,trim=0cm 10cm 4cm 4cm,clip]{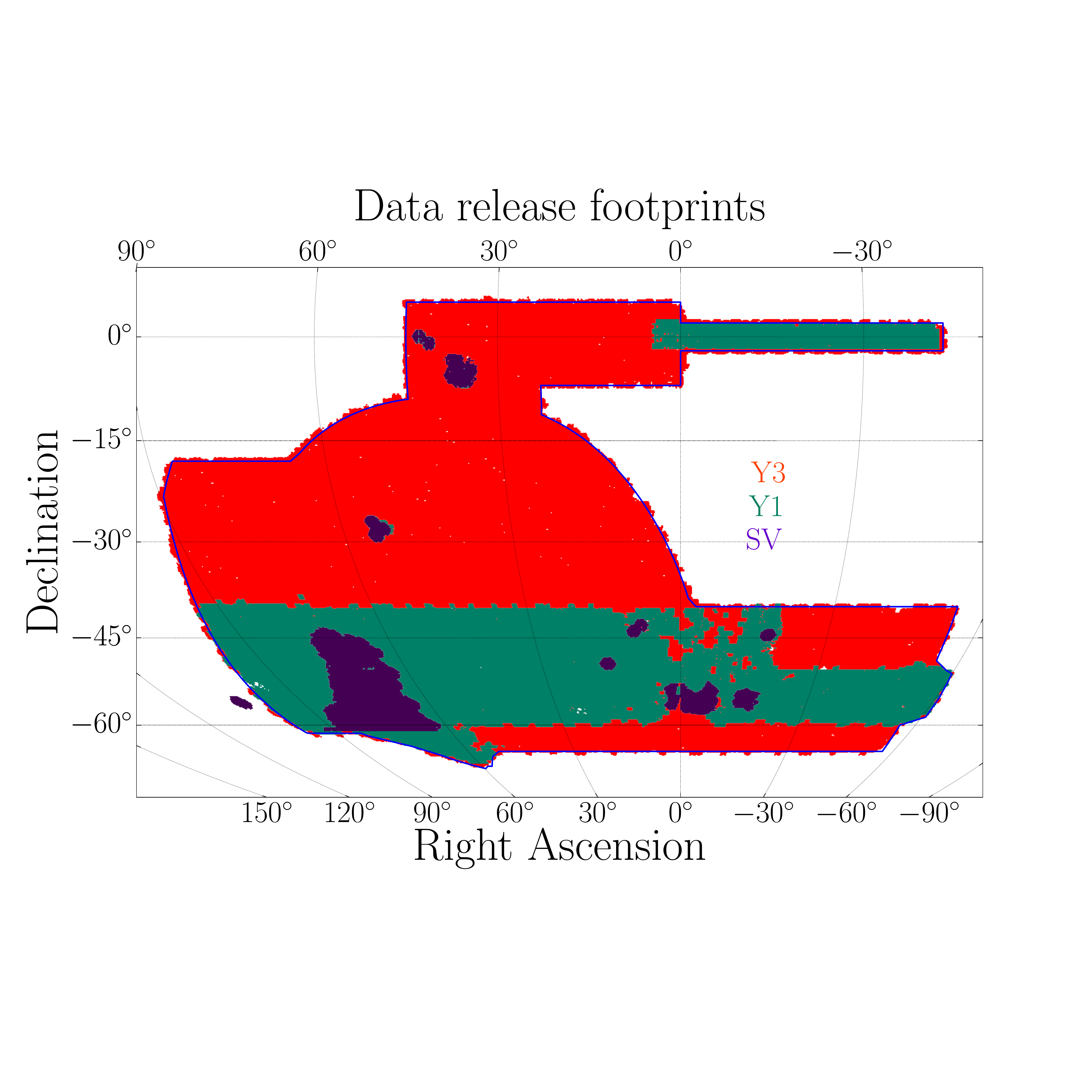}
    \caption{(Left) Area covered to a certain magnitude limit in 2\asec diameter apertures for each data release in the $i$ band. 
    (Right) Survey map for the Science Verification (SV), Year 1 (Y1) and Year 3 (Y3) data sets (which overlaps mostly the previous ones). Year 6 data covers a very similar footprint as Y3 to a greater depth.}
\label{fig:svy1y3y6}
\end{figure*}

Below, we summarize commonalities and differences in input data, image processing, and catalog generation between \gold and previous releases. 
\figref{svy1y3y6} shows the progression in areal coverage and depth from the Science Verification phase through the completed six seasons of DES.

\subsubsection{Differences relative to \yonegold}

\begin{itemize}
\item The Wide Survey area has increased by a factor of 2.7, from $1786 \deg^2$ to $4946 \deg^2$ with simultaneous coverage in $griz$. The exact choice of survey property selections for specific science analyses can modify the final footprint size.
\item Coadd depth has slightly increased with respect to \yonegold. Y3 focused on expanding the area and uniformity of the Y1 data set, and thus the increase in depth was fairly small. 
\item The Y3 astrometric calibration is performed exclusively based on 2MASS \citep{Skrutskie:2006} instead of UCAC-4 \citep{Zacharias:2013} used in \yonegold, for the first single-epoch pass. \gaia \citep{Gaia:2016, gaiadr2} catalogs were not available during development of the Y3 Coadd processing; however, these catalogs will be used in future DES data processing campaigns (\secref{astrometry}). 
\item The Y3 photometric calibration adopts the Forward Global Calibration Method \citep[FGCM;][]{Y3FGCM} as the default algorithm for this purpose, as described in \citet{dr1} (\secref{photometry}).
\item Improved pipelines have led to some changes in the flagging of objects. This is especially true with the introduction of the \var{IMAFLAGS\_ISO} flag. This is a \sextractor output that provides an `OR' of all flags set in the image over all pixels in the objects' isophote, which enables the identification of image artifacts and affected objects.
\end{itemize}

\subsubsection{Differences relative to DR1}

\begin{itemize}

\item In \gold, the morphological and photometric measurements are based on the Multi-Object Fitting pipeline \citep[\mof,][]{y1gold} and its variant, the Single-Object Fitting pipeline (\sof; \secref{mofsof}).
\item  In \gold, zeropoint estimates incorporate Year 4 imaging (which was already available as single epoch images at the time of making the coadded Y3 catalogs, see \appref{gray_zeropoint} for details). This was due to poor sky conditions during Year 3. \gold adds chromatic and SED-dependent interstellar extinction corrections based on a spectral template for each individual coadd object (\secref{photometry}). 
\item \gold includes an updated set of flags relative to \yonegold to indicate various measurement anomalies (\secref{flagsgold}).
\item \gold includes updated photometric redshifts produced with the \bpz \citep{Benitez:2000}, \dnf \citep{dnf}, and \ann \citep{annz2} algorithms (\secref{photozs}).
\item The catalog includes a flag to indicate whether a given object lies within the \gold footprint used for Y3 cosmology analyses, instead of making any kind of fixed selection over the extracted sources.
Accordingly, all objects from Y3 processing are included in \gold. This approach allows alternative footprint definitions needed for specific science cases (\secref{footprint}).
\item The survey masks are now separated into astrophysical foregrounds (e.g., bright stars and large nearby galaxies) and `bad' regions with recognized data processing issues (\secref{masks}).
\item \gold includes maps of survey properties, such as airmass, seeing, and sky brightness, generated from combinations of image-level measurements (\secref{surveyproperties} and \appref{sps}).
\end{itemize}

\subsubsection{Differences relative to \yonegold and DR1}

\begin{itemize}

\item \gold star-galaxy separation is performed using \mof and \sof quantities, as recommended in \citet{y1sgsep} (\secref{stargalaxy} and \appref{extended}).

\end{itemize}

\section{Data processing}
\label{sec:dataprocessing}

The DESDM framework processes raw data acquired by DECam and produces the calibrated images and catalogs used for science. In this section, we review the overall system, and refer readers to \citet{desdm} for a detailed description of the pipeline used for \gold.

\subsection{Single- and multi-epoch image processing}
\label{sec:imageproc}

Individual DECam exposures must be detrended for diverse instrumental signatures. 
This Single-Epoch processing stage produces calibrated images and catalogs, which are made available periodically at the NSF's National Optical-Infrared Astronomy Research Laboratory archive\footnote{\url{https://astroarchive.noao.edu/}}. The single-epoch calibrated images are the basis of the shape catalogs \citep*{y3-shapecatalog}, and are referenced later by the pipeline to fit the multi-epoch photometry for \mof, \sof and \metacal codes (see \secref{mofsof}).

The \gold data set is based on the imaging products from a subsequent stage, coaddition, which has a fainter object detection limit due to the combination of the single epoch images.  At the same time, the weight maps that are produced during the single-epoch processing are used to build depth maps using the \mangle \citep{mangle} software.  

\subsection{Catalog generation}
\label{sec:catalog_generation}

Base object detections are obtained using \sextractor with the settings described in \citet{desdm}, tuned for an efficient extraction of $\SNR\sim10$ objects from the $r+i+z$ coadd (or detection) images. For these objects we measure various quantities with several pipelines. The base catalog for \gold is identical to the DR1 catalog, i.e, approximately 399 million objects. However, the definition of the \gold footprint in \secref{footprint} removes $\sim 11$ million objects that lie in areas where $griz$ coverage criteria are not met. Multi-epoch image `postage stamps'  \citep[i.e., MEDS files;][]{Jarvis:2015} are created at this stage for each source, and used for a variety of purposes, including the multi-epoch fitting pipelines (\secref{mofsof}). 

\subsection{Single- and multi-object fitting pipeline on multi-epoch data}
\label{sec:mofsof}

\citet{y1gold} described the advantages of performing a multi-object, multi-epoch, multi-band fit (\mof) to the object shape to determine the morphology and flux, and we refer the reader to that paper for details of this process, based on the \ngmix software \citep{Sheldon:2014}. In \gold, we introduce a variant, called \sof, that simplifies the fitting process by eliminating the multi-object light-subtraction step, speeding up the processing time by a factor of a few, with negligible impact in performance (as shown in \secref{photozs}). In addition, \sof has fewer objects with fit failures.

Both \mof and \sof employ \ngmix to fit objects using reconstructed Point Spread Functions (PSFs) modeled as mixtures of three Gaussians at the coordinates in the MEDS files where an object was detected in the coadds. For each object, there are as many of these stamps as there are epochs and bands observed at those coordinates. The fitting is performed for several objects simultaneously, identified with a friends-of-friends algorithm. In a first step, a bulge-plus-disk model is fit to each object in the group separately (masking the pixels containing other objects). This way we obtain the single-object fit quantities (\sof). In subsequent iterations, we can subtract the flux from these Gaussian mixture models obtained from the neighbors for each particular source (multi-object fit, \mof). \ngmix-based photometry generally provides a tighter reconstruction with respect to \sextractor quantities. 

DES Y3 cosmology uses \metacal photometry (\citet*{zuntz}, \citet{SheldonHuff:2017, metacal}) for source galaxies in weak-lensing analyses, as described in \citet*{y3-shapecatalog}. Similarly to the \mof and \sof pipelines, the \metacal photometry is measured from all epochs and bands, but using a simplified Gaussian model for the PSF, and with an artificial shear applied to the images to obtain four different versions of the photometry. A fifth set of measurements corresponds to the un-sheared version (see \citet*{zuntz} for more details).

Photometric redshifts are generated from the \sof photometry within the DESDM pipeline, using the Bayesian photometric redshift code, BPZ \citep{Benitez:2000} that provides several point estimates and uncertainty estimates. The fluxes and magnitudes computed from the \sextractor- and \ngmix-based pipelines are also the basis for other photometric redshift estimates, including those from \dnf \citep{dnf}, \ann \citep{annz2}, and \sompz \citep*{y3-sompz} algorithms, which are used for various purposes in DES Y3 cosmology (see \secref{photozs}.  ).

\subsection{Deep Field data set creation}
\label{sec:deepfields}

As mentioned in \secref{strategy}, the SN Survey repeatedly observed 10 fields to identify transient phenomena that can later be ascribed to SNe Ia \citep{Kessler:2015}. Taking advantage of these data and in parallel to the \gold data set, \citet*{y3-deepfields} have constructed the Deep Field data set to complement and enhance the main survey in the Y3 cosmology analyses. Up to 90 images of the same patches of the sky have been stacked to achieve a depth of $griz =$ [\maglimdeepg, \maglimdeepr, \maglimdeepi, \maglimdeepz] respectively for $\SNR = 10$ in $2\asec$ diameter apertures. A modified pipeline with new software to handle the higher source density and addition of near-infrared images has resulted in an 8-band catalogue ($ugrizJHKs$) of 1.7 M objects over a total area of $\footprintareadeep \deg^{2}$, after artifact masking. The Deep Field processing and data products are fully described in \citet*{y3-deepfields}. One of its applications is the creation of the \balrog simulations \citep{y3-balrog}, in which Deep Field sources are injected into the Wide Survey coadd images in order to understand the DES transfer function, among other uses such as for checks of the \gold catalog (see \secref{depth}).

\section{Astrometric and Photometric Calibration}
\label{sec:astrophoto}

We describe here the astrometric and photometric calibration performance of the \gold catalog.
Details of the pixel-level instrument response model and associated astrometric and photometric calibrations are presented in \citet{Bernstein:2017a} and \citet{Bernstein:2017}. Description of the relative photometric calibration pipeline can be found in \citet{Y3FGCM}.

\subsection{Astrometric Solution}
\label{sec:astrometry}

DES Y3 astrometric solution was found via the \SCAMP software using 2MASS \citep{Skrutskie:2006} stars, as described in \citet{desdm}. As a first pass an initial solution is found per exposure. During coaddition, overlapping images of the same reference objects can be used to refine the solution, using simultaneously the whole catalog of objects from multiple exposures falling within that `tile' (DESDM's sky unit for coadd processing; $0.73\degree \times 0.73\degree$). 

\subsection{Astrometric performance}

\begin{figure*}
	\includegraphics[width=\textwidth]{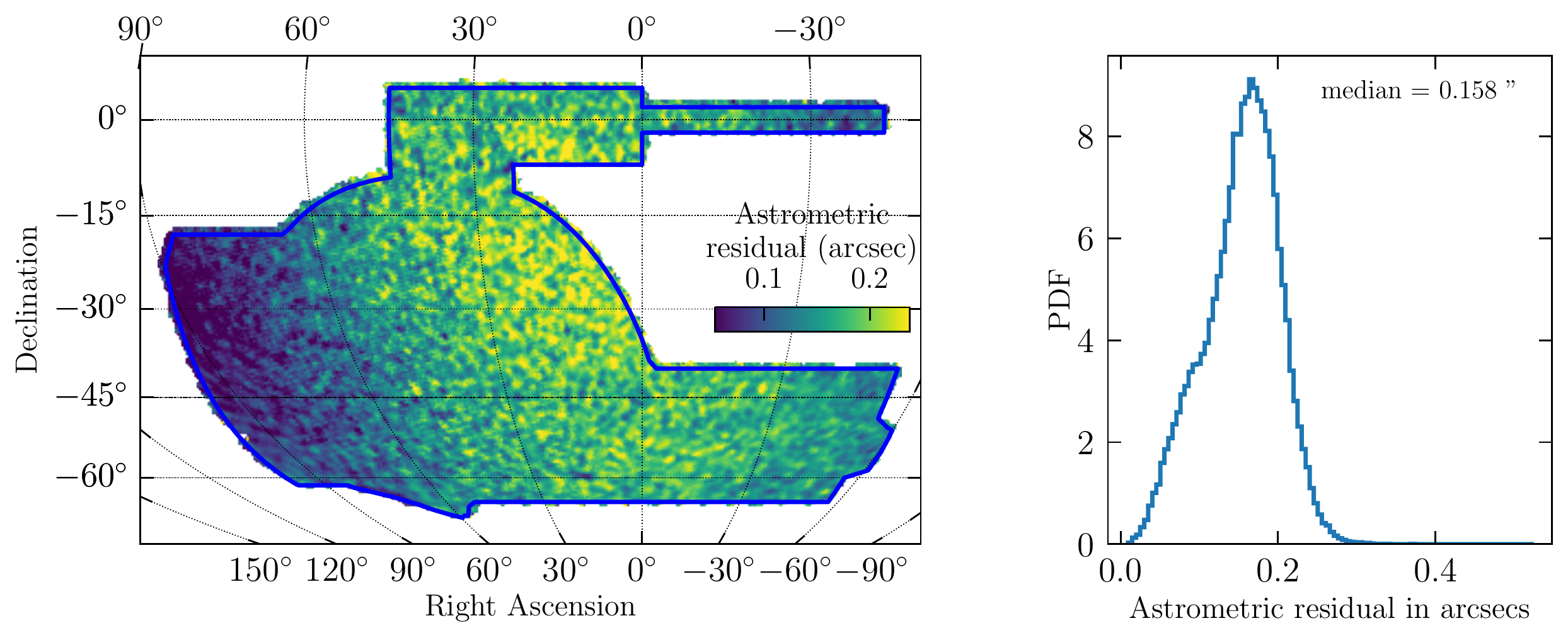}
    \caption{Astrometric residuals of \gold vs \gaia's DR2 objects, computed as the median value of the modulus of the displacement vectors between the matched stars of both catalogs.}
\label{fig:astrometry_vs_gaiadr2}
\end{figure*}

We present updated absolute astrometric accuracy measurements for the coadd catalog \citep{dr1} relative to the \gaia DR2 catalog \citep{gaiadr2} as an external reference. 
A $0.5 \asec$ matching radius is used against all \gaia's raw sources\footnote{ \url{http://cdn.gea.esac.esa.int/Gaia/gdr2/}}. The result of the comparison is shown in \figref{astrometry_vs_gaiadr2}. This analysis shows a median of \astroabs \mas between \gold and \gaia DR2 positions, with median of \astrorel \mas between reobservations by DES. 

A systematic trend, already noted in \citet{dr1}, is visible in the astrometric residuals across the survey footprint. This is at least in part due to the unaccounted proper motion effect from the 2MASS star catalog used as a reference in the solution. Celestial coordinate corrections can be obtained by fitting two 2D polynomials to the residual distribution in each coordinate, which are not included in the \gold catalog celestial position columns but will be made available upon public release of the \gold data set as separate coordinates for each object. This sub-arcsecond precision correction is estimated to be negligible for Y3 cosmology results. More recent DES processing uses the \gaia DR2 catalog as a reference (see \citealt{bernardinelli} for an example using \gaia DR1). Upon release of the \gold data set, solutions from the \texttt{WCSFit} software \citep{Bernstein:2017a} for sub-arcsecond corrections to astrometry will be made available.

\subsection{Photometric Calibration and Corrections}
\label{sec:photometry}

The \gold photometric calibration is based on the Forward Global Calibration Method (FGCM) introduced by \citet{Y3FGCM}. 
FGCM calibrates the entire survey using a forward modeling approach that incorporates atmospheric and instrumental behavior, obtaining the best-fit parameters of such a model, rather than performing a global minimization of the fluxes from detected stars with respect to a network of secondary standards \citep[the latter was done in \yonegold; see][]{y1gold}.

Our objective is to report top-of-the-Galaxy\footnote{``Top-of-the-Galaxy'' refers to the source spectrum incident at the Milky Way before reddening by interstellar dust. ``Top-of-the-atmosphere'' refers to the source spectrum incident at the top of the Earth's atmosphere after reddening by interstellar dust. The majority of faint halo stars detected by DES are expected to be located beyond most of the total dust column \citep{2008ApJ...673..864J}, and thus correcting the inferred top-of-the-atmosphere spectra assuming the full dust column provides a good approximation of their intrinsic spectra.} energy fluxes in $griz$ (AB magnitudes) corresponding to the particular spectral energy distribution (SED) of each individual coadd object as observed through the DES Y3A2 Standard Passband \citep{dr1} with a precision of several \mmag (Y3A2 being the internal release version). 
We aim to account for all photometric calibration effects possible and study their impact on photometric redshifts, and test the FGCM methodology for future applications that require \mmag precision \citep[e.g.,][]{desc_2018_srd}.
To achieve a sub-percent photometric calibration, we include chromatic corrections that account for differences in the system response that arise from observing objects with different SEDs through passbands that vary with environmental conditions and instrument coordinates \citep{Li:2016}.
Our implementation in \gold includes three steps: (1) zeropoints computed from FGCM fits for each and every CCD image for use in image coaddition and transient analyses,
(2) chromatic corrections corresponding to per-object SED templates, and (3) interstellar extinction corrections that optionally include the per-object SED-dependence.
We briefly describe the fiducial calibration steps here and provide the detailed formalism in \appref{photometric_calibration}.

Prior to coaddition, each CCD image was assigned an FGCM zeropoint assuming that the bright stars used as calibration sources share the SED of a spectrophotometric standard, specifically the G star C26202 \citep{Bohlin:2014}.
For \gold, we update the zeropoints incorporating Year 4 imaging for the purpose of calibration only. This was necessary due to bad sky conditions throughout the third year preventing adequate uniformity in the calibration, and faulty GPS data that spoiled the FGCM solutions. We additionally made improvements to aperture corrections, updates to DES Y3A2 Standard Bandpass, and other technical modifications to the FGCM procedure \citep{fgcmupd}. This correction corresponds to the \texttt{DELTA\_MAG\_Y4} quantities. These updates are possible without a complete coaddition thanks to the scheme described in \appref{photometric_calibration}.

Next, we associate a spectral template with each individual coadd object based on the preliminary coadd photometry. We use the \citet{Pickles:1998} stellar spectral library for high-confidence stars, and the COSMOS SED library \citep{Ilbert2009} for galaxies and ambiguous objects.
We fit SED templates together with preliminary photo-$z$ estimates.Using these spectral templates, we compute per-object chromatic corrections to obtain top-of-the-atmosphere calibrated fluxes as observed through DES Y3A2 Standard Bandpass. This correction corresponds to the \texttt{DELTA\_MAG\_CHROM} quantities. 

Finally, we calculate per-object SED-dependent interstellar extinction corrections using the same SED templates for several dust reddening maps, including those of \citet{sfd98}, \citet{planck13}, and \citet{lenz17}. Additional details are provided in \appref{extinction}. This correction corresponds to the \texttt{A\_SED} quantities.

While we have focused on chromatic corrections for precision photometry, we note that per-object SED templates could be used for other data processing steps that are sensitive to chromatic effects, such as PSF modeling and differential chromatic refraction \citep{meyers_2015_chromatic,eriksen_2018_chromatic,carlsten_2018_chromatic}.
The general procedure developed here may be applicable to other imaging surveys with increasingly stringent systematic error budgets, such those of Rubin Observatory and \textit{Euclid} \citep[e.g.,][]{galametz_2017_sed}.

\subsection{Photometric performance}
\label{sec:photometry_performance}

Summary statistics for the relative photometric calibration of \gold are reported in \tabref{summary}.
We refer the reader to \cite{dr1} for detailed information on the calibration of DES photometry to a physical (AB magnitude) scale, which was assessed via repeated observations of a CALSPEC standard star, C26202 \citep{Bohlin:2014}.

\begin{figure*}
	\includegraphics[width=\textwidth]{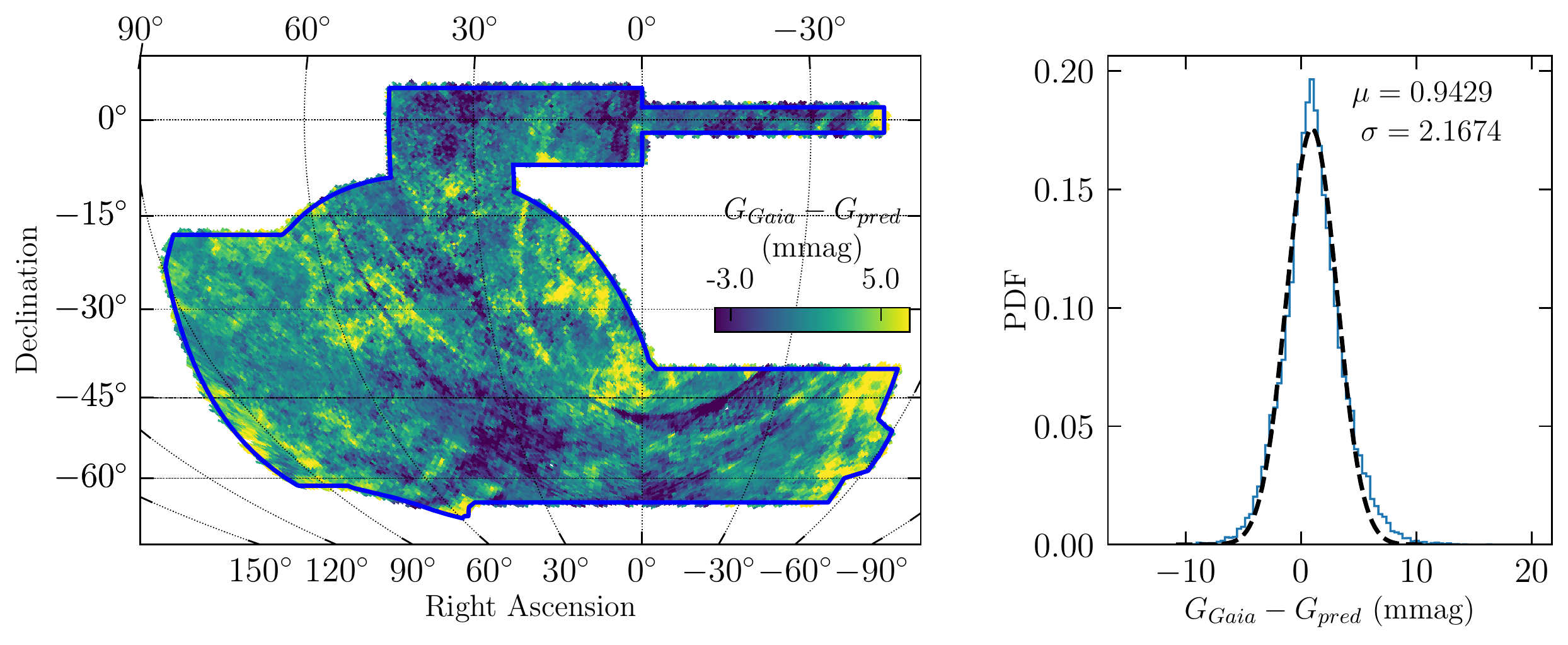}
    \caption{Photometric residuals of DES stars versus \gaia DR2 counterparts, transforming DES fluxes to \gaia's $G$-band (see footnote in \secref{photometry_performance}). Some of the arc-like spatially correlated residual features match the \gaia scanning pattern.}
\label{fig:photometry_vs_gaiadr2}
\end{figure*}

\figref{photometry_vs_gaiadr2} shows updated results for the top-of-the-atmosphere photometric uniformity measured against \gaia DR2. 
Relative to \citet{dr1}, use of Y4 zeropoints, improvements in the FGCM pipeline, and application of chromatic corrections have decreased the scatter of photometric residuals versus \gaia photometry (transformed to \gaia's $G$ band\footnote{This transformation is performed using a subset of common stars between \gaia and DES FGCM standards as a training set for a random forest which is built to transform stellar photometry from one system to the other. This transformation uses DES $gri$ magnitudes and colors as features in the training, and is valid for the interval $0.5 < g - i < 1.5$}) from $\sim7\mmag$ to $< 3 \mmag$. 
Importantly, DES and \gaia photometric calibration pipelines are completely independent from each other, implying that at least one of these surveys has photometric uniformity of better than $\sim 1.4 \mmag$.

The top-of-the-atmosphere calibration in \gold has reached a level of precision such that the treatment of interstellar extinction is now a limiting source of systematic uncertainty affecting the photometry of most DES objects.
Differences between varying prescriptions for interstellar extinction corrections are typically $\gtrsim10 \mmag$ for object colors, even in the low extinction regime that characterizes the DES footprint (see \appref{extinction} for details).
Whereas inclusion of additional overlapping exposures in the coadd tends to improve uniformity and average down differences between the observed passband and the Standard Bandpass, such that chromatic corrections are reduced, uncertainty in both the normalization of the dust opacity as well as chromatic effects of interstellar extinction persist.

When comparing the primary photometric methods for point-like (\sofpsfmag) and extended objects (\sofmag) for high-confidence stars, we find an average systematic offset in each of the $griz$ bands that varies at the level $0.02 \magn$ between bands. 
The \sofpsfmag agrees well with the \sextractor PSF photometry used by FGCM for photometric calibration of the survey in all of the $griz$ bands. 
Accordingly, there might be a systematic color offset for galaxies at the $0.02 \magn$ level. 
We do not expect this color offset to substantially affect photometric redshift methods that are trained and evaluated consistently, however, template-fitting methods might be impacted. 
Several dedicated studies have been performed to validate the \photoz distribution of samples used for DES Y3 cosmology.

\section{Depth}
\label{sec:depth}

The depth of \gold can be quantified using several approaches, as detailed in \citet{dr1}. 
Here, we focus on the effective depth obtained using the \sof photometry which is unique to \gold, and on measurements of the detection completeness of the galaxy population.

\subsection{Depth from \sof photometry for galaxy analyses} 
In order to have a more accurate description of $\SNR = 10$ depth for galaxy photometry (using the \sof model magnitudes), we follow the procedure described in \citet{Rykoff:2015} in which a model is trained on a coarse \nside = 1024 \healpix pixelization using several survey properties as features. The depth within the coarse pixels is estimated by fitting the magnitude versus magnitude error function.
This fit is done only for extended objects with $\sofmash > 1$ (with a median $\sim1.9$ pixel semi-major axis length) to capture the depth for galaxy-like sources. The model is then applied to pixels in default survey property map resolution (\nside = 4096) to produce the reference depth map for that photometry.

This results in the following estimates, again taking the median of the histogram distribution: $griz =$ [$\maglimsofgapp$, $\maglimsofrapp$,  $\maglimsofiapp$, $\maglimsofzapp$]. These values are a more accurate representation of the galaxy photometry since a selection of galaxies with good properties is used to obtain the magnitude limit estimates. These depth estimates include chromatic corrections and the extinction model described in \secref{photometry}. For comparison, the \mof $\SNR = 10$ depth from \yonegold in $griz =$ [$\maglimmofyoneg$, $\maglimmofyoner$, $\maglimmofyonei$, $\maglimmofyonez$].

\subsection{Detection completeness}

An alternative to the signal-to-noise threshold depth measurement is characterization of the object detection completeness as a function of magnitude. We use the Hyper Suprime Camera Subaru Strategic Program Data Release 2 \citep[HSC-SSP DR2,][]{hscdr2}, which reaches a depth of $i\sim26.2$ for point sources at $\SNR = 5$, for the Wide Survey, significantly deeper than the DES Year 3 data set. For \gold data, we have at our disposal additional techniques that can be used as crosscheck, namely deeper observations with DECam through Deep Fields processing \citep*{y3-deepfields} and the \balrog simulations \citep{y3-balrog}.

We use as a common mask for both data sets the \gold footprint and foregrounds mask, defined in \secref{masks} coupled with the HSC star masks from the latest iteration (S18A\footnote{\url{https://hsc-release.mtk.nao.ac.jp/doc/index.php/bright-star-masks-2/}}). Similarly, we combine the \gold masks with the deep field data sets, which incorporates its own set of masks. We perform a $0.5\asec$ matching between each catalog and \gold in this region with these constraints. The \balrog data set contains its own self-matching with similar characteristics, and the catalog already includes objects that have been detected with a flag. 

\begin{figure*}
    \centering
    \subfigure[SN-X3]{\includegraphics[width=0.9\columnwidth]{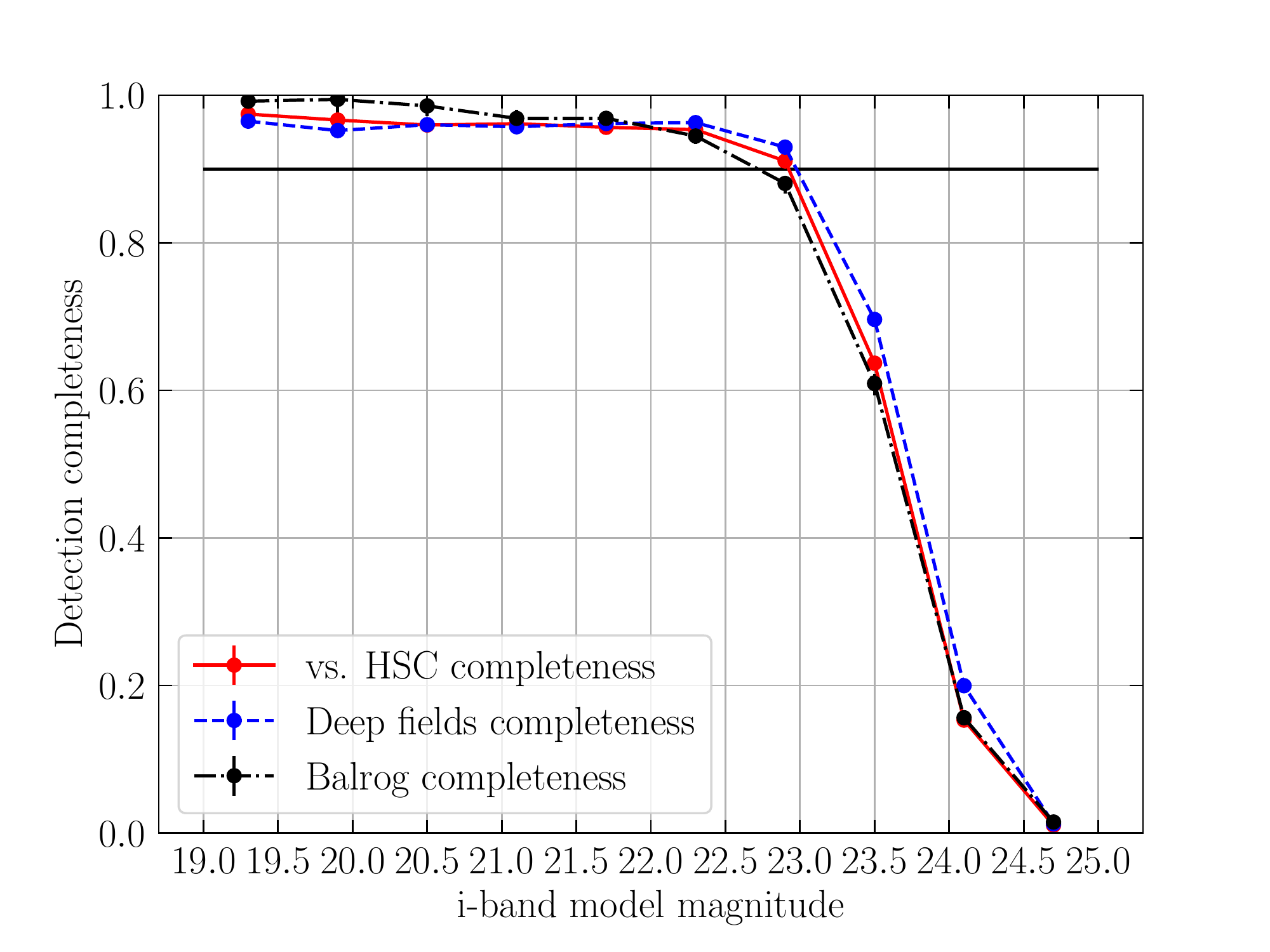}\label{fig:completeness_snx3}}
    \subfigure[HSC-W05]{\includegraphics[width=0.9\columnwidth]{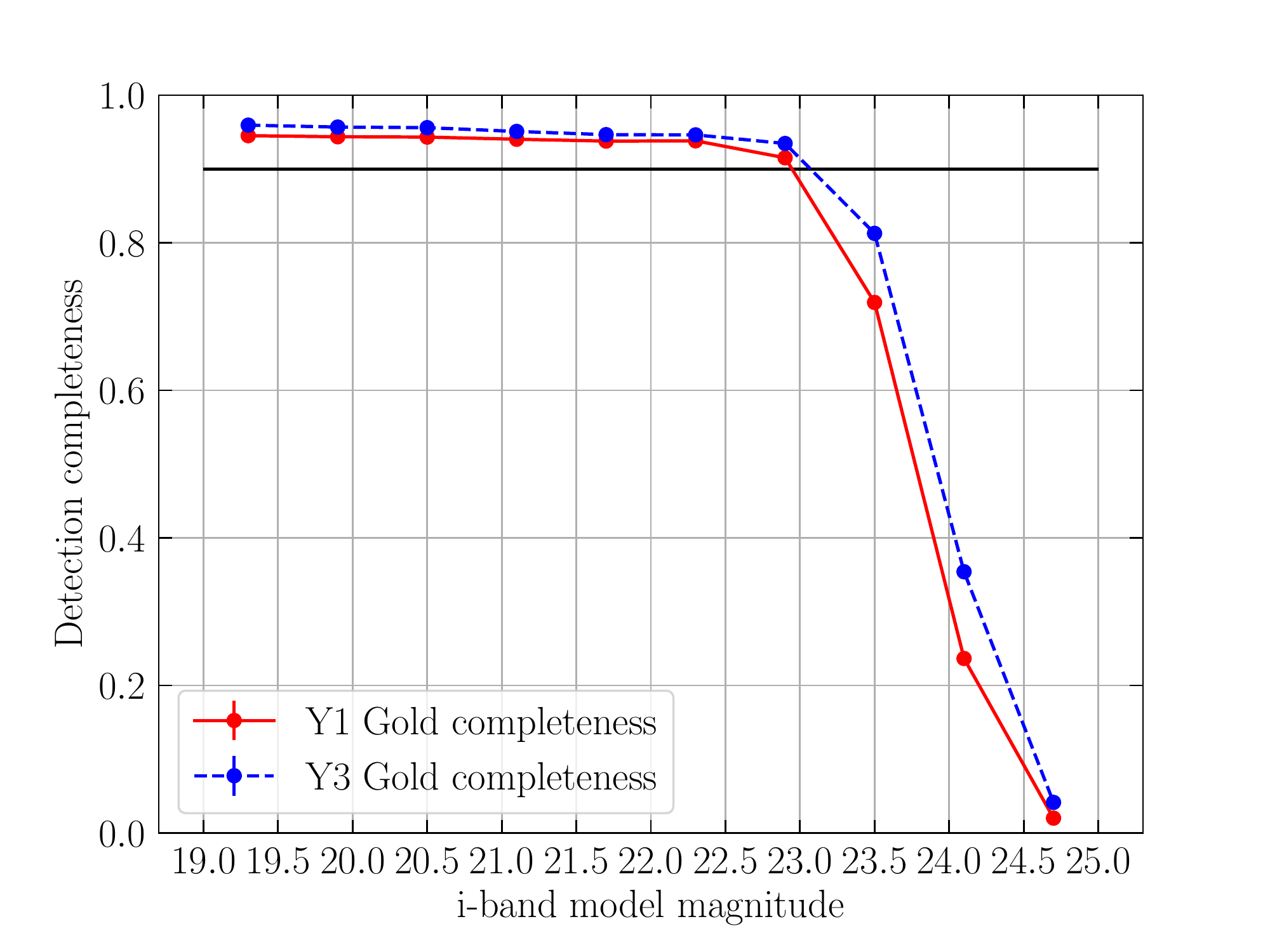}\label{fig:completeness_w05}}
   \caption{(left) Detection completeness for extended objects in the SN-X3 field ($\alpha, \delta \sim 36.5^{\circ}, -4.5^{\circ}$; approximately $3 \deg^2$ in area) in the $i$ band, comparing three methodologies: using a deeper external data set (HSC-SSP DR2, $\sim 360$ thousand matches to \gold), the Deep Field catalog in that region, and simulations from \balrog processing, which inject realistic images onto coadded Y3 images. Similar agreement is obtained in other bands; (right) Detection completeness for extended objects in the HSC-SSP W05 field ($\alpha \sim 330^{\circ} - 360^{\circ},\delta \sim 0^{\circ}$; approximately $90 \deg^2$ in area of overlap) in the $i$ band, comparing \yonegold and \gold versus the wide field HSC-SSP DR2 data set ($\sim 4.9$ million matches to \gold). NB that this region is wider and more representative of \gold than the comparison shown in \figref{completeness_snx3} and is 0.23 magnitudes deeper. Errors are $95\%$ containment errors computed using a Bayesian approach for efficiencies as detailed in \citet{Paterno:2004cb} but cannot be visualized as they are small compared to the data markers themselves. The black solid horizontal line represents the $90\%$ level for visual reference.}
   \label{fig:completeness}
\end{figure*}

Results (for extended objects) are shown in \figref{completeness_snx3}, where we see good agreement between the various methodologies, and \figref{completeness_w05} where a comparison with \yonegold completeness is shown. We note that the completeness presented here is different from that computed in \citet{Kessler:2019}: completeness for the SNIa cosmology analysis was computed as a function of supernova peak $i$ band magnitude rather than as a function of its host galaxy magnitude (Figure 4 of \citet{Kessler:2019}).

\subsection{Stellar obscuration}

\begin{figure*}
    \centering
	\subfigure[SN-X3]{\includegraphics[width=0.9\columnwidth]{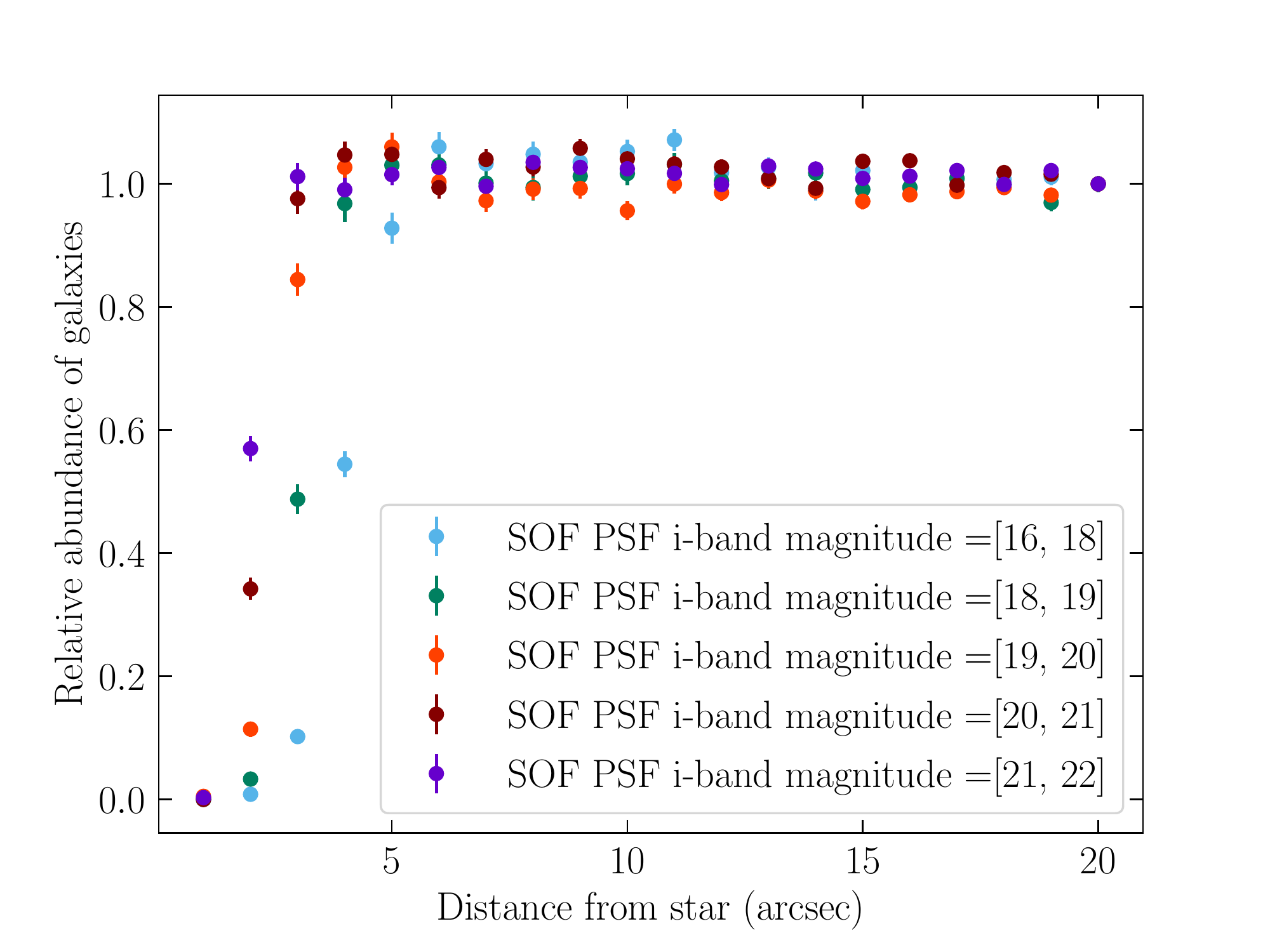}\label{fig:stellar_obscuration}}
	\subfigure[Near Galactic plane]{\includegraphics[width=0.9\columnwidth]{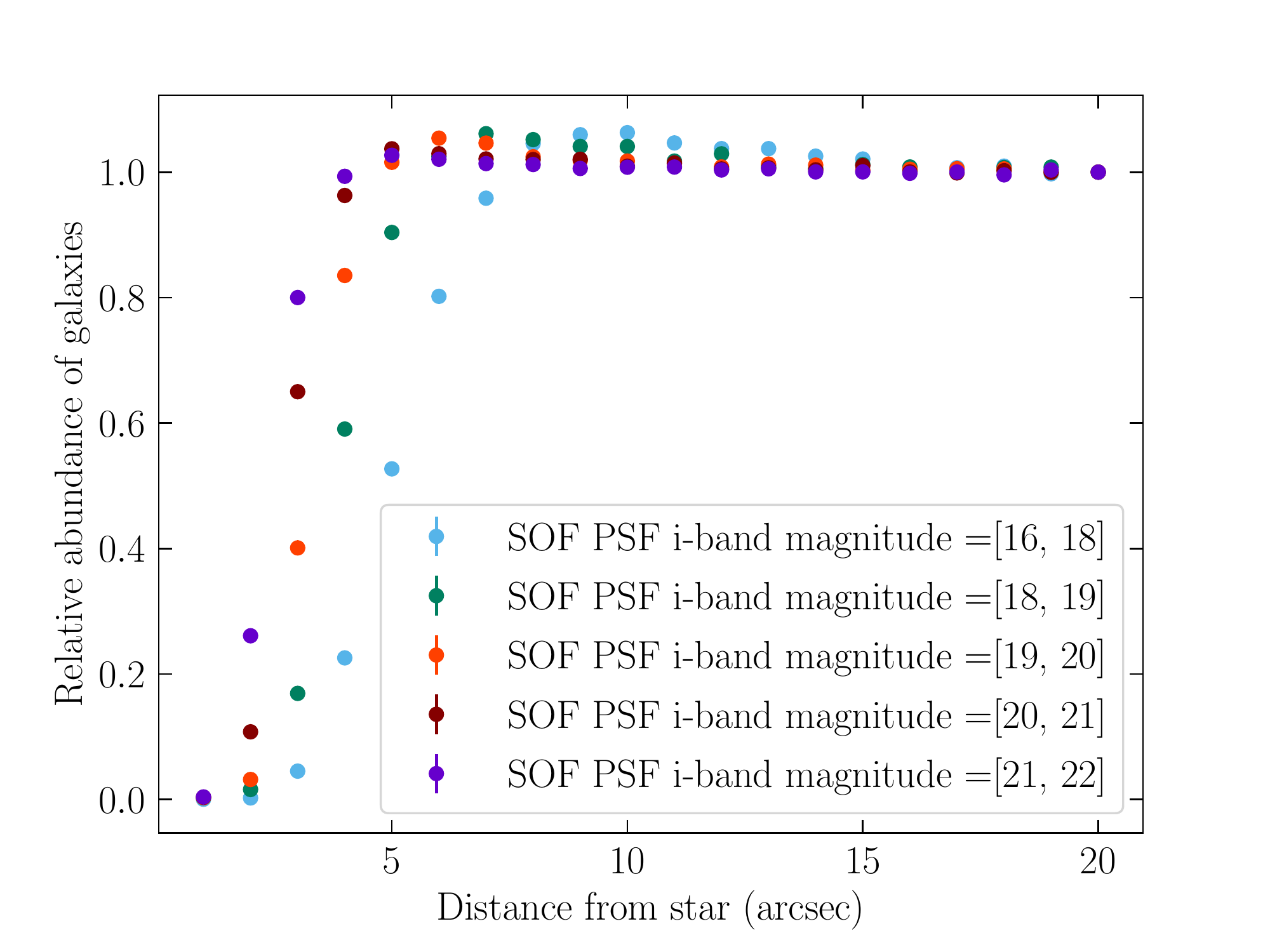}\label{fig:stellar_obscuration_f300}}
    \caption{(left) Stellar obscuration in the SN-X3 field as a function of distance from the star, expressed as the deficit of galaxy density with respect to the density of galaxies at 20 \asec, binned within several intervals of \sofpsfmag. (right) Same in a region at $b$ = [-38, -31] degrees, closer to the Galactic plane. The obscuration effect is slightly larger in radius, due to overlap with other obscuring stars.}
    \label{fig:stellar_obscuration_vvds}
\end{figure*}

\citet{ross11} noted the effect of obscuration around point sources as a systematic effect for clustering, and quantified the impact through measurements of the underdensity of galaxies around these sources. A similar measurement for DES has been done around VVDS sources and is shown in \figref{stellar_obscuration_vvds}, and for a region closer to the Galactic plane in \figref{stellar_obscuration_f300}.  The obscuring radius is slightly larger in the case of the field near the plane, which will impact the galaxy distribution, and is addressed using correction weighting, as developed in \citet{y3-galaxyclustering} and \citet{elvinpoole}. Alternatively, or in addition to this approach, a mask can be built around stellar objects (\sofmash = 0, $i<20$) with a 5 \asec radius to remove them together with the exclusion radius found here. 

\section{Object Characterization}
\label{sec:characterization}

In this section, we report on several additional flags and labels computed for each object in the catalog.

\subsection{Object classification}
\label{sec:stargalaxy}

The \mof and \sof pipelines provide a better measurement of the extension of a given object, as compared with coadd quantities, as shown in \citet{y1sgsep}, where 
insufficient modeling of PSF variations and discontinuities can have an important impact in the selection of objects with similar characteristics even when close to each other.

For \gold we have created a set of \var{EXTENDED\_CLASS} classifiers that group objects according to their consistency with a point-like morphology, with a higher value corresponding to more spatially extended shapes (details in \appref{extended}). 
Here we summarize the performance, including completeness and purity characterization for stars and galaxies, for our default  classifier \sofmash. 

\begin{figure}
	\includegraphics[width=\columnwidth]{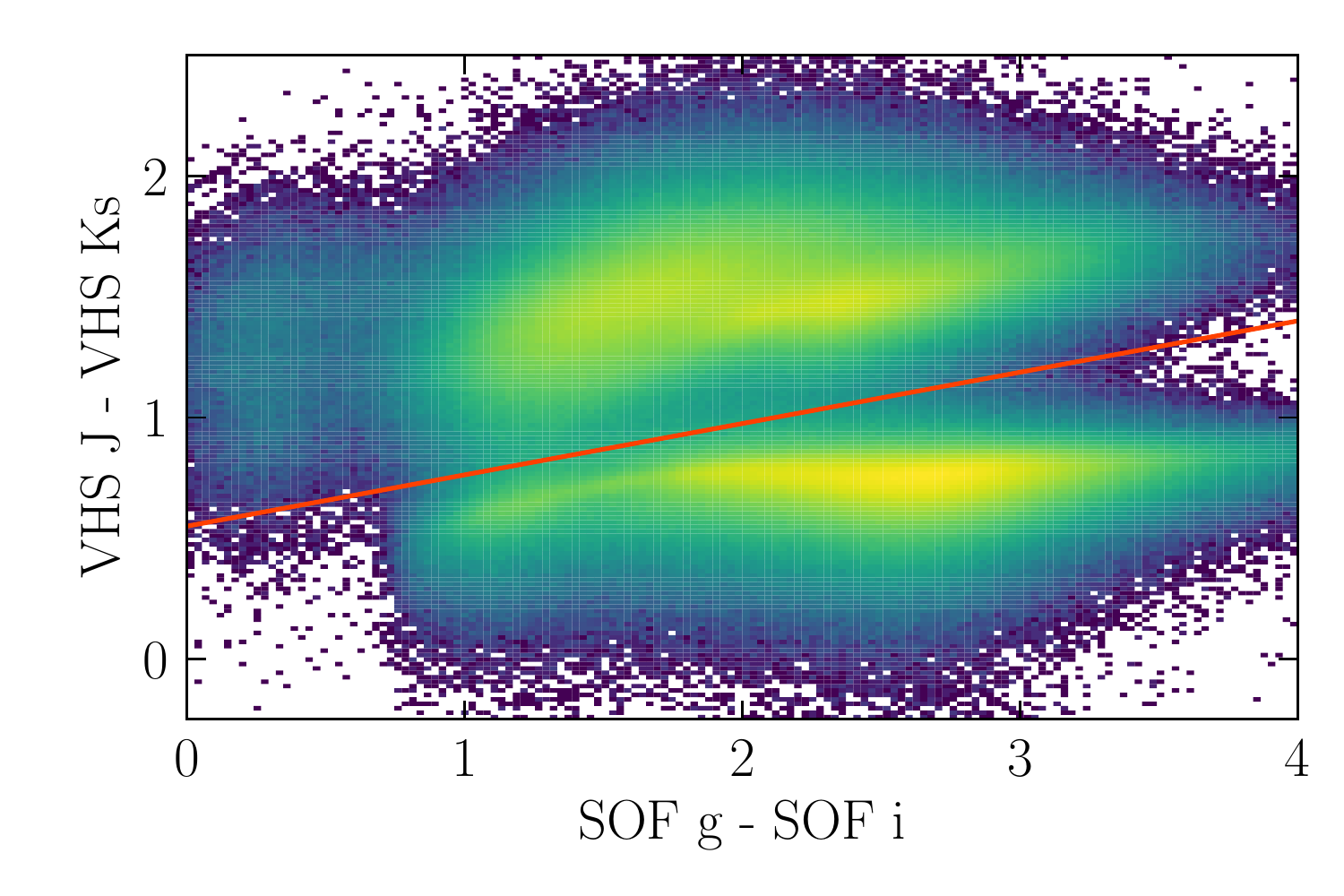}
    \caption{Optical DES vs infrared VHS color distribution for star classification. The objects above the dividing line have galaxy or QSO spectra. This color-based `truth' classification for Galactic and extragalactic populations is possible where VHS data are available and matched to DES sources, and is therefore limited to approximately $i<21$.}
\label{fig:ir_optical_classification}
\end{figure}

We can validate the bright end of the classification using additional infrared data from VHS \citep{vhs} as demonstrated in \citet{baldry} and \citet{y1sgsep}. We perform a 0.5\asec astrometric match to overlapping VHS sources, and define a stellar vs non-stellar classification based on $g-i$ DES optical color versus $J-Ks$ VHS color. The result of such separation in this space is seen in \figref{ir_optical_classification}.

\begin{figure*}
    \centering
	\subfigure[Truth from VHS colors]{\includegraphics[width=0.9\columnwidth]{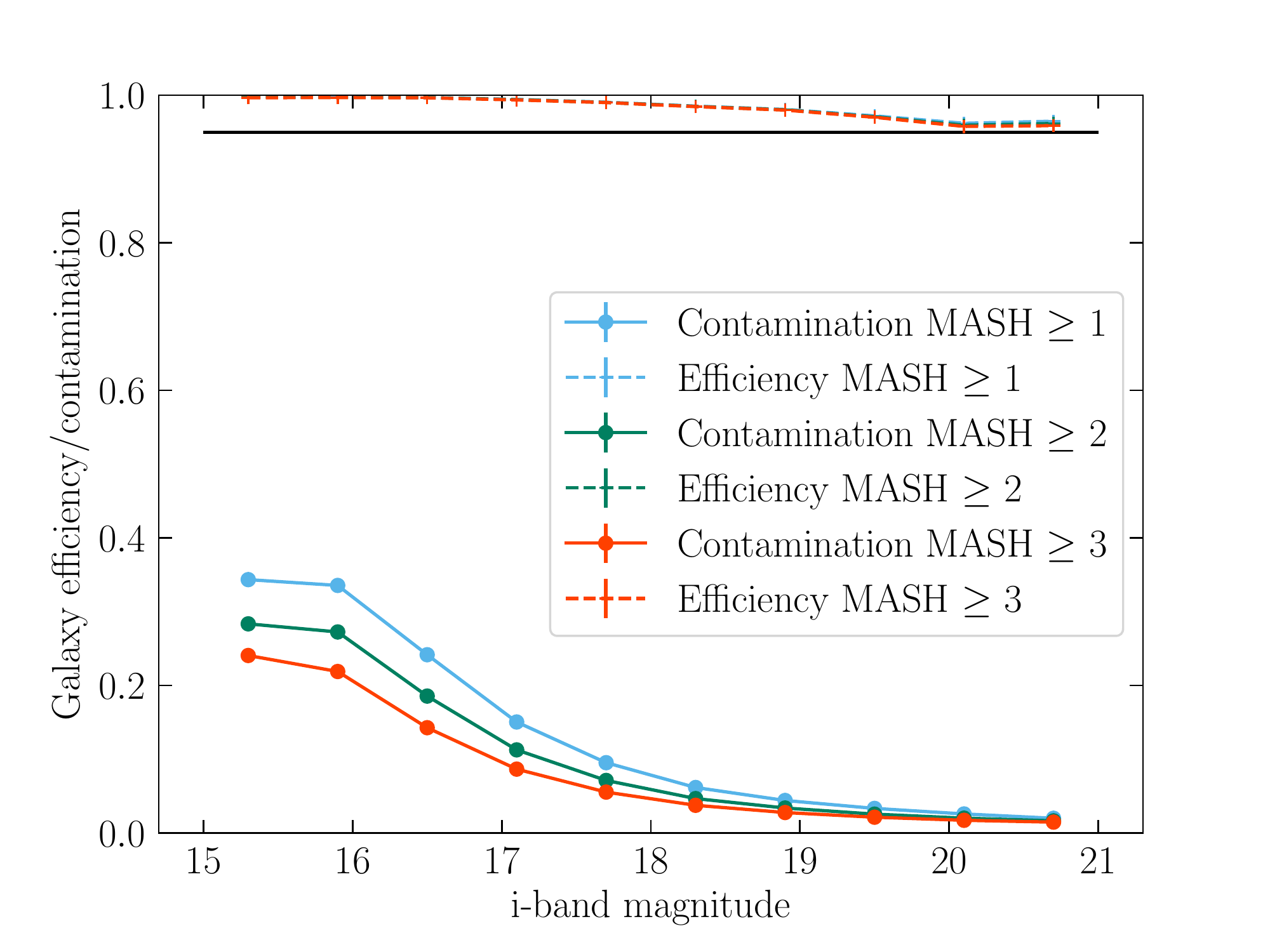}\label{fig:effpur_vs_vhs}}
	\subfigure[Truth from HSC morphology]{\includegraphics[width=0.9\columnwidth]{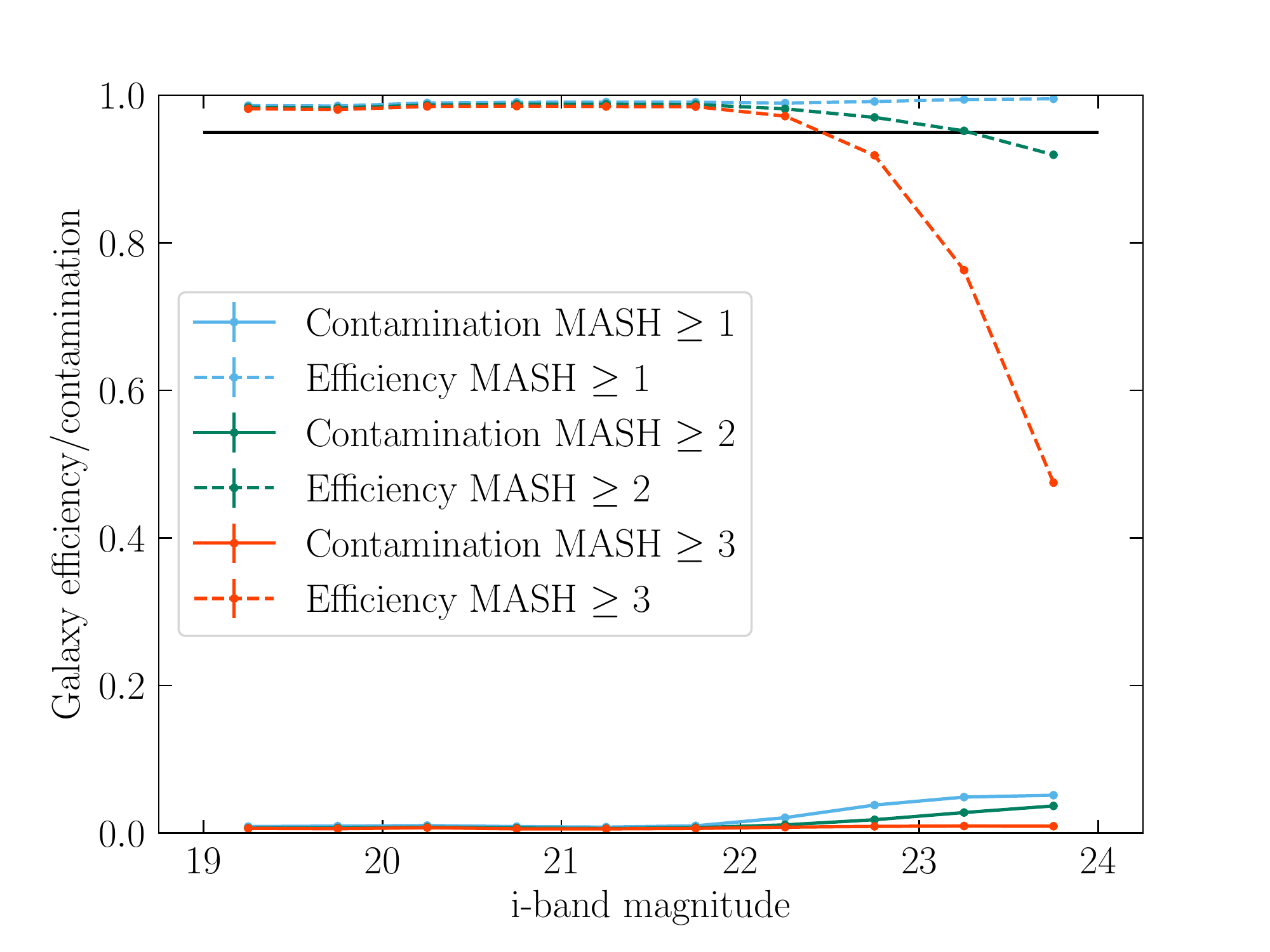}\label{fig:effpur_vs_hsc_dr2}}
    \caption{(left) Efficiency and contamination of point sources in different extended source samples using the VHS color selection as a reference. A large contamination at the bright end can be avoided by using infrared color selection and/or faint magnitude selection when the former is absent; (right) same measurements using the HSC-SSP DR2 catalog as a reference (see \secref{catalog_generation}). Errors are $95\%$ containment errors computed using a Bayesian approach for efficiencies as detailed in \citet{Paterno:2004cb}, and cannot be visualized at this scale as they are $\sim 0.3\%.$}
    \label{fig:effpur}
\end{figure*}

Using this clean color-based classification scheme as a `truth' reference, it is possible to evaluate the quality of the \sofmash classification at the bright end of the magnitude distribution (approximately from 15 to 21 in the $i$ band, where a significant number of matched VHS objects are available). From this comparison, we can deduce two useful performance indicators for galaxy samples that are relevant for cosmology analyses: purity (also called precision or positive predictive value) or equivalently, contamination as (1 - purity), and efficiency (also called completeness, or true positive rate). \figref{effpur_vs_vhs} shows these results for a match to the VHS catalog over the overlapping footprint. 
Near the saturation threshold of DES, we see that up to $\sim30\%$ of objects classified as morphologically extended have colors that are more consistent with being stars. 
Some fraction of these objects might be double stars, and should be eliminated from galaxy samples.
The galaxy samples used for DES Y3 cosmology do not include this population of bright objects due to flux and/or color selections. 

We can also use deeper surveys with good seeing and/or space-based imaging to provide a morphological reference to validate the star-galaxy classifier for fainter objects. We used the HSC-SSP DR2 catalog \citep{hscdr1} in W02, overlapping the SN-X3 field.  In \figref{effpur_vs_hsc_dr2} we show the efficiency and purity of an extended source sample versus point sources. In this measurement, we estimate a $2\%$ systematic error at fainter magnitudes due to classification errors in the reference catalog itself (as compared with space imaging). 

From these figures, we can estimate that the galaxy sample as defined by $\sofmash = 3$ in the range of $i=[19,22.5]$ will contain a contamination smaller than $2\%$. This range for example will contain most of the lens sample for the DES cosmology analyses. 

\begin{figure}
	\includegraphics[width=\columnwidth]{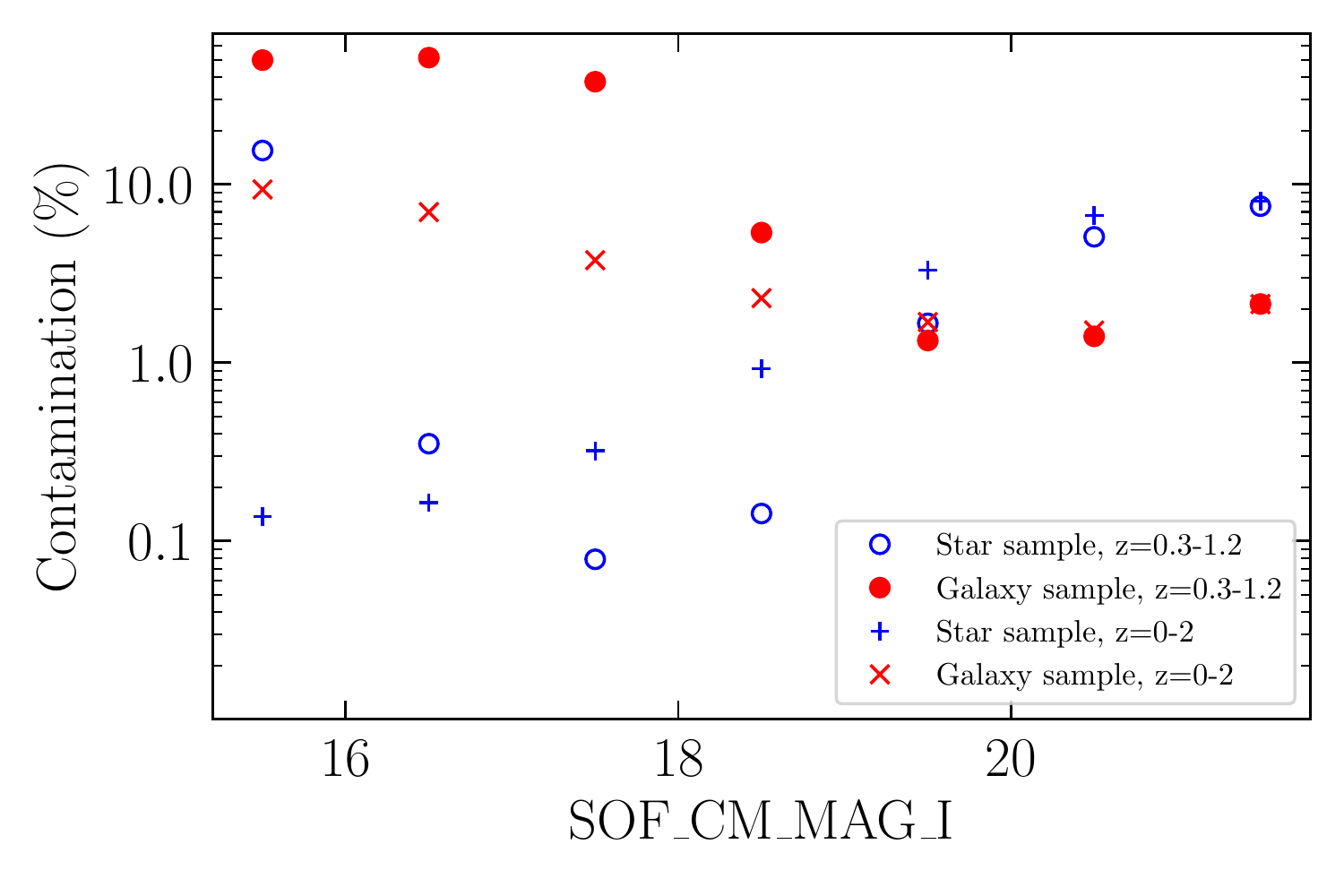}
    \caption{Impurity level in stellar and galaxy samples as classified by \sofmash ( = 0 or = 3 respectively) for two photometric redshift ranges defined by \dnf. Every object in the \gold sample has a photometric redshift computed for them, including stars.}
\label{fig:contamination}
\end{figure}

In \figref{contamination} we show the contamination level (1-purity) for the stellar and galaxy samples. The photometric redshift range considered is very important to consider when accounting for contamination from a stellar or extragalactic component. Stars will have a photometric redshift assigned as well and tend to accumulate at photo-$z\gtrsim0.5$. This can impact the galaxy sample specifically at bright magnitudes, where the true star to galaxy ratio is higher, in this moderate photo-$z$ range \citep[see][]{baosample}. The apparent extendedness of the contaminating stellar population is likely attributed to double stars in many cases. For the cosmology studies showcased in \citet{desy1cosmo} and Y3 analyses, the `source' and `lens' samples avoid this contamination through specific shape measurement codes and by removing bright objects, respectively. If one is interested in this bright regime however, additional color constraints or more sophisticated shape selections can help separate extended sources and double stars that have been merged into a single detection. 

The \sofmash classifier was built for general application based on the best options studied in  \citet{y1sgsep}. Analyses in regions around foreground objects (such as globular clusters or the Magellanic Clouds) would test the performance of these morphological classifiers and/or build improved classifications with color information \citep{y1sgsep}.

\subsection{Object quality flag: \flagsgold}
\label{sec:flagsgold}

We use \flagsgold to indicate unusual characteristics of individual objects, including fitting failures and measurement anomalies.
Flagged objects can be excluded as appropriate for a given analysis using bitwise operations.
See \tabref{flagsgold} for a description of the various bits available per object.

\begin{deluxetable*}{c c l}
\tablewidth{0pt}
\tabletypesize{\tablesize}
\tablecaption{ \gold \flagsgold bit flag variable \label{tab:flagsgold}}
\tablehead{
\colhead{Flag Bit} & \colhead{Number of objects affected} & \colhead{Description} 
}
\startdata
1 & 14185334 & \var{MOF\_FLAGS} != 0 or \var{MOF\_FLAGS} = NULL, flag raised by \MOF processing \\
2 & 6555347 & $\var{SOF\_FLAGS} != 0$, flag raised by \SOF processing \\
4 & 1532648 & $\var{SOF\_FLAGS == 1}$ or $\var{SOF\_FLAGS > 2}$, flags  for PSF fit failures \\
8 & 746568  & Any SExtractor $\var{FLAGS\_[GRIZ]} > 3$ \\
16 & 3091171 & Any of $\var{IMAFLAGS\_ISO\_[GRIZ]} != 0$. \tablenotemark{$\dagger$} \\
32 & 152999 & Bright blue artifacts in the images \\
64 & 62653 & Bright objects with unphysical colors, possible transients \\
\enddata
\tablenotetext{\dagger}{The $\var{IMAFLAGS\_ISO}$ flag is set as an \texttt{OR} condition among the multiple pixels on multiple epochs composing the image, regarding a processing flag being set, according to the definition in \citet{desdm}.}
\end{deluxetable*}

\subsection{Photometric redshifts}
\label{sec:photozs}

Three standard photometric redshift codes were run on \mof and \sof photometries of \gold objects ($griz$).  We provide in this section a description of the estimates provided with \gold, together with figures of merit describing their performance against an extensive collection of spectra, described in \citet{Gschwend:2018}. The reference catalog includes $\sim 2.2\times10^5$ spectra matched to DES objects from 24 different spectroscopic catalogs, most notably SDSS DR14 \citep{sdssdr14}, DES's own follow-up through the OzDES program \citep{ozdes}, and VIPERS \citep{VIPERS:2014}. Half of the spectra have been used for training the machine learning methods, and the other half for the tests shown in this section. In all cases, point estimates and probability distribution functions of the samples can be computed. It is important to remark that cosmology analyses making use of \gold will often employ other approaches more suitable for the task at hand (see, e.g., \citet*{y3-sompz}), with their own set of validation procedures. In this work, we present three approaches available in \gold, with a measurement of bias and dispersion as a function of spectroscopic redshift using an extensive spectroscopic catalog.

\subsubsection{Bayesian Photometric Redshifts, BPZ}

The \bpz code uses a template-fitting approach where a collection of galaxy SEDs are fit to the measured fluxes. The original code from \citet{Benitez:2000} has been modified for efficiency of execution as described in \citet*{y1photoz} and incorporated into the DESDM system.

\bpz has the capability of providing estimates for the redshifts from our knowledge of galaxy spectra, up to high redshifts if needed, by modeling adequately their spectral evolution, thereby alleviating the need for expensive and biased measurements of spectroscopic sources for training sets. On the other hand, biases can arise in this case from inadequate calibration or incompleteness of the template base itself.

\subsubsection{Directional Neighborhood Fitting photometric redshifts, DNF}

\dnf \citep{dnf}, creates an approximation of the redshift of objects through a nearest-neighbors fit of a hyperplane in color and magnitude space using a reference, or training set, from a large spectroscopic database.

\dnf also provides a second redshift estimation as the nearest-neighbor in the reference sample.  This second estimation allows the method to replicate the science sample photo-$z$ distribution, $N(z)$,under the assumption of training sample representativeness (see \citet*{hartley2020impact} for an exploration of this fact in machine learning codes). Galaxies without close references in the training sample are tagged in this code.

This kind of solution offers an automatic incorporation of the subtleties of the photometric behavior of the system. In addition, some of the degeneracies in the photometry-redshift association can be detected as large differences between the two photo-z's provided by the method. However the lack of representativeness of the training set as mentioned above is one of the major drawbacks of this kind of methodologies.

\subsubsection{Machine Learning methods for photometric redshifts, ANNz2}

\texttt{ANNz2} \citep{annz2} provides an alternative training--based photo-$z$ estimate. \texttt{ANNz2} is an updated version of the neural network code \texttt{ANNz} \citep{annz}, and it differs from the latter by using several additional machine learning methods beyond Artificial Neural Networks (ANNs), such as Boosted Decision Trees (BDTs) and $k$-Nearest Neighbours ($k$NN) algorithms. 

For the Y3 Gold photo-$z$ catalog, \texttt{ANNz2} was run in randomized regression mode with 50 BDTs, using the same spectroscopic sample utilized for \dnf, randomly split into training, and validation and testing sets. 
The estimate provided in the catalog results from the BDT with the best performance on the testing sample.
The uncertainty is estimated through a $k$NN method, which takes into account the distance of galaxies in the Y3 sample from training galaxies in color--space. As with DNF,  an incomplete training set can introduce biases that need to be calibrated or accounted for.

\subsubsection{Photo-$z$ Performance metrics}

\begin{figure}
	\includegraphics[width=\columnwidth]{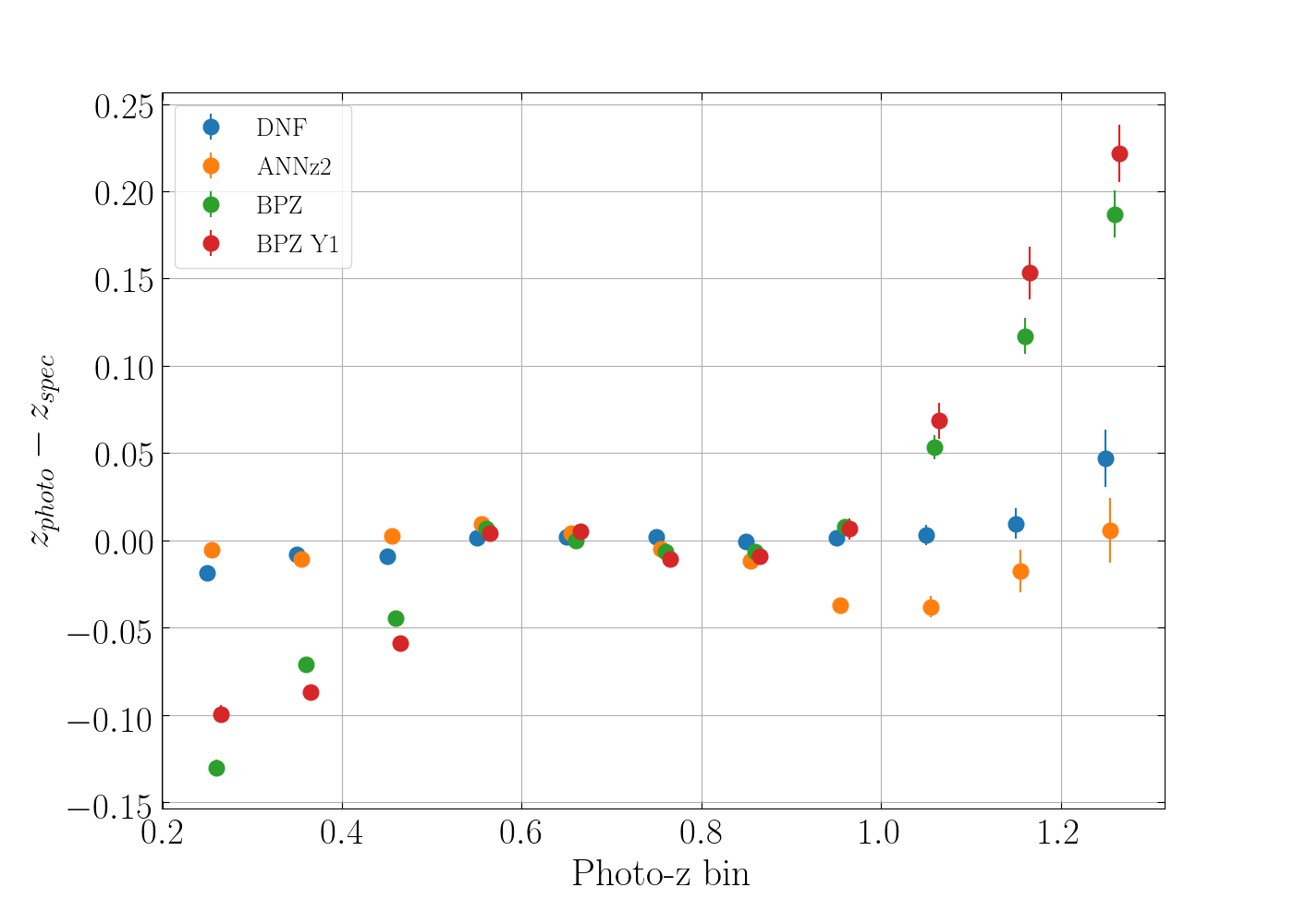}
	\includegraphics[width=\columnwidth]{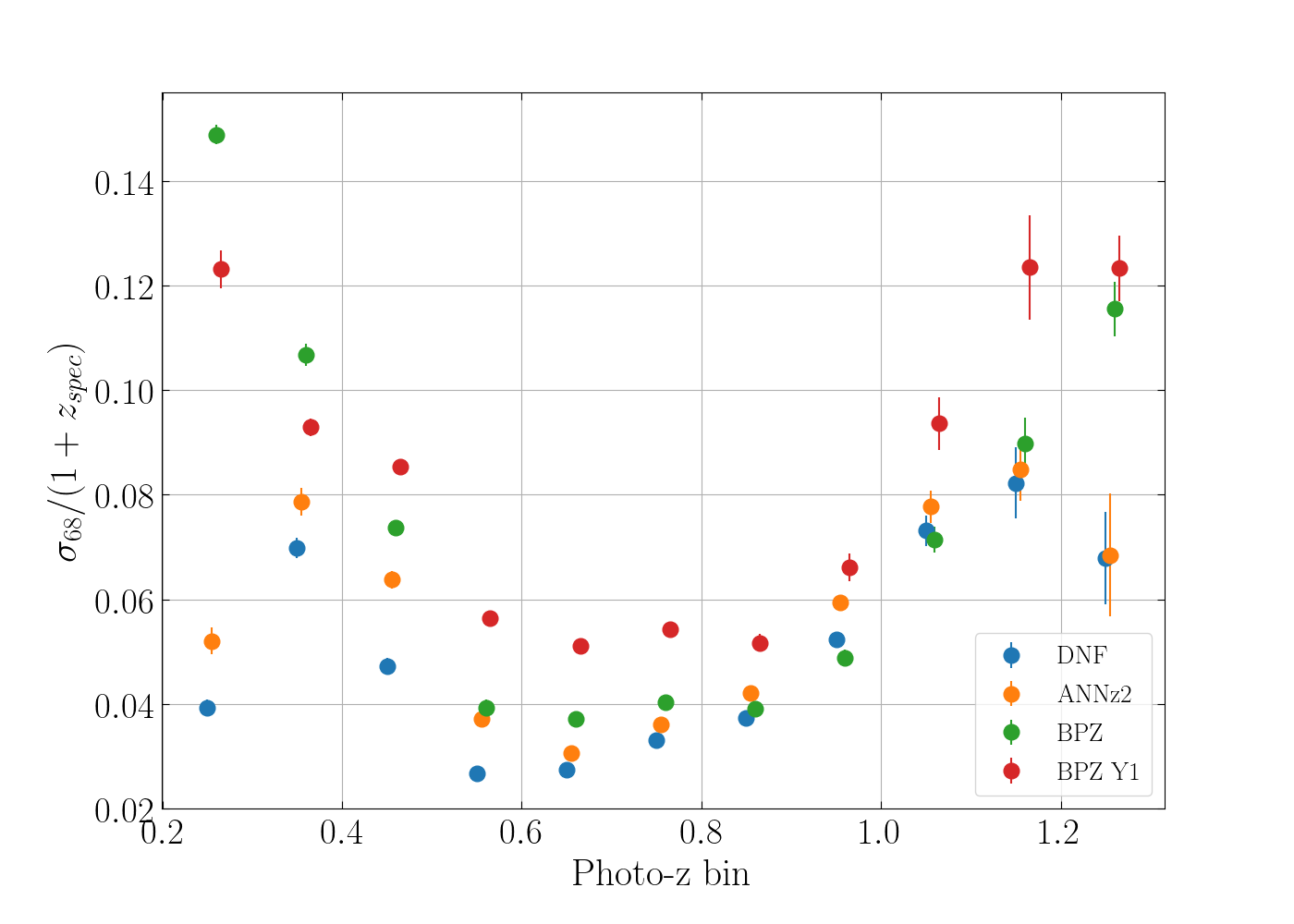}
    \caption{The residual of the point photometric redshift estimates (top) and the $68\%$ quantile error using a test sample from a collected spectroscopic catalog \citep{Gschwend:2018}, from \bpz, \dnf and \ann (as well \yonegold \bpz). The lack of $u$-band limits precision at low redshifts. The training set is common between both machine learning photo-z codes. NB, that although the test spectroscopic sample is quite extensive, collecting more than 100000 spectra, it is not a statistically representative sample of \gold.}
\label{fig:metricspz}
\end{figure}

\begin{figure}
	\includegraphics[width=\columnwidth]{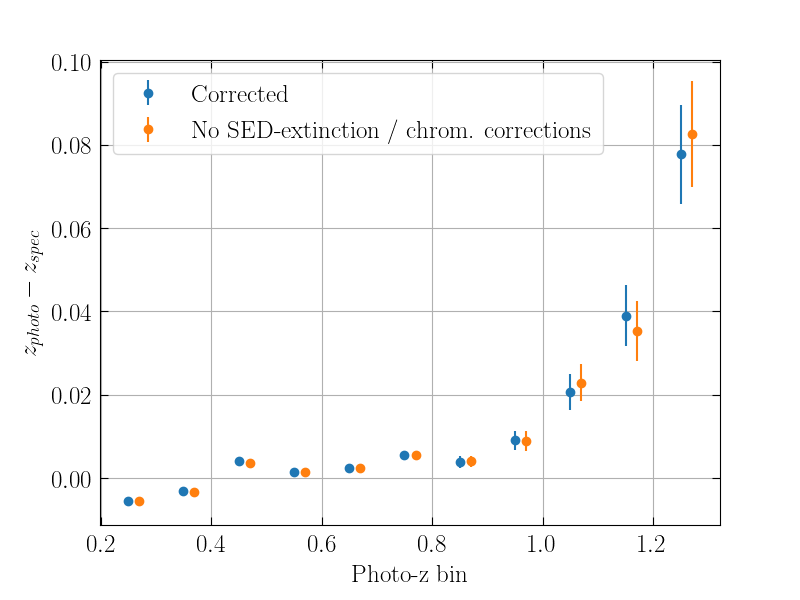}
	\includegraphics[width=\columnwidth]{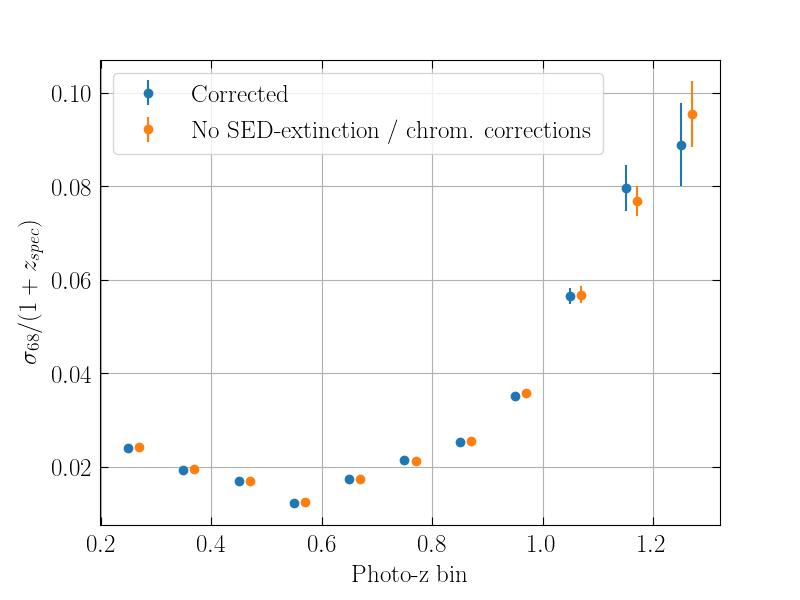}
    \caption{The residual of the point photometric redshift estimates (top) and the $68\%$ quantile error over (1+z) (bottom) for \dnf, with and without chromatic corrections. It can be seen that the impact in performance is negligible.}
\label{fig:metricspz_cc}
\end{figure}

As a standard check on the performance of these photo-$z$ codes, we present some quality metrics against the spectroscopic data set compiled as described in \citet{Gschwend:2018}. \figref{metricspz} shows that the reconstructed estimation of the redshift is, in general, more accurate with \dnf. At low redshifts ($z<0.5$), the \bpz run available in \gold shows poor performance, due to the adaptation of templates for better performance at high redshifts. In addition, we encounter some difficulties related to the lack of $u$-band to break some degeneracies among galaxy types and redshift at $z\sim0.4$. We also show in \figref{metricspz_cc} that the impact of incorporating the chromatic corrections to photometric calibrations is negligible. 

The point photo-$z$ estimates shown here are mainly used for binning galaxy samples. In Y3 cosmology, this is the done for the magnitude limited sample \citep{y3-2x2maglimforecast} and the BAO sample \citep{y3-baosample}. \redmagic uses its own point estimate from the red-sequence template fitting (with a similar performance as DNF for those galaxies) as described in \citet*{Rozo:2016}. The fiducial binning and redshift distributions of the source sample for the combined weak lensing and large scale structure analysis are described and validated in \citet*{y3-sompz}.  

\begin{figure}
\centering
\includegraphics[width = \columnwidth]{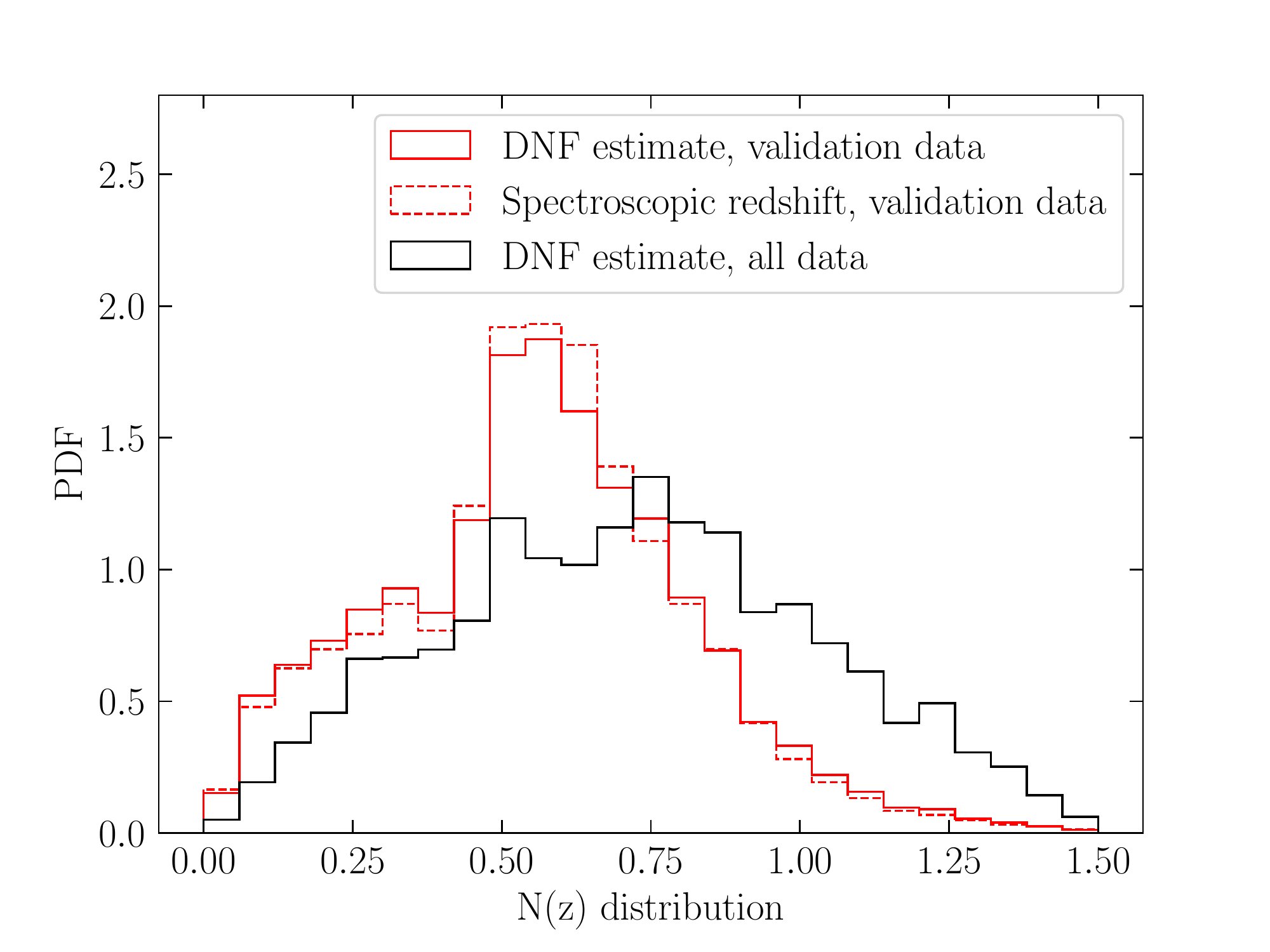}
\caption{N(z) comparison between DNF and the spectroscopic validation sample, as well as the distribution for a random sampling of \gold. Selection included some basic quality cuts on \flagsgold = 0 and \flagsforeground < 2, extended object selection (\sofmash = 3) and $i$-band magnitude range cut in the interval [17.5, 23.5].}
\label{fig:nzcomp}
\end{figure}

In \figref{nzcomp} we show the comparison of the estimate of \dnf against the spectroscopic redshift distribution on the validation sample, for illustration purposes. A qualitative agreement of the N(z) estimate can be readily seen for the validation set used in this work.

\section{Ancillary maps}
\label{sec:maps}

As with \yonegold, the \gold map products use \healpix \citep{Gorski:2005} as their base code, usually with an \nside = 4096 resolution (approximately 0.738 square arcminutes per pixel or 0.86 \amin across). 

\subsection{Footprint}
\label{sec:footprint}

The \gold footprint is a geometric mask used to select regions of the survey with good coverage in multiple bands. 
While the complete \gold object catalog contains all objects measured in the Y3A2 coadd processing (same objects as DR1 release; \secref{dataprocessing}), only the subset of objects located within the \gold footprint are considered as part of the DES Y3 cosmological analyses.
We use the \flagsfootprint variable to ensure consistency between the object catalog and \gold footprint.

The minimum requirement for an object to be part of \gold is summarized as follows:
\begin{itemize}
    \item At least 1 exposure on each band $griz$ is required in the object's \healpix pixel from the \numimage map (\secref{surveyproperties}).
    \item At least $50\%$ of overlapping coverage for each band is required in the \fracdet $griz$ map for that pixel as well (\secref{surveyproperties}).
    \item The object itself must have a value for the \nitermodel variable greater than zero for $griz$, that is, it must have been successfully fit to a model by \sextractor for the light profile in each of these bands.
\end{itemize}

In summary, these conditions require that the object must be in a \healpix region with certain minimum observations, and that the object itself has been observed in the 4 bands in which \sof photometry is computed. Considering the detection fraction in each of the \healpix pixels, the footprint area amounts to $\footprintareaapp \deg^2$.

\subsection{Additional masks}
\label{sec:masks}

For most cosmology analyses, we apply two additional geometric selections beyond the minimal `footprint' observability requirements  (\secref{footprint}). These masks correspond to distinct types of effects: regions where nearby astrophysical objects hinder studies of distant galaxies (called `foreground regions'), and areas that are considered faulty from the measurement point of view, due to some deficiency in the source extraction or photometric measurement \citep[`bad regions';][]{scienceportal}. 

In addition, we produce maps of survey properties and observing conditions (e.g., sky brightness, image quality) extracted from the set of single-epoch images that overlap each position in the survey.

\subsubsection{Foreground mask}
\label{sec:foregroundmask}

\begin{deluxetable}{c c l}
\tablewidth{0pt}
\tabletypesize{\scriptsize}
\tablecaption{ \gold Foreground Region Mask \label{tab:foregrounds}}
\tablehead{
\colhead{Flag Bit} & \colhead{Area} & \colhead{Description} \\
 & ($\deg^2$)  & 
}
\startdata
1   & 220.59
& 2MASS moderately bright star regions ($8 < J < 12$) \\
2   & 22.63
& Large nearby galaxies (HyperLEDA catalog) \\
4   & 91.12
& 2MASS bright star regions ($5< J < 8$) \\
8  & 100.61
& Region near the LMC \\
16  & 86.51
& Yale bright star regions \\
32  & 0.53 & Globular clusters \\ 
64 & 61.13 & Brightest stars \\
\enddata
\tablecomments{
Foreground mask for \gold. The masked area from the \gold catalog is calculated using the coverage fraction of the pixels that are removed from the footprint by each mask. The rationale for each mask can be found in \secref{foregroundmask}.
}
\end{deluxetable}

%% 1   & 33.04 & High density of astrometric discrepancies \\
%% 2   & 133.1 & 2MASS moderate star regions ($8 < J < 12$) \\
%% 4   & 5.40  & RC3 large galaxy region ($10 < B < 16$) \\
%% 8   & 31.6  & 2MASS bright star regions ($5< J < 8$) \\
%% 16  & 98.9  & Region near the LMC \\
%% 32  & 19.75 & Yale bright star regions ($-2 < V < 5.6$) \\
%% 64  & 1.37  & High density of unphysical colors \\ 
%% 128 & ...   & Not used \\
%% 256 & 0.70  & Milky Way globular clusters \\
%%Bad regions are ordered from most least impactful (lowest bit) to the most impactful (highest bit). 

\begin{figure*}
	\includegraphics[width=0.9\textwidth]{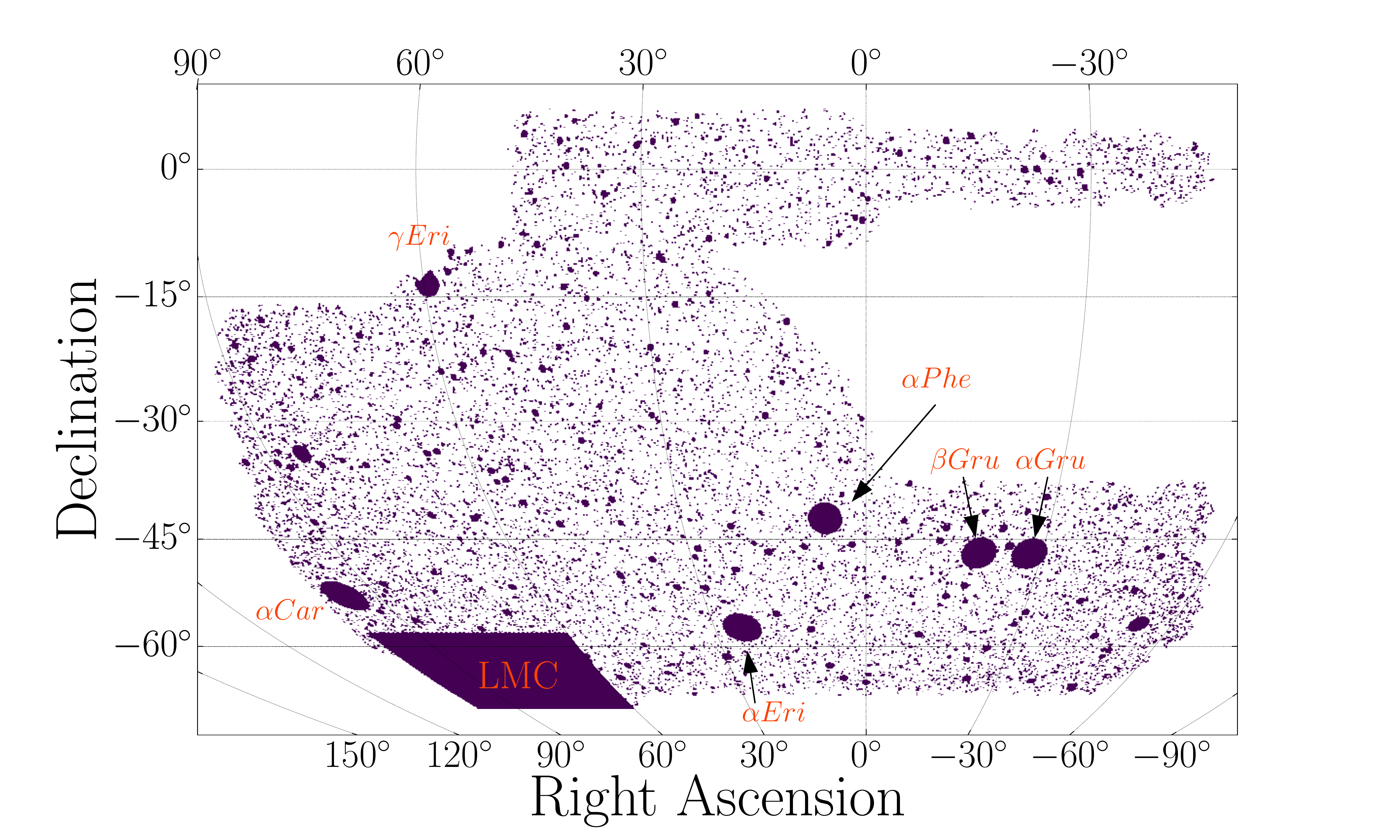}
    \caption{The foreground mask for \gold, including all astrophysical objects which could hamper cosmological analyses (see text for individual descriptions). The total area removed amounts to \foregroundareaapp $deg^2$. NB that the mask corresponding to bit=1 (faint 2MASS stars) is not shown for clarity. Some of the largest masked individual masked areas (Large Magellanic Cloud, very bright stars) are pointed out as well with a text label.}
\label{fig:foreground}
\end{figure*}

\tabref{foregrounds} summarizes the mask bits and regions described in this section.
\figref{foreground} shows the foreground mask. If a \gold object is located within a \healpix pixel that is part of one or more of the regions indicated in \tabref{foregrounds}, the bit flag variable \flagsforeground is set using the corresponding bits. These are defined as follows:

\begin{itemize}
    \item \textbf{Bit 1, 2MASS moderately bright stars:} includes regions around stars with a $J$ magnitude from the 2MASS \citep{Skrutskie:2006}  catalog in the range $8 < J <12$.
    \item \textbf{Bit 2, large nearby galaxies:} this bit selects areas around large, nearby galaxies found in the HyperLEDA\footnote{http://leda.univ-lyon1.fr/} catalog \citep{hyperleda}.
    \item \textbf{Bit 4, 2MASS bright stars:} same as bit 1 above, but including stars in the range $4 < J < 8$.
    \item \textbf{Bit 8, region near the LMC:} this mask avoids the area with an overabundance of stars around the Large Magellanic Cloud, which can easily overwhelm the galaxy catalog, or create heavy obscuration for cosmology analyses.
    \item \textbf{Bit 16, Yale bright star catalog \citep{yale}:} approximately 1000 objects from the catalog overlap with the \gold footprint. A linear function has been implemented to create a mask as a function with $V$-band magnitude from the catalog.
    \item \textbf{Bit 32, Globular clusters:} the list includes five globular clusters with magnitude $V < 10$, using the radius provided in the NGC2000 catalog\footnote{\url{https://heasarc.gsfc.nasa.gov/W3Browse/all/ngc2000.html}}. These are NGC 1261, NGC 1851, NGC 7089, NGC 288 and NGC 1904.
    \item \textbf{Bit 64, Very bright stars:} these are 11 stars that produce a large scattered light artifacts due to their brightness that goes beyond the image masking and exclusion listing set up for the rest of the stars. A large radius is defined around them to remove areas with large densities of bright objects with anomalous colors. These stars are listed, including the exclusion radius, in \tabref{famousstars}.
\end{itemize}

\begin{deluxetable}{c c c}
\tablewidth{0pt}
\tabletypesize{\scriptsize}
\tablecaption{ Very bright stars exclusion list \label{tab:famousstars}}
\tablehead{
\colhead{Name} & \colhead{$\alpha$,$\delta$} & \colhead{Radius (deg.)} }
\startdata
$\alpha$ Phe   & ($6.5708$, $-42.3061$) & $2.0$\\
$\alpha$ Eri   & ($24.4288$, $-57.2367$) & $2.0$\\
$\gamma$ Eri   & ($59.5075$, $-13.5086$) & $1.5$\\
$\alpha$ Hyi   & ($29.6925$, $-61.5697$) & $0.5$\\
$\alpha$ Col   & ($84.9121$, $-34.0741$) & $1.0$\\
$\alpha$ Car   & ($95.9879$, $-52.6958$) & $2.0$\\
$\alpha$ Pav   & ($306.41214$, $-56.7350$) & $1.0$\\
$\alpha$ Gru   & ($332.0583$, $-46.9611$) & $2.0$\\
$\beta$ Gru   & ($340.6671$,$-46.8847$) & $2.0$\\
Pi1 Gru & ($335.6829$, $-45.9478$) & $0.5$\\
P Dor & ($69.1900$, $-62.0775$) & $0.5$\\
\enddata
%\tablecomments{}
\end{deluxetable}

\subsubsection{Bad regions mask}
\label{sec:badregionsmask}

\tabref{badregions} summarizes the mask bits and regions described in this section. 
As with the foregrounds, if a \gold object is within a \healpix pixel that is part of one or more of the regions indicated in \tabref{badregions}, the bit flag variable \flagsbadregions is set using the corresponding bits. These are described below:

\begin{itemize}
    \item \textbf{Bit 1, coadd PSF failure regions}: The coaddition process produces a discontinuous PSF function across the footprint that will corrupt \sextractor quantities which depend on the PSF such as MAG\_PSF and SPREAD\_MODEL \citep{Desai:2012, Bouy:2013}. 
    Analyses using these \sextractor quantities, should mask out these regions.
    \item \textbf{Bit 2, tiles with errors in \mof processing}: 66 DESDM `tiles' failed to finish processing the \mof pipeline. These problematic tiles are all associated with foreground objects and/or dense regions.
    \item \textbf{Bit 4, high density of anomalous colors}: This mask is intended to remove reflections in the images, bad coverage of foreground galaxies and a few satellite trails remaining in the images, using a selection on high density of objects with extreme colors (with any color $g-r$, $r-i$, $i-z$ outside of the range [-2, 3]).
    
\end{itemize}

\begin{deluxetable}{c c l}
\tablewidth{0pt}
\tabletypesize{\scriptsize}
\tablecaption{ \gold Bad Region Mask \label{tab:badregions}}
\tablehead{
\colhead{Flag Bit} & \colhead{Area} & \colhead{Description} \\
 & ($\deg^2$)  & 
}
\startdata
1   & 42.18 & Coadd PSF failure regions \\
2   & 28.43 & Tiles with errors in \mof processing \\
4   & 5.95  & High density of anomalous colors \\
\enddata
\tablecomments{
Bad regions mask for \gold. The masked area from the \gold catalog is calculated using the coverage fraction of the pixels that are removed from the footprint by each mask. The rationale for each mask can be found in \secref{badregionsmask}. 
}
\end{deluxetable}

\subsection{Survey properties}
\label{sec:surveyproperties}

We track the spatial variation of several observing conditions (see \tabref{observingconditions}) across the survey footprint using \mangle polygon masks (\secref{imageproc}). Given that each location in the survey contains the information from a stack of images, a statistic (e.g., mean, minimum, maximum) is used to summarize this information as a scalar value for that location \citep{Leistedt:2016}. This step is explained in~\citet{desdm}. 

The detailed geometry of the survey given by \mangle is transformed into \healpix maps for simplification and homogenization. In addition, bleed-trail and bright star masks for each band produced by DESDM are compacted into a single detection fraction \healpix map (\texttt{FRACDET}), giving the effective coverage at each \healpix pixel for each band. Furthermore, using the bleed-trail and bright star masks for any choice of bands $grizY$ we can also produce a combined detection fraction map. The \healpix maps were produced using the DES Science Portal \citep{scienceportal}; the process is described in \appref{sps}.

DESDM delivered 27 survey properties for \gold, together with bleed-trail and bright star masks for each band, totalling 135 \mangle products for the entire survey (see \appref{sps}). 
 \gold provides pixelized versions of these survey property maps at \healpix $\nside=4096$ resolution in \code{NESTED} ordering, as well as other useful maps used in cosmology analyses such as de \mof, \sof and \texttt{MAG\_AUTO} depth maps described in \secref{depth}, a stellar density map computed using `secure' stars, according to the \sofmash classifier, and interstellar extinction maps.

\section{Using \gold}
\label{sec:usage}

The \gold data set will be released as was done with \yonegold as an online resource, available at \drurl. This release includes the catalog itself, along with the maps detailed in \secref{maps} in \healpix format.

The \gold data set used for Y3 cosmology analyses has internally been labeled as version 2.2. This version contains 399 million objects and 446 columns, which include the object ID, position, measured photometry and associated errors in each band using a variety of algorithms, shape information and errors, photometric redshifts and related quantities, and several flags (described in previous sections). We also provide the interstellar extinction in the direction of each object, as estimated from three different reddening maps \citep[][see \appref{extinction}]{sfd98, lenz17, planck13}.

In the online documentation, we provide usage notes for the current \gold version (to be updated for any subsequent versions produced). Some general recommendations are listed below. 

\begin{itemize}
    \item The fundamental selection for \gold is to \textbf{select objects with \flagsfootprint = 1}, as described in \secref{footprint}.
   \item In general, the areas identified in \secref{foregroundmask} can present various problems in terms of photometry, junk objects, obscuration, etc., so the \flagsforeground $= 0$ selection is generally recommended for extragalactic studies.
   \item `Bad' regions coming from internal processing or data taking issues (\secref{badregionsmask}) will vary depending on the choice of photometric pipeline. A \sof-based analysis can be safely done with \flagsbadregions $< 4$ whereas a \mof-based one should restrict to \flagsbadregions $< 2$. A \sextractor-based analysis should use the bitwise AND operation (\flagsbadregions \& 101 = 0). 
   \item As explained in \secref{flagsgold}, \flagsgold allows for a selection of good quality objects, by summarizing various flags and signatures of poor reconstructions in a single bitmask. However, \gold provides the component flags from the different processes that were executed over the objects for more refined measurements. Typically, a \sof-based galaxy sample would use the bitwise AND operation (\flagsgold \& 1111110 = 0).
   \item \textbf{Photometry is provided as computed after FGCM calibration is applied}, after atmospheric and instrumental corrections have taken place (i.e., top-of-the-atmosphere photometry). \textbf{By default, cataloged magnitudes are not corrected for Galactic extinction.} However a further zero-point correction based on Y4 imaging (with better quality) was computed prior to this release. In addition, as described in \secref{photometry}, a minor modification to take into account the spectral shape of the sources in the calibration plus the correction for Galactic extinction, has to be applied to obtain the final top-of-the-Galaxy fluxes. \textbf{Only the magnitudes and fluxes containing the \texttt{CORRECTED} suffix include these minor adjustments as well as Galactic extinction}. For example, in the case of magnitudes, this computation is:
    \begin{align}
    \begin{split}
    \texttt{MAG\_CORRECTED} &={} \texttt{MAG} \\ 
    & + \texttt{DELTA\_MAG\_Y4} \\
    & + \texttt{DELTA\_MAG\_CHROM} \\
    & - \texttt{A\_SED\_SFD98}
    \end{split}
    \label{eqn:gold_corrected_photometry}
    \end{align}
    (other extinction corrections may be applied as appropriate).
    \item Only \sextractor $Y$-band photometry is available, as tests showed that incorporating this band into the overall multi-object fit degraded the performance in the rest of the bands.
    \item \textbf{The default recommended star-galaxy separation method to identify stars and galaxies is \texttt{EXTENDED\_CLASS\_MASH\_SOF}}. It is based in morphological quantities as described in \secref{stargalaxy}. This method employs \texttt{EXTENDED\_CLASS\_SOF} as the main classifier for an object (see \tabref{extended}) but defaults to \texttt{EXTENDED\_CLASS\_WAVG}, available for the brighter objects, or \texttt{EXTENDED\_CLASS\_COADD} in case none of the others have been computed (in which case their values are set to a `sentinel' value). For cosmology analyses, the selection $\sofmash=3$ is recommended, as it shows very low stellar contamination up to the magnitude limit, with a decrease in galaxy selection efficiency only beyond $i>22.5$. By exploring different ranges of \var{EXTENDED\_CLASS} values, users can identify an appropriate sample for their science case. A default value of $-9$ is assigned when there is insufficient data available to compute the \var{EXTENDED\_CLASS} variable.
    \item At low redshifts ($z<0.5$), the \bpz run available in this catalog shows poor metrics (\secref{photozs}), therefore we recommend the usage of \dnf or \ann over \bpz in general.
\end{itemize}

In \tabref{queries}, some example queries are shown for illustration purposes, to reflect the usage of flags and specific \gold columns for a few typical situations.

\begin{deluxetable*}{l l}
\tablewidth{0pt}
\tabletypesize{\tablesize}
\tablecaption{ Example selections from the \gold catalog, provided for illustration purposes. \label{tab:queries}}
\tablehead{
\colhead{Sample}  & \colhead{Selection from \gold columns} 
}
\startdata
High purity galaxy sample (\sof) & \makecell{$\flagsfootprint = 1$ AND $\flagsforeground = 0$ AND \\ $(\flagsgold \& 1111110) = 0$ AND $\sofmash = 3$ AND \\ $\texttt{SOF\_CM\_MAG\_CORRECTED\_I} = [18,22.5]$} \\\\
High purity galaxy sample (\mof) & \makecell{$\flagsfootprint = 1$ AND $\flagsforeground = 0$ AND \\ $(\flagsgold \& $1111110$) = 0$ AND $(\flagsbadregions \& 110) = 0 $ \\ AND $\mofmash = 3$ AND $\texttt{MOF\_CM\_MAG\_CORRECTED\_I} = [18,22.5]$} \\\\
High purity galaxy sample (\sextractor) & \makecell{$\flagsfootprint = 1$ AND $\flagsforeground = 0$ AND \\ $(\flagsgold \& 1111110) = 0$ AND $\flagsbadregions = 0$ \\ AND $\texttt{EXTENDED\_CLASS\_COADD} = 3$ AND $(\texttt{MAG\_AUTO\_I} +$ \\ $+ \texttt{DELTA\_MAG\_Y4\_I} +  \texttt{DELTA\_MAG\_CHROM\_I} - \texttt{A\_SED\_SFD98\_I}) = [18,22.5$]} \\\\
Basic object detections for subsequent shear studies & \makecell{$\flagsfootprint = 1$ AND $\flagsforeground = 0$ AND \\ $(\flagsgold \& 1111000)$ AND $(\flagsbadregions \& 110) = 0$}\\\\ 
High purity stellar sample (\sof) & \makecell{$\flagsfootprint = 1$ AND $\flagsforeground = 0$ AND \\ $(\flagsgold \& 1111100) = 0$ AND $\sofmash \leq 2$ \\ AND $(\texttt{SOF\_PSF\_MAG\_R} + \texttt{DELTA\_MAG\_Y4\_R} +$ \\ $+\texttt{DELTA\_MAG\_CHROM\_R} - \texttt{A\_SED\_SFD98\_R}) = [16,23$] } \\\\
High completeness stellar sample (\sof) & \makecell{$\flagsfootprint = 1$ AND $\flagsforeground = 0$ AND \\ $(\flagsgold \& 1111100) = 0$ AND $\sofmash \leq 2$ \\ AND $(\texttt{SOF\_PSF\_MAG\_R} + \texttt{DELTA\_MAG\_Y4\_R} +$ \\ $+\texttt{DELTA\_MAG\_CHROM\_R} - \texttt{A\_SED\_SFD98\_R}) = [16,23]$ } \\\\
Red galaxy sample & \makecell{$\flagsfootprint = 1$ AND $(\flagsforeground \& 11111100) = 0$ AND \\ $(\flagsgold \& 1111110) = 0$ AND $\sofmash = 3$ AND \\ $\texttt{SOF\_CM\_MAG\_CORRECTED\_I} = [17.5,22]$ AND \\
$\texttt{SOF\_CM\_MAG\_CORRECTED\_I} - \texttt{SOF\_CM\_MAG\_CORRECTED\_Z} +$ \\ $2*(\texttt{SOF\_CM\_MAG\_CORRECTED\_R}-\texttt{SOF\_CM\_MAG\_CORRECTED\_I}) > 1.7$}
\enddata
\tablecomments{
Here `\&' corresponds to the bitwise AND operation.
}

\end{deluxetable*}

\section{Conclusions}
\label{sec:conclusions}

The \gold data set is the basic resource for cosmology using the Wide Survey of DES. It constitutes one of the largest galaxy catalogs to date, and is the basis of a new set of results exploring the robustness of the $\Lambda$CDM model and its alternatives in exquisite detail. Beyond serving the immediate needs of the DES Collaboration, we hope that \gold stimulates further analyses by the astronomy and cosmology community at large (as demonstrated by \citealt{asgari} and \citealt{cheng}, for example).
Data access tools and documentation are publicly available at \drurl.
We highlight several notable features (\tabref{summary}) of this data set:

\begin{itemize}
    \item Sky coverage of nearly $5000 \deg^2$ in five photometric bands, $grizY$, at optical and near-infrared wavelengths;
    \item $<3$\mmag homogeneity using multi-epoch photometry and the FGCM calibration model;
    \item Depth of $\maglimsofgapp$, $\maglimsofrapp$, $\maglimsofiapp$, $\maglimsofzapp$ $\magn$ in $griz$ for extended objects at $\SNR \sim 10$;
    \item 399M measured objects of which $\sim226$M are extended objects marked as `good' (very high galaxy purity up to $i=22.5$), prior to any flux or signal-to-noise selection;
    \item Approximate coverage of $z\sim0.2-1.2$ in photometric redshift
\end{itemize}

Looking forward, the next major DES data processing campaign involves the full set of observations from the complete six seasons of DES, and an associated second public data release (DES DR2).
DES Y6 data roughly double the integrated exposure time over most of the footprint (see \figref{footmap}).
In addition, several upgrades have been implemented in the science pipelines, including a lower S/N threshold for object detection, \gaia DR2 astrometric calibration, PSF modeling upgrades, and enhanced algorithms for the photometry of blended objects.
The next generation of ground-based imaging surveys, including the Rubin Observatory LSST, will require more stringent control of systematic uncertainties associated with galaxy measurement and survey characterization \citep[e.g.,][]{desc_2018_srd}, motivating continued use of DES as a proving ground for new data reduction techniques and data products to support cosmological analyses.

\section*{Acknowledgments}

KB acknowledges support from the U.S. Department of Energy, Office of Science, Office of High Energy Physics, under Award Numbers DE-SC0020278 and DE-SC0017647.

Funding for the DES Projects has been provided by the U.S. Department of Energy, the U.S. National Science Foundation, the Ministry of Science and Education of Spain, the Science and Technology Facilities Council of the United Kingdom, the Higher Education Funding Council for England, the National Center for Supercomputing Applications at the University of Illinois at Urbana-Champaign, the Kavli Institute of Cosmological Physics at the University of Chicago,  the Center for Cosmology and Astro-Particle Physics at the Ohio State University,
the Mitchell Institute for Fundamental Physics and Astronomy at Texas A\&M University, Financiadora de Estudos e Projetos, Funda{\c c}{\~a}o Carlos Chagas Filho de Amparo {\`a} Pesquisa do Estado do Rio de Janeiro, Conselho Nacional de Desenvolvimento Cient{\'i}fico e Tecnol{\'o}gico and the Minist{\'e}rio da Ci{\^e}ncia, Tecnologia e Inova{\c c}{\~a}o, the Deutsche Forschungsgemeinschaft and the Collaborating Institutions in the Dark Energy Survey. 

The Collaborating Institutions are Argonne National Laboratory, the University of California at Santa Cruz, the University of Cambridge, Centro de Investigaciones Energ{\'e}ticas, Medioambientales y Tecnol{\'o}gicas-Madrid, the University of Chicago, University College London, the DES-Brazil Consortium, the University of Edinburgh, 
the Eidgen{\"o}ssische Technische Hochschule (ETH) Z{\"u}rich, Fermi National Accelerator Laboratory, the University of Illinois at Urbana-Champaign, the Institut de Ci{\`e}ncies de l'Espai (IEEC/CSIC), the Institut de F{\'i}sica d'Altes Energies, Lawrence Berkeley National Laboratory, the Ludwig-Maximilians Universit{\"a}t M{\"u}nchen and the associated Excellence Cluster Universe, the University of Michigan, NFS's NOIRLab, the University of Nottingham, The Ohio State University, the University of Pennsylvania, the University of Portsmouth, SLAC National Accelerator Laboratory, Stanford University, the University of Sussex, Texas A\&M University, and the OzDES Membership Consortium.

Based in part on observations at Cerro Tololo Inter-American Observatory at NSF's NOIRLab (NOIRLab Prop. ID 2012B-0001; PI: J. Frieman), which is managed by the Association of Universities for Research in Astronomy (AURA) under a cooperative agreement with the National Science Foundation.

The DES data management system is supported by the National Science Foundation under Grant Numbers AST-1138766 and AST-1536171.The DES participants from Spanish institutions are partially supported by MICINN under grants ESP2017-89838, PGC2018-094773, PGC2018-102021, SEV-2016-0588, SEV-2016-0597, and MDM-2015-0509, some of which include ERDF funds from the European Union. IFAE is partially funded by the CERCA program of the Generalitat de Catalunya. Research leading to these results has received funding from the European Research Council under the European Union's Seventh Framework Program (FP7/2007-2013) including ERC grant agreements 240672, 291329, and 306478. We acknowledge support from the Brazilian Instituto Nacional de Ci\^encia e Tecnologia (INCT) do e-Universo (CNPq grant 465376/2014-2). A.C.R. acknowledges financial support from the Spanish Ministry of Science, Innovation and Universities (MICIU) under grant AYA2017-84061-P, co-financed by FEDER (European Regional Development Funds) and by the Spanish Space Research Program ``Participation in the NISP instrument and preparation for the science of EUCLID" (ESP2017-84272-C2-1-R).

The Hyper Suprime-Cam (HSC) collaboration includes the astronomical communities of Japan and Taiwan, and Princeton University. The HSC instrumentation and software were developed by the National Astronomical Observatory of Japan (NAOJ), the Kavli Institute for the Physics and Mathematics of the Universe (Kavli IPMU), the University of Tokyo, the High Energy Accelerator Research Organization (KEK), the Academia Sinica Institute for Astronomy and Astrophysics in Taiwan (ASIAA), and Princeton University. Funding was contributed by the FIRST program from Japanese Cabinet Office, the Ministry of Education, Culture, Sports, Science and Technology (MEXT), the Japan Society for the Promotion of Science (JSPS), Japan Science and Technology Agency (JST), the Toray Science Foundation, NAOJ, Kavli IPMU, KEK, ASIAA, and Princeton University. 

Based [in part] on data collected at the Subaru Telescope and retrieved from the HSC data archive system, which is operated by Subaru Telescope and Astronomy Data Center at National Astronomical Observatory of Japan.

This paper makes use of software developed for the Vera Rubin Observatory Legacy Survey of Space and Time (LSST). We thank the LSST Project for making their code available as free software at  \url{http://dm.lsst.org}. \texttt{healsparse} was developed under the Rubin Observatory Legacy Survey of Space and Time Dark Energy Science Collaboration (DESC) using LSST DESC resources. 

This work has made use of data from the European Space Agency (ESA) mission \gaia (\url{https://www.cosmos.esa.int/gaia}), processed by the \gaia Data Processing and Analysis Consortium (DPAC, \url{https://www.cosmos.esa.int/web/gaia/dpac/consortium}). Funding for the DPAC has been provided by national institutions, in particular the institutions participating in the \gaia Multilateral Agreement.

This manuscript has been authored by Fermi Research Alliance, LLC under Contract No. DE-AC02-07CH11359 with the U.S. Department of Energy, Office of Science, Office of High Energy Physics.

\vspace{5mm}
\facility{Blanco (DECam)} 

\software{\sextractor \citep{Bertin:1996}, \PSFEx \citep{Bertin:2011}, \scamp \citep{Bertin:2006}, \swarp \citep{Bertin:2002,Bertin:2010}, \mangle \citep{Hamilton:2004,mangle}, \healpix \citep{Gorski:2005} \footnote{\url{http://healpix.sourceforge.net}}  , \easyaccess  \citep{easyaccess} ,\code{astropy} \citep{Astropy:2013}, \code{matplotlib} \citep{Hunter:2007}, \code{numpy} \citep{numpy:2011}, \code{scipy} \citep{scipy:2001}, \code{healpy},\footnote{\url{https://github.com/healpy/healpy}} \code{fitsio}\footnote{\url{https://github.com/esheldon/fitsio}}, \ngmix \citep{Sheldon:2014}\footnote{\url{https://github.com/esheldon/ngmix}}},\code{TOPCAT} \citep{topcat}, \code{healsparse}\footnote{\url{https://healsparse.readthedocs.io/en/latest/}}, \code{athena}\footnote{\url{http://www.cosmostat.org/software/athena}}. 

\newpage

\appendix

\numberwithin{figure}{section}
\numberwithin{table}{section}

\section{Unified Approach for Chromatic and Interstellar Extinction Corrections}
\label{app:photometric_calibration}

We present the detailed formalism to apply SED-dependent photometric corrections to each source in the \gold release. 
Building upon the work of \citet[][]{Y3FGCM}, our framework accounts for both chromatic corrections associated with the DECam bandpass (instrument and atmosphere) and interstellar extinction.
We consider first the corrections for individual exposures, and then the corrections for multi-epoch photometry.

\subsection{Single-Epoch Corrections}

Working forward along the path of light, the top-of-the-Galaxy (TOG) source spectrum incident at the Milky Way $F_\nu^{\rm TOG}(\lambda)$ is reddened by interstellar dust before arriving at the Earth.
Consider a reddening law with optical index $a \tau(\lambda)$,  where $\tau$ is normalized to $1\micron$. 
Let $a = f(\mathcal{E})$  be  a  normalization  factor  for  the reddening law, where $\mathcal{E} \approx E(B - V)$ in the SFD98 prescription, but in general is some scaling from an external map providing the dust optical depth normalization. 
The dust-reddened top-of-the-atmosphere (TOA) source spectrum is $F_\nu^{\rm TOG}(\lambda) e^{-a \tau (\lambda)}$. 

The analog-to-digital (ADU) counts registered by the camera for a given band $b \in \{ grizY \}$ is proportional the TOA source spectrum weighted by the transmission of the observed bandpass $S_b^{\rm obs}(\lambda)$ integrated over wavelength.

\begin{equation}
\mathrm{ADU}_b = \frac{A \Delta t}{g h} \int_0^\infty F_\nu^{\rm TOG}(\lambda) e^{-a \tau (\lambda)} S_b^{\rm obs} (\lambda) \lambda^{-1} d \lambda.
\end{equation}

\noindent The instantaneous system throughput varies as a function of focal plane location and environmental conditions. The effective aperture $A$, exposure time $\Delta t$, gain $g$, and Planck's constant $h$ appear as multiplicative factors.

We define three flux measurements of interest and the relationships between these measurements. The first is the TOA source spectrum as seen through the observed bandpass, i.e., the flux directly measured on the camera \citep{Fukugita:1996}:

\begin{align}
m_b^{\rm TOA,obs} &= -2.5 \log_{10} \left[ \frac{\int_0^\infty F_\nu^{\rm TOG}(\lambda) e^{-a \tau (\lambda)} S_b^{\rm obs} (\lambda) \lambda^{-1} d \lambda}{\int_0^\infty F^{\rm AB} S_b^{\rm obs} (\lambda) \lambda^{-1} d \lambda} \right] \\
&= -2.5 \log_{10} \left[ \frac{gh\mathrm{ADU}_b}{A \Delta t F^{\rm AB} \int_0^\infty S_b^{\rm obs} (\lambda) \lambda^{-1} d \lambda} \right] \\
&= -2.5 \log_{10} (\mathrm{ADU}_b) + 2.5 \log_{10}(\Delta t) + 2.5 \log_{10}\left( \int_0^\infty S_b^{\rm obs} (\lambda) \lambda^{-1} d \lambda \right) + 2.5 \log_{10}\left( \frac{A F^{\rm AB}}{gh} \right).
\end{align}

\noindent The magnitude is normalized relative to the AB scale with $F^{\rm AB} = 3631 \jy$ \citep{Oke:1983}.
For a known observed bandpass, the measured $\mathrm{ADU}_b$ uniquely determines $m_b^{\rm TOA,obs}$. 
In \gold, the observed bandpass is provided by FGCM \citep{Y3FGCM} for each individual CCD image together with the zeropoint in the AB magnitude system 

\begin{equation}
\mathrm{ZP}^{\rm AB} = 2.5 \log_{10}\left( \frac{A F^{\rm AB}}{gh} \right).
\end{equation}

Second, we define the TOA source spectrum as seen through the DES standard bandpass:

\begin{equation}
m_b^{\rm TOA,std} = -2.5 \log_{10} \left[ \frac{\int_0^\infty F_\nu^{\rm TOG}(\lambda) e^{-a \tau (\lambda)} S_b^{\rm std} (\lambda) \lambda^{-1} d \lambda}{\int_0^\infty F^{\rm AB} S_b^{\rm std} (\lambda) \lambda^{-1} d \lambda} \right].
\end{equation}

\noindent The standard DES bandpass is defined as the instrument throughput averaged over CCDs and multiplied by the standard atmosphere. 
The difference between the TOA source spectrum seen through the observed and standard bandpass is the chromatic correction:

\begin{equation}
\delta m_b^{\rm chrom} = m_b^{\rm TOA,std} - m_b^{\rm TOA,obs} = -2.5 \log_{10} \left[ \frac{\int_0^\infty S_b^{\rm obs} (\lambda) \lambda^{-1} d \lambda}{\int_0^\infty S_b^{\rm std} (\lambda) \lambda^{-1} d \lambda} \right] + 2.5 \log_{10} \left[ \frac{\int_0^\infty F_\nu^{\rm TOG}(\lambda) e^{-a \tau (\lambda)} S_b^{\rm obs} (\lambda) \lambda^{-1} d \lambda}{\int_0^\infty F_\nu^{\rm TOG}(\lambda) e^{-a \tau (\lambda)} S_b^{\rm std} (\lambda) \lambda^{-1} d \lambda} \right].
\label{eqn:chromatic_correction}
\end{equation}

Third, we define the TOG source spectrum observed through the standard bandpass:

\begin{equation}
m_b^{\rm TOG,std} = -2.5 \log_{10} \left[ \frac{\int_0^\infty F_\nu^{\rm TOG}(\lambda) S_b^{\rm std} (\lambda) \lambda^{-1} d \lambda}{\int_0^\infty F^{\rm AB} S_b^{\rm std} (\lambda) \lambda^{-1} d \lambda} \right].
\end{equation}

The difference between the TOA and TOG source spectrum observed through the standard bandpass is the interstellar extinction correction:

\begin{equation}
\delta m_b^{\rm extinction} = m_b^{\rm TOA,std} - m_b^{\rm TOG,std} = - 2.5 \log_{10} \left[ \frac{\int_0^\infty F_\nu^{\rm TOG}(\lambda) e^{-a \tau (\lambda)} S_b^{\rm std} (\lambda) \lambda^{-1} d \lambda}{\int_0^\infty F_\nu^{\rm TOG}(\lambda) S_b^{\rm std} (\lambda) \lambda^{-1} d \lambda} \right].
\end{equation}

\noindent This expression allows computation of SED-dependent interstellar extinction corrections. Our expectation is that most science users will primarily use $m_b^{\rm TOG,std}$ because this quantity is straightforward to compute for a given intrinsic source spectrum and allows for more direct comparisons of source photometry across the survey. Summarizing the results above, we can write the overall transformation from raw ADU counts to the chromatically correct and de-reddened magnitude as

\begin{equation}
m_b^{\rm TOG,std} = m_b^{\rm TOA,obs} + \delta m_b^{\rm chrom} - \delta m_b^{\rm extinction}.
\label{eqn:single-epoch_correction}
\end{equation}

\noindent The photometric corrections above require an SED template for each object, as described in \appref{sed_templates}.

\subsection{Multi-epoch Corrections}

We now generalize \eqnref{single-epoch_correction} to the case of multi-epoch photometry.
For the purpose of this derivation, we assume that the coaddition weighting is constant on a per-object rather than a per-pixel basis. 
While this assumption is not precisely correct, most of the scaling in the \swarp coaddition is from the image-based zeropoint weighting, rather than the local pixel-scale weighting. 
For \code{WAVG} catalog-coadd quantities this assumption is correct because the weighting is done explicitly on the object level.

Suppose we have $N$ observations of an object in band $b$ that are enumerated with the index $i$. 
To simplify the subscripts, in this subsection we neglect the subscript $b$ and assume we are working in band $b$. 
The raw multi-epoch magnitude $\langle m^{\rm obs} \rangle$ is given by the weighted sum of individual measurements

\begin{equation}
\langle m^{\rm obs} \rangle = \frac{\sum w_i m_i^{\rm obs}}{\sum w_i},
\end{equation}

\noindent where $w_i$ are the individual weights. For the weighted-average quantities, we use inverse-variance weights $w_i = 1/ \sigma^2_i$, where $\sigma_i$ is the single-epoch photometric error.
For the coadd quantities, the weights are the median of the weight plane per amplifier for each observation. We can then apply per-observation-epoch photometric corrections $\delta m_i$ to obtain the multi-epoch photometric corrected magnitude $\langle m^{\rm corrected} \rangle$ as follows:

\begin{align}
\langle m^{\rm corrected} \rangle &= \frac{\sum w_i (m_i^{\rm obs} + \delta m_i)}{\sum w_i} \\
&= \frac{\sum w_i m_i^{\rm obs}}{\sum w_i} + \frac{\sum w_i \delta m_i}{\sum w_i} \\
&= \langle m^{\rm obs} \rangle + \frac{\sum w_i \delta m_i}{\sum w_i}.
\end{align}

For \gold, we applied three per-observation-epoch corrections corresponding to a per-object chromatic correction, a gray zeropoint correction, and a zeropoint correction to shift to the AB magnitude scale:

\begin{equation}
\delta m_i = \delta m_i^{\rm chrom} + \delta m_i^{\rm ZP,gray} + \delta m_i^{\rm ZP,AB}.
\end{equation}

The per-observation-epoch chromatic corrections $\delta m_i^{\rm chrom}$ come from \eqnref{chromatic_correction}. 
The gray zeropoint corrections $\delta m_i^{\rm ZP,gray}$ are described in \appref{gray_zeropoint}. 
The AB magnitude zeropoint corrections $\delta m_i^{\rm ZP,AB}$ arise due to an internal bookkeeping convention. 
Before  we  perform  the  coaddition,  each  individual  image  must  be  given  a  zeropoint. 
With FGCM, zeropoints are SED-dependent but we do not know the per-object SEDs ahead of time, nor can we perform the coaddition with varying zeropoints. 
The ``native'' FGCM SED is a flat $F_{\nu}(\lambda)$ spectrum in wavelength (AB magnitudes), but very few objects have this color. 
Therefore, we decided to make the coadds based on a reference spectrum of a G star (our absolute calibrator star, C26202). 
In the database, the \var{FGCM\_ZPT} value that is used for the coadds is shifted to the AB magnitude spectrum and the shift is recorded as \var{FGCM\_COADD\_ZPTSHIFT}.
$\delta m^{\rm ZP,AB}_i$ is obtained from the database as \var{FGCM\_COADD\_ZPTSHIFT}. 

To apply these multi-epoch corrections to objects in \gold, we computed the weighted average of each per-observation correction. 
Finally, the interstellar extinction correction $\delta m_b^{\rm extinction}$ can be applied per-object (\appref{extinction}) rather than per-observation because we have defined the interstellar extinction correction in terms of the standard bandpass.
The chromatic correction term (\eqnref{chromatic_correction}) includes reddening. However, any changes in the assumed reddening law or reddening map would only cause second-order effects (especially in the low-extinction regime that aptly describes most of the DES footprint), so we decided to keep the $e^{-a \tau (\lambda)}$ factor in \eqnref{chromatic_correction} fixed to the fiducial prescription so that chromatic and interstellar extinction effects can be computed and tested independently. The multi-epoch magnitude for the TOG object spectrum observed through the standard bandpass is

\begin{equation}
\langle m^{\rm TOG,std} \rangle = \langle m^{\rm obs} \rangle + \langle \delta m_i^{\rm ZP} \rangle + \langle \delta m^{\rm chrom} \rangle - \delta m_b^{\rm extinction}.
\end{equation}

\noindent \tabref{photometric_correction} summarizes the multi-epoch photometric corrections in the \gold catalog, which can be applied as shown in \eqnref{gold_corrected_photometry}. 
We combine the multi-epoch corrections for the gray and AB magnitude zeropoint corrections $\langle \delta m_i^{\rm ZP} \rangle$ as \texttt{DELTA\_MAG\_Y4} in the \gold table:

\begin{equation}
\langle \delta m_i^{\rm ZP} \rangle = \frac{\sum w_i (\delta m_i^{\rm ZP,gray} + \delta m_i^{\rm ZP,AB})}{\sum w_i}.
\label{eqn:multiepoch_zeropoint}
\end{equation}

\begingroup
\renewcommand{\arraystretch}{2.} 
\begin{deluxetable}{ccc}
\tablewidth{0pt}
\tabletypesize{\tablesize}
\tablecaption{Summary of multi-epoch photometric corrections\label{tab:photometric_correction} 
}
\tablehead{
\colhead{Correction} & \colhead{\gold Column} & \colhead{Expression}
}
\startdata
Gray and AB Zeropoint & \texttt{DELTA\_MAG\_Y4} & $\langle \delta m_i^{\rm ZP} \rangle = \frac{\sum w_i (\delta m_i^{\rm ZP,gray} + \delta m_i^{\rm ZP,AB})}{\sum w_i}$ \\
Chromatic & \texttt{DELTA\_MAG\_CHROM} & $\langle \delta m^{\rm chrom} \rangle = \frac{\sum w_i \delta m_i^{\rm chrom}}{\sum w_i}$ \\
Interstellar Extinction & \texttt{A\_SED\_SFD98} & $\delta m_b^{\rm extinction}$ \\
\enddata
\end{deluxetable}
\endgroup

\subsection{Updated Gray Zeropoint Corrections in \gold}
\label{app:gray_zeropoint}

There are several improvements to the ``gray'' SED-independent zeropoints \citep{fgcmupd} between the initial DES DR1 release and the \gold release:

\begin{itemize}
    
    \item Aperture corrections in \gold are performed internally during the calibration rather than as an afterburner step. \citet{Bernstein:2017_photometry} found that photometric residuals between individual exposures within the same night could be primarily accounted for by improved aperture corrections. The \gold calibration is based on \sextractor \magpsf photometry from \code{FINALCUT} processing and normalized to the flux measured within a $6\arcsec$ diameter aperture (\magapereight).
    
    \item The initial photometric calibration used for DR1 was based on a preliminary version of the DES Standard Bandpass. The updated \gold calibration is now fully consistent with the DES Y3A2 Standard Bandpass publicly released with DR1.
    
    \item The observation strategy used during the first three years of DES concentrated observations in two distinct halves of the footprint during the first and second years, respectively. It was only during the third year of DES that both halves were routinely observed within the same night. However, the third year of DES encountered unusually poor weather conditions. The \gold photometric calibration incorporates a fourth year of observations to improve the uniformity across the full footprint (for the purpose of photometric calibration only; no Y4 imaging was included in the coadd).
    
    \item We did not use the Global Positioning System as input to the water vapor term in FGCM for the \gold calibration, as this GPS input was compromised during a period of the Y1-Y3 observations, and led to spatially coherent photometric residuals in $z$-band over a small region of the footprint.
    
    \item Technical improvements to the fitting procedure in the FGCM code have improved the overall stability of the calibration.

\end{itemize}

\begin{figure}
    \centering
    \includegraphics[width=0.5\columnwidth]{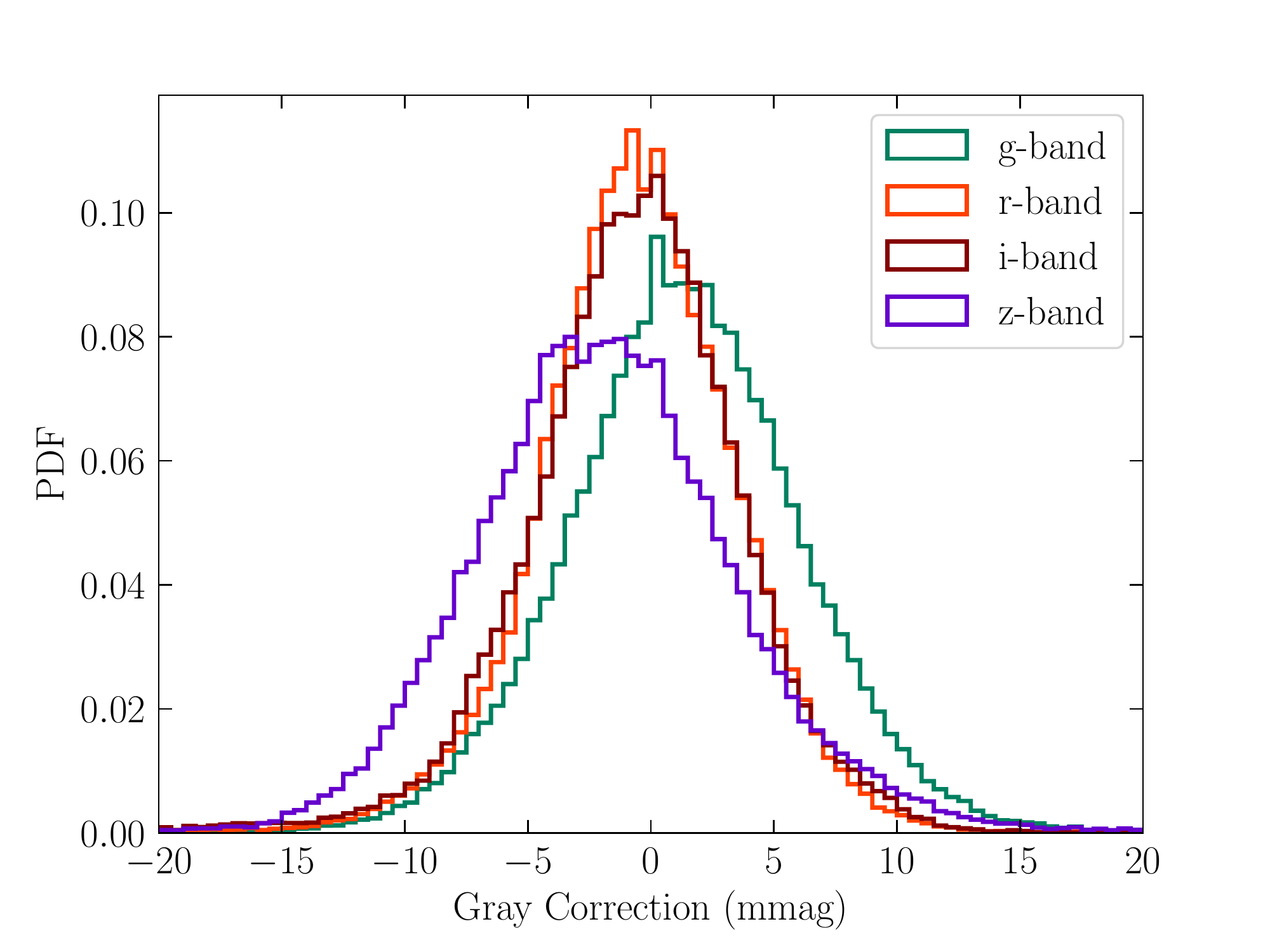}
    \caption{Distribution of multi-epoch zeropoint corrections ($\langle \delta m_i^{\rm ZP} \rangle$) that apply to the \gold release, updated since the DR1 release.}
\label{fig:gray_correction}
\end{figure}

The distribution of updated zeropoint corrections $\langle \delta m_i^{\rm ZP} \rangle$ that include both the AB magnitude and updated gray zeropoint corrections per \eqnref{multiepoch_zeropoint}, is plotted in \figref{gray_correction}.

\subsection{Estimating Per-Object Template SEDs: $F_{\nu}(\lambda)$}
\label{app:sed_templates}

We compute photometric corrections for every coadd object as a function of its spectral type, as defined by their colors, to account for the differences relative to the FGCM reference spectrum.

Except in special cases (in particular SNe Ia), we generally do not know and/or do not want to assume an intrinsic spectrum of a given source. 
Therefore, we must decide what source spectrum to use when computing chromatic corrections.
One could take an empirical approach and derive a linearized source spectrum directly from DES data, but this is problematic in bands at the boundaries of DES wavelength coverage (i.e., $g$ and $Y$) and for drop-outs, since the color is not well constrained.  
We instead identify a best-fit realistic spectrum as a first step of the chromatic corrections described above.

We divide sources into two sets: (1) clearly identified stars, and (2) galaxies + ambiguous sources, which will mainly be faint galaxies. 
For the secure stars, flux measurements in two or more bands are sufficient to identify a template source spectrum, since the stellar locus is narrow and approximately monotonic in color. 
We use the~\citet{Pickles:1998} stellar spectral library taken from the \BIGMACS code\footnote{\url{https://github.com/patkel/big-macs-calibrate}}, augmented with the bluest spectral templates from the original library. 
The \BIGMACS library does not cover the full range of stellar colors, however, its template library has some important advantages since the spectral resolution is increased relative to the initial library, reducing the scatter considerably for the reddish M stars. 
Secure stars are selected as follows:

\footnotesize
\begin{lstlisting}[language=sql]
WHERE (((mag_auto_r BETWEEN 5. AND 22.0) 
AND abs(wavg_spread_model_r) < 0.003) 
OR ((mag_auto_i BETWEEN 5. AND 22.0) 
AND abs(wavg_spread_model_i) < 0.003) 
OR
((mag_auto_r BETWEEN 5. AND 20.0) 
AND abs(spread_model_r) < 0.005 
AND abs(spread_model_i) < 0.005));
\end{lstlisting}
\normalsize

For galaxies, we use the COSMOS SED library and run the \lephare photo-$z$ code to identify a best-fit spectral template and redshift for each individual source. 
The initial fit uses the standard DES bandpass and fiducial reddening  correction (\appref{extinction}). 
Even if the initial best-fit spectrum is not fully accurate, the shape will be constrained at the level allowed by DES data alone. 
At this stage, the specific value of the best-fit galaxy redshift is actually not important, so long as the best-fit spectral \emph{shape} is  approximately correct.

The COSMOS galaxy SED library was chosen based on agreement between the colors of the templates and measured colors of the \gold galaxy sample, as seen in \figref{chro_cosmos}. 
Various tests indicated that alternative choices of SED library provided sub-optimal color matching, as they did not overlap some regions of color space occupied by the \gold galaxies.
We verified, using a random sub-sample, that the colors of the best-fit SEDs are not biased relative to the \gold colors, as seen in \figref{chro_residual}. 
The shape of the SEDs correctly represents the $griz$ colors of sources for all spectral types in the galaxy sample.

\begin{figure}
    \centering
	\includegraphics[width=0.5\columnwidth]{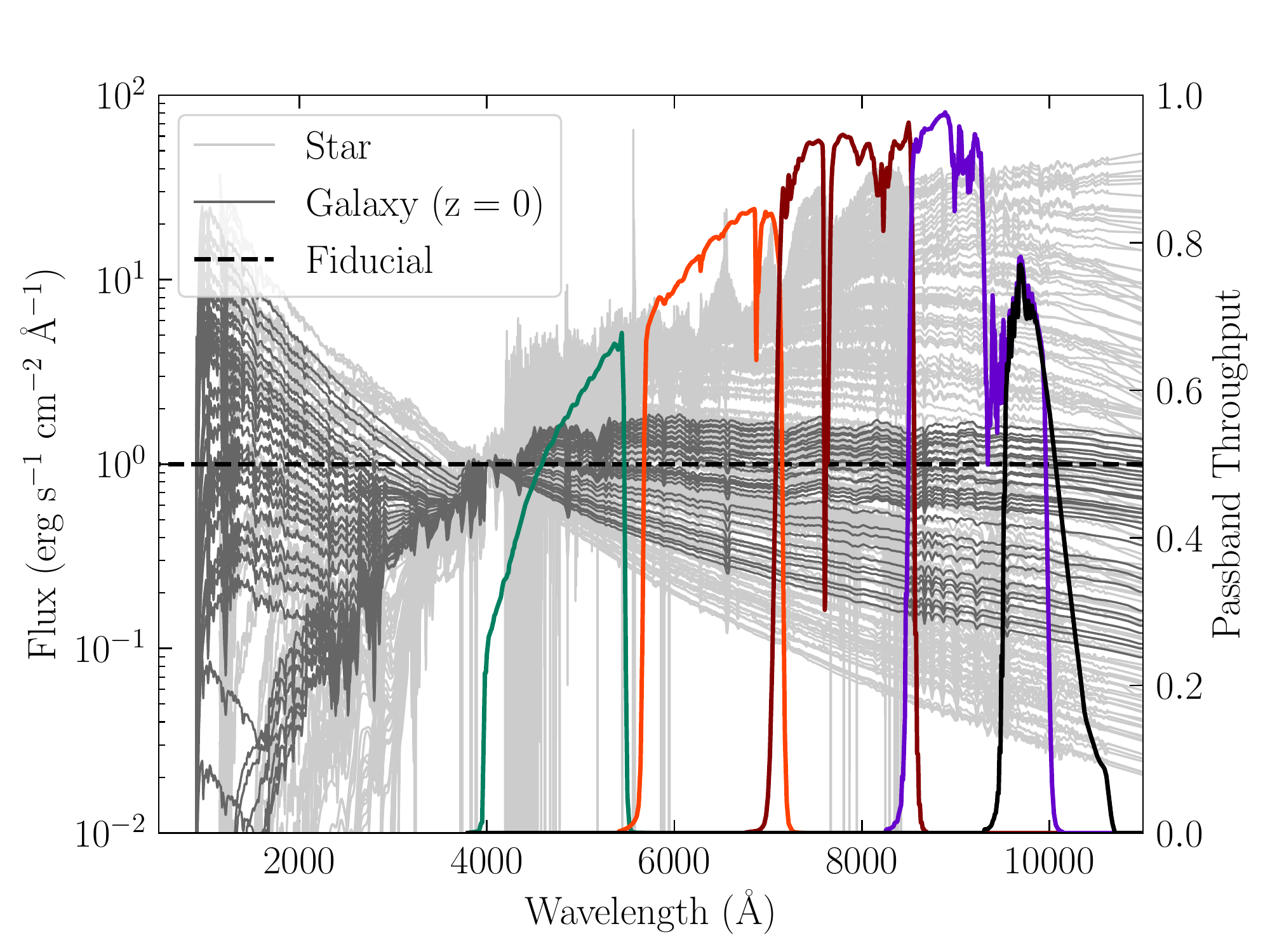}
    \caption{Spectral libraries used for chromatic corrections and SED-dependent interstellar extinction corrections. The galaxy and stellar SEDs are compared to a constant $F_{\nu}(\lambda)$ spectrum, used as the reference for the fiducial interstellar extinction correction.
    For galaxies, we use the COSMOS SED collection from ~\citet{Ilbert2009}, and the ~\citet{Pickles:1998} library for stars. In addition, we also show the DES spectral passbands covering from 4000 A to 12000 A.}
\label{fig:chro_speclib}
\end{figure}

\begin{figure}
    \centering
	\includegraphics[width=0.5\columnwidth]{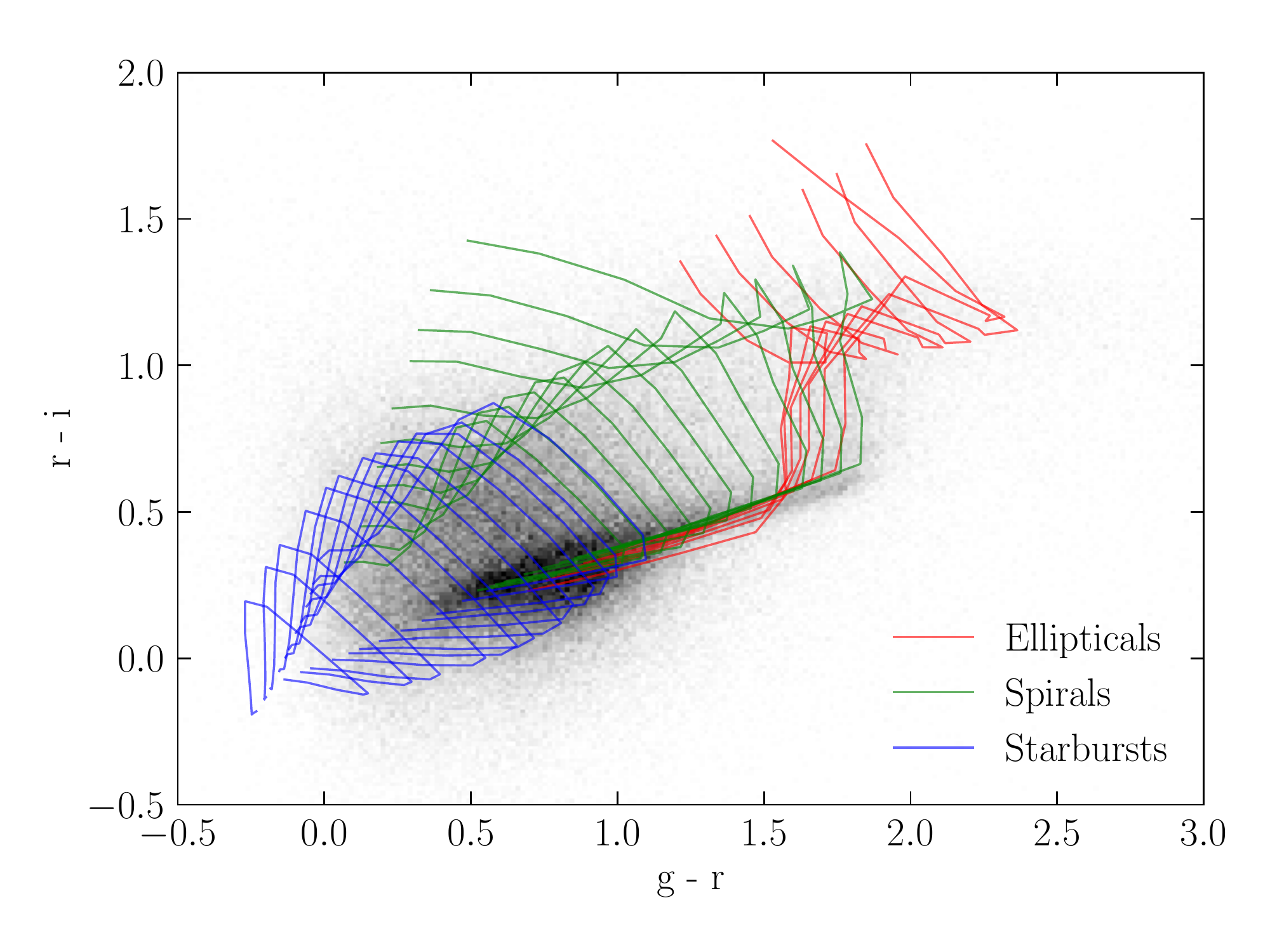}
    \caption{DES \gold galaxy photometry (\sof; black points) compared to predicted DECam colors for 31 COSMOS SED tracks \citep{Ilbert2009}. Each track represents a range in redshift, and is colored by galaxy type. Of the various galaxy SED libraries considered, COSMOS had the highest overlap with the observed \gold galaxy color locus.}
\label{fig:chro_cosmos}
\end{figure}

\begin{figure}
    \centering
	\includegraphics[width=0.5\columnwidth]{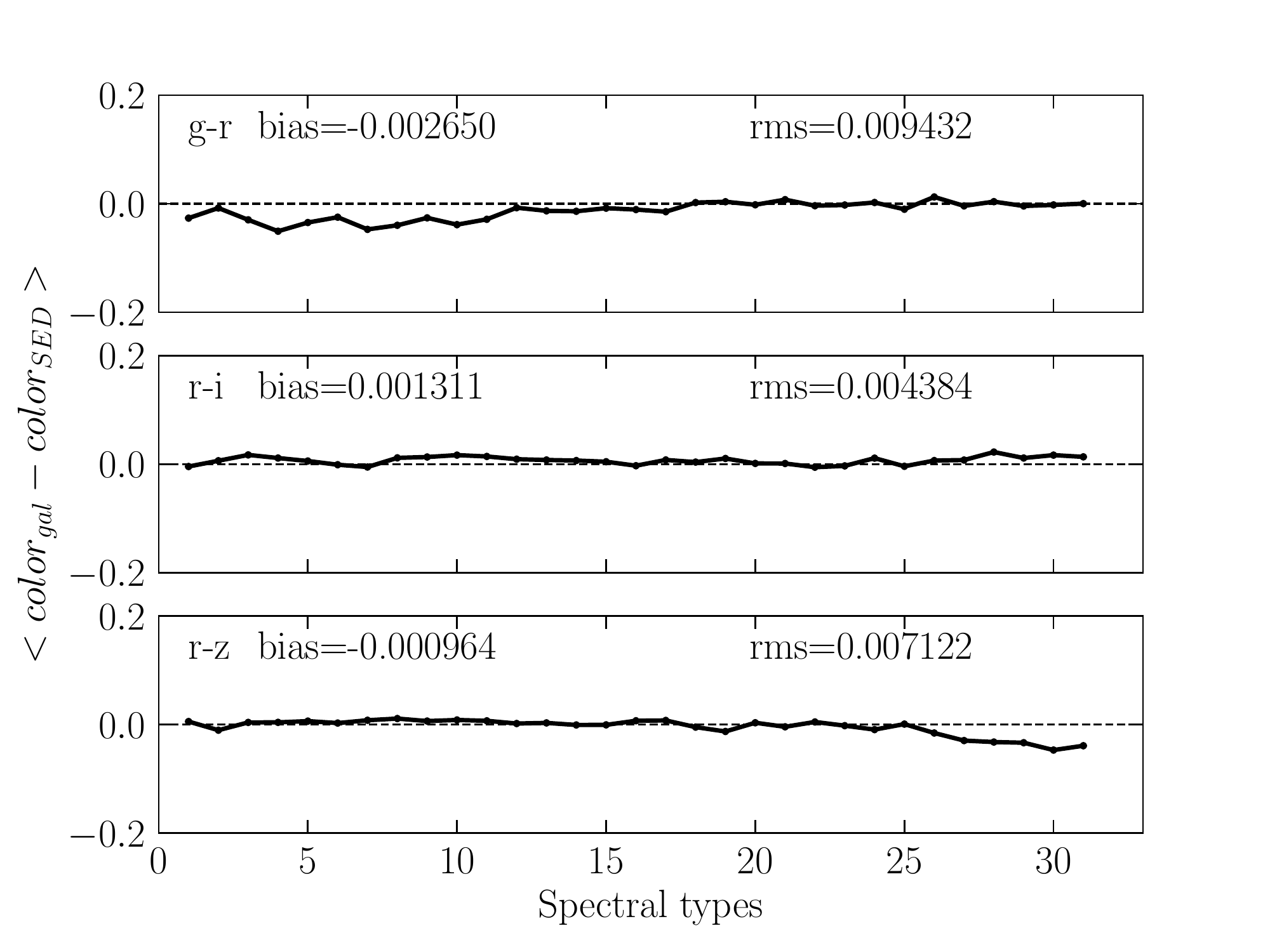}
    \caption{Mean difference between the colors of DES \gold and the colors of the best-fit SED as a function of SED, for a sub-sample of galaxies. In general, there is an excellent agreement.}
\label{fig:chro_residual}
\end{figure}

\figref{chro_speclib} shows the range of SED templates considered for both stars and galaxies, as well as flat $F_{\nu}(\lambda)$ spectrum, used as a reference for the fiducial interstellar extinction correction (\appref{extinction}).

\begin{figure}
    \centering
    \includegraphics[width=0.5\columnwidth]{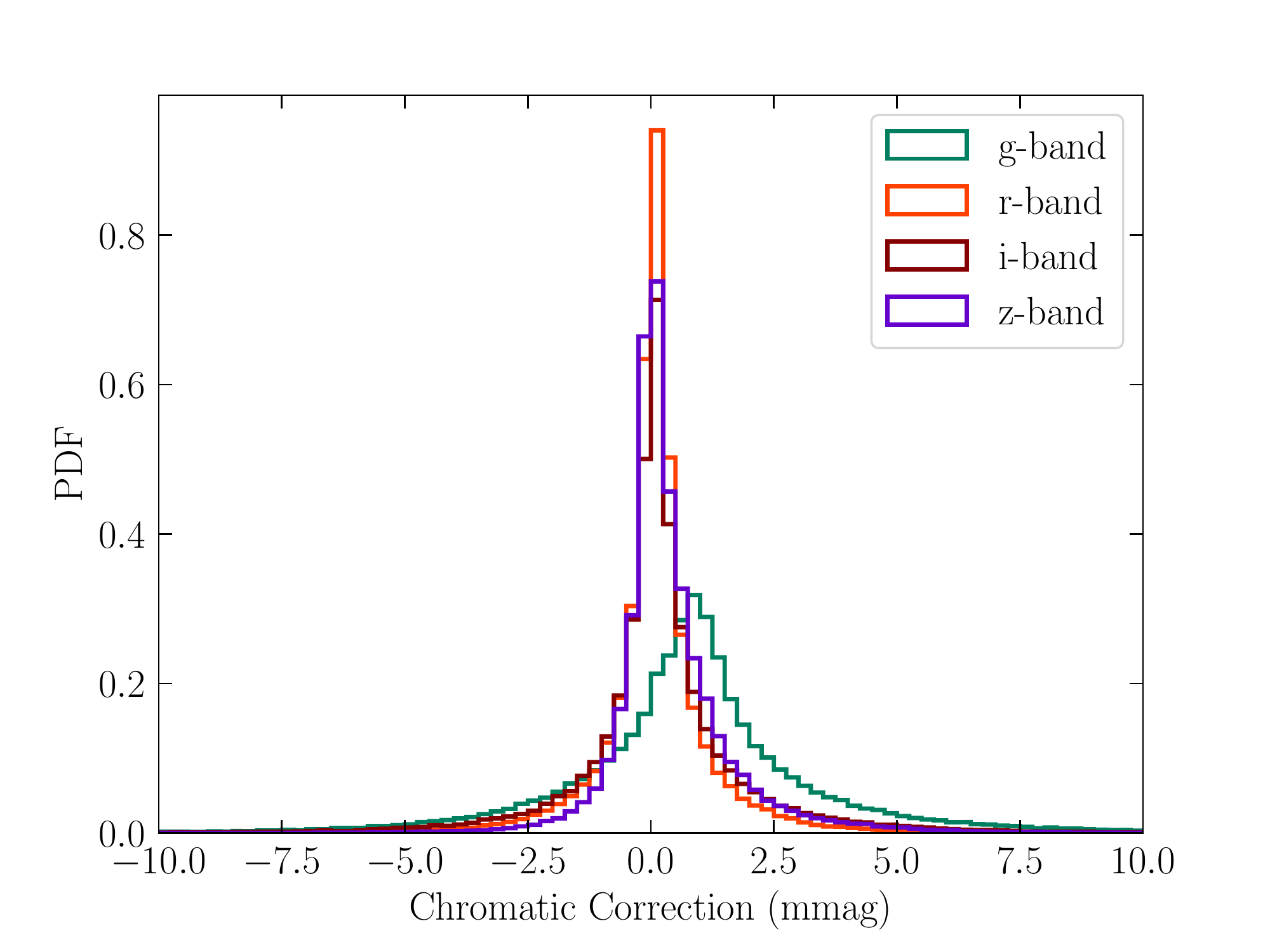}
    \caption{Distribution of chromatic corrections for $griz$ bands for a sub-sample of \gold. Compared with the gray zero-point corrections (\figref{gray_correction}), these are of smaller amplitude.}
\label{fig:chromatic_correction}
\end{figure}

In \figref{chromatic_correction} we show the distribution of chromatic corrections for the $griz$ bands. 
Even though chromatic corrections improve the photometric calibration and are therefore applied to \gold, its effect is typically at the \mmag level.
Also, we measured a negligible effect when we measured the effect of chromatic corrections on the recovered cosmological parameters on supernova science~\citep{Lasker:2019}, or as we saw in internal tests when estimating the photometric redshifts against a validation sample (\secref{photozs}). 

Chromatic corrections are needed when two conditions are both met (1) the observed passband differs from the Standard Passband and (2) the object SED is different from the reference SED.
Most objects differ from the flat $F_{\nu}(\lambda)$ reference spectrum adopted for Y3 processing, and chromatic corrections can be tens of \mmag in individual DECam exposures \citep{Y3FGCM}.
However, chromatic corrections are reduced in the coadd as the number of exposures increases because one typically averages over observing conditions and the observed passband approaches the Standard Passband. 
By contrast, SED-dependent effects do not average down for interstellar extinction.

\subsection{Interstellar Extinction Corrections}
\label{app:extinction}

In general, both $\tau(\lambda)$ and $a$ vary between lines of sight through the Galaxy. 
For our fiducial interstellar extinction correction, we will treat the reddening law $\tau(\lambda)$ as invariant with respect to Galactic coordinates over the DES footprint.

To obtain de-reddened (TOG) photometry, per-object corrections corresponding to four interstellar extinction models are delivered with \gold: one ``fiducial'' SED-independent interstellar extinction based on the $E(B-V)$ reddening map of \citet[SFD98]{sfd98}, and three SED-dependent models based on the reddening maps of SFD98, \citet{planck13}, and \citet{lenz17}, respectively.
The reddening maps of SFD98 and \citet{planck13} estimate the dust column density based on thermal emission, whereas \citet{lenz17} use the 21 cm emission of neutral hydrogen in our Galaxy as a dust proxy.

For the fiducial model, we assume a flat reference spectrum in $F_\lambda(\lambda)$ (i.e., constant value in units of ergs~s$^{-1}$~cm$^{-2}$~\AA$^{-1}$), which is roughly centered within the color space of stellar and galaxy SED templates (see \figref{chro_speclib}).
For the three SED-dependent models, we use the same per-object SED template identified for chromatic corrections.
We use the DES Y3A2 Standard Bandpass, and for each model, we adopt a \citet{fitzpatrick} reddening law with $R_V = 3.1$, consistent with the $E(B-V)$ map usage recommendations.
We consider a low-extinction limit for which the correction is linear with respect to $E(B-V)$ values.
Following \citet{sfdrecal}, we rescale the SFD98 reddening map by a factor $N=0.78$.

\begin{figure}
    \centering
    \includegraphics[width=0.5\columnwidth]{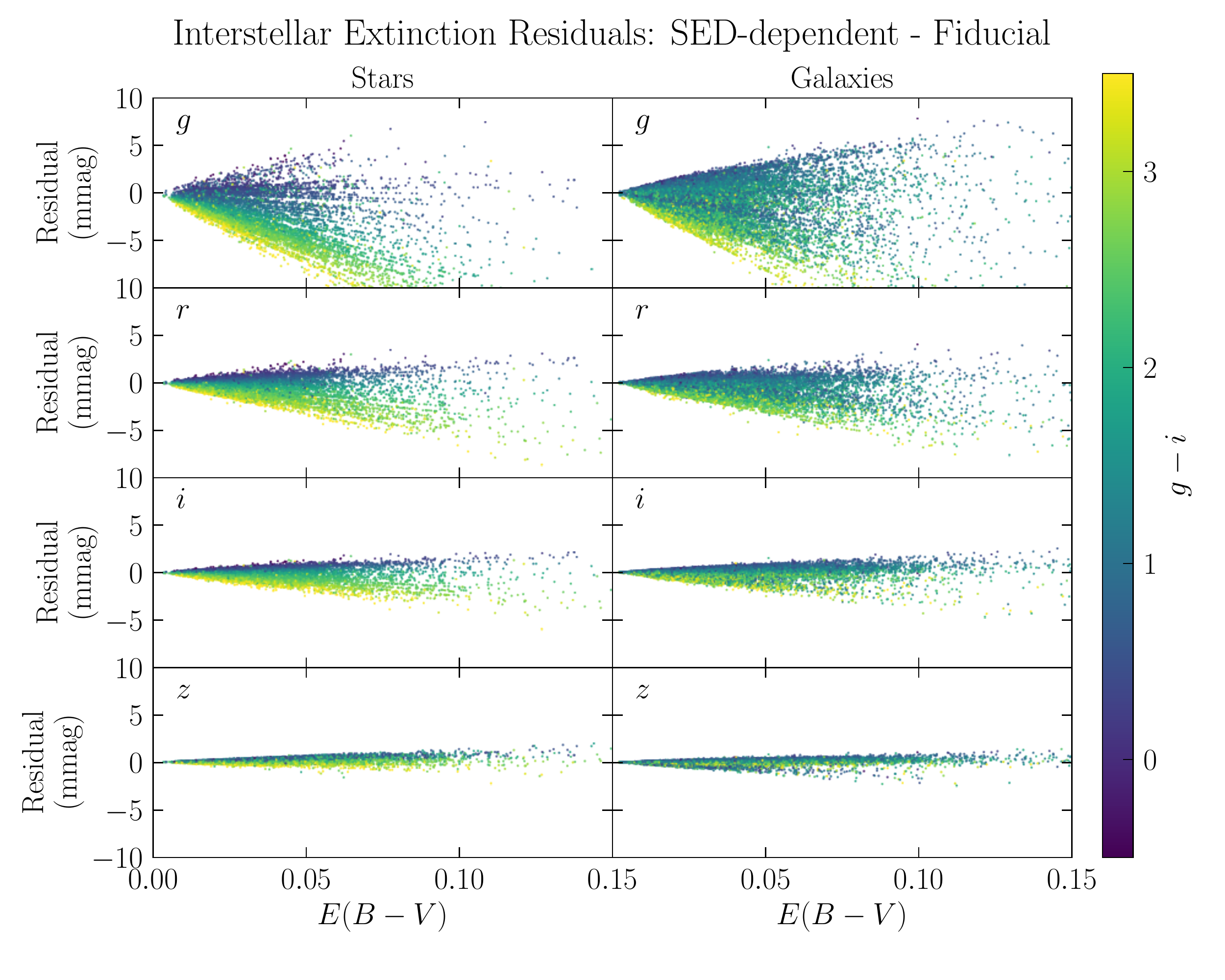}
\caption{SED-dependent interstellar extinction corrections by band.}
\label{fig:reddning_sed-dependence_band}
\end{figure}

\begin{figure}
    \centering
    \includegraphics[width=0.5\columnwidth]{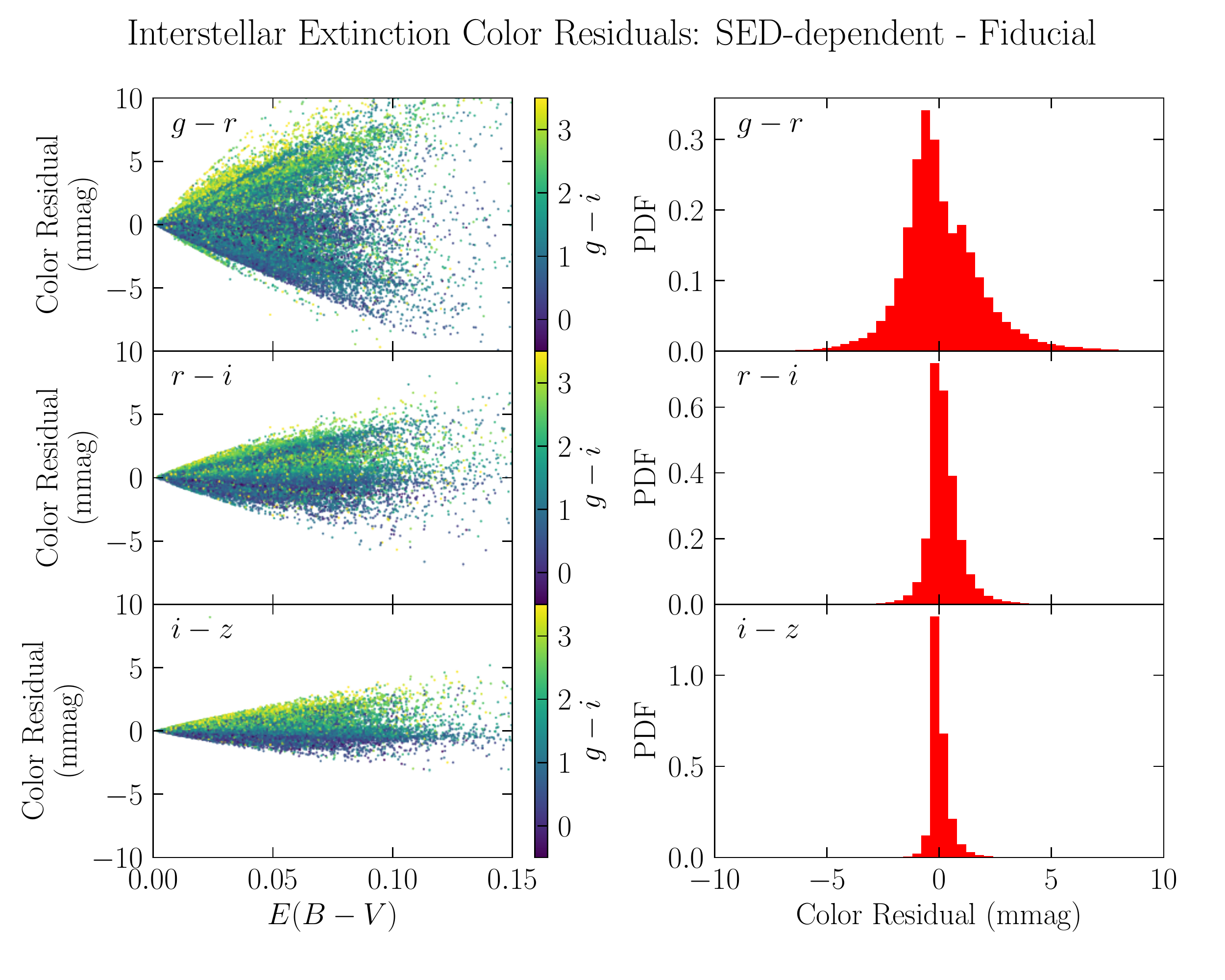}
\caption{SED-dependent interstellar extinction corrections by color.}
\label{fig:reddning_sed-dependence_color}
\end{figure}

\begin{figure}
    \centering
    \includegraphics[width=0.5\columnwidth]{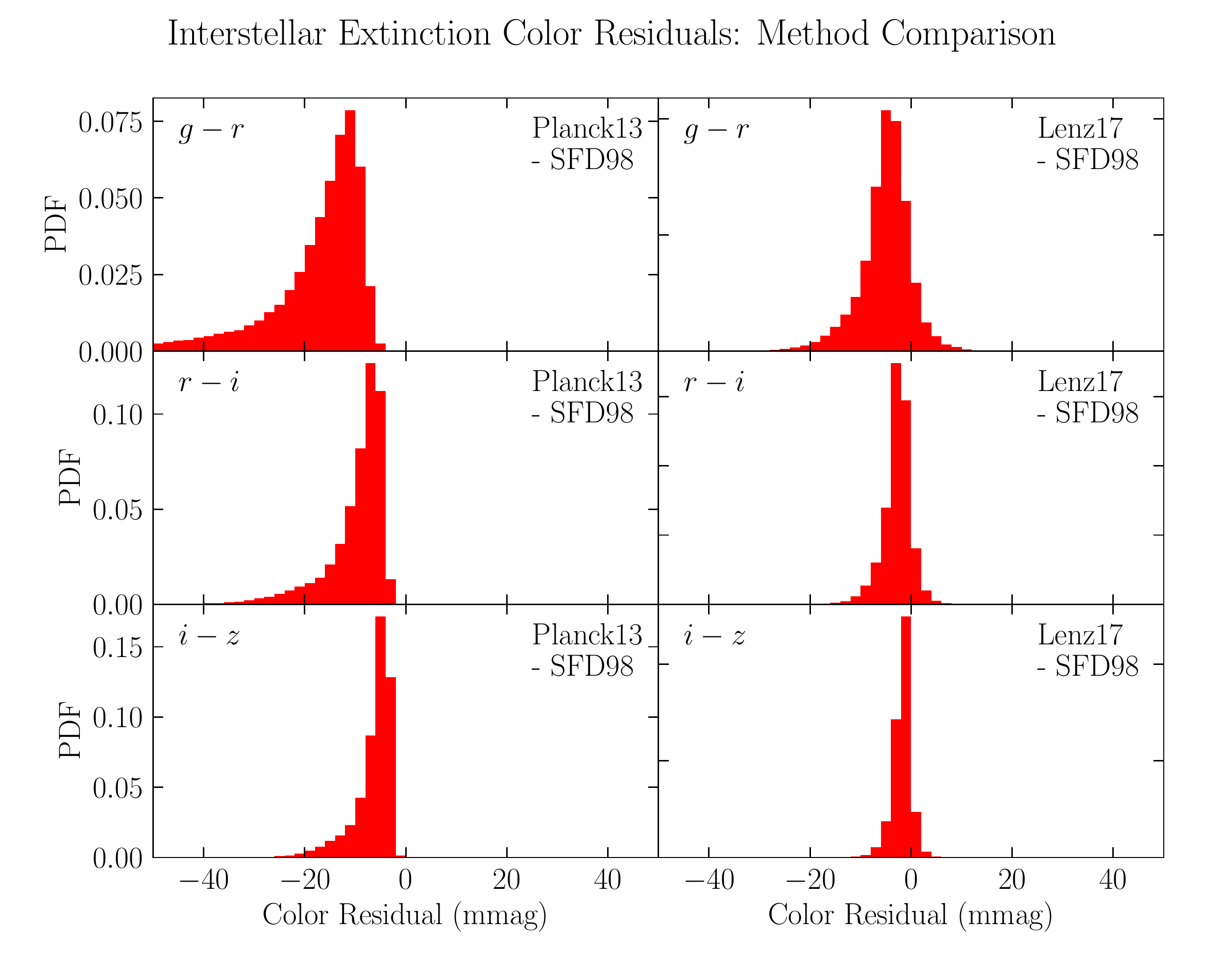}
\caption{Color residuals between several extinction $E(B-V)$ maps \citep{lenz17, planck13} and our fiducial choice, SFD98 \citep{sfd98}.}
\label{fig:reddning_method_dependence_color}
\end{figure}

The values of the SED-dependent extinction correction for stars and galaxies are shown in \figref{reddning_sed-dependence_band} and \figref{reddning_sed-dependence_color}.
\figref{reddning_method_dependence_color} shows the distribution of color residuals between several choices of reddening maps relative to SFD98.

\section{The \texttt{EXTENDED} object classifiers}
\label{app:extended}

The \texttt{EXTENDED} object classifiers were specifically designed for the \gold release. Different flavors correspond to different usages of shape-related quantities from the \gold data set, including \sextractor variables, \ngmix-based, or both. All of these classifiers are built according to the same logical structure, using the following equation:

\begin{equation}
    \texttt{EXTENDED\_CLASS} = \sum_{i=1}^{3}\:[\,var + E_i \cdot \:varerr > th_i]
\label{eq:extended}
\end{equation}

\noindent where $var$ and $varerr$ correspond to a specific morphological variable and its error, and the values $E_i$, $th_i$ are obtained according to the performance of the classifier against deeper imaging. For a given object, each time the condition is met in the summation in Equation \ref{eq:extended}, a unit is added to \texttt{EXTENDED\_CLASS}, therefore obtaining an integer value between 0 and 3. The parameters $E_i$, $th_i$ are chosen so that larger numbers correspond to more secure extended objects, whereas lower numbers correspond to more likely point-like objects. $\texttt{EXTENDED\_CLASS} = 0$ indicates high-confidence stars and QSOs. When $var$ cannot be computed for the particular object, a default value of $\texttt{EXTENDED\_CLASS} = -9$ is assigned. In \tabref{extended} we provide the specific parameters used for each classifier in \gold. The `\texttt{MASH}' variants default to \texttt{EXTENDED\_CLASS\_COADD} for those objects with unavailable \sof or \mof information.

\begin{deluxetable}{c c c c c}[h]
\tablewidth{0pt}
\tabletypesize{\tablesize}
\tablecaption{ \texttt{EXTENDED\_CLASS} detailed description, including input variables and parameter values \label{tab:extended}}
\tablehead{
\colhead{Classifier name} & \colhead{$var$} & \colhead{$varerr$} & \colhead{$E_{1,2,3}$} & \colhead{$th_{1,2,3}$} 
}
\startdata
\texttt{EXTENDED\_CLASS\_SOF} & \texttt{SOF\_CM\_T} & \texttt{SOF\_CM\_T\_ERR} & ($5$,$1$,$-1$) & ($0.1$,$0.05$,$0.02$) \\
\texttt{EXTENDED\_CLASS\_MOF} & \texttt{MOF\_CM\_T} & \texttt{MOF\_CM\_T\_ERR} & ($5$,$1$,$-1$) & ($0.1$,$0.05$,$0.02$) \\
\texttt{EXTENDED\_CLASS\_COADD} & \texttt{SPREAD\_MODEL} & \texttt{SPREADERR\_MODEL} & ($3$,$1$,$-1$) & ($0.005$,$0.003$,$0.001$) \\
\texttt{EXTENDED\_CLASS\_WAVG} & \texttt{WAVG\_SPREAD\_MODEL} & \texttt{WAVG\_SPREADERR\_MODEL} & ($3$,$1$,$-1$) & ($0.005$,$0.003$,$0.001$) \\
\enddata
\tablecomments{See Equation \ref{eq:extended} for details on the expression. \texttt{EXTENDED\_CLASS\_SOF} and \texttt{EXTENDED\_CLASS\_MOF} have essentially the same performance. \texttt{SPREAD\_MODEL} and \texttt{SPREADERR\_MODEL} are \sextractor outputs described in \citep{Desai:2012, Bouy:2013}.
}
\end{deluxetable}

\begin{figure}
    \centering
	\includegraphics[width=0.5\columnwidth]{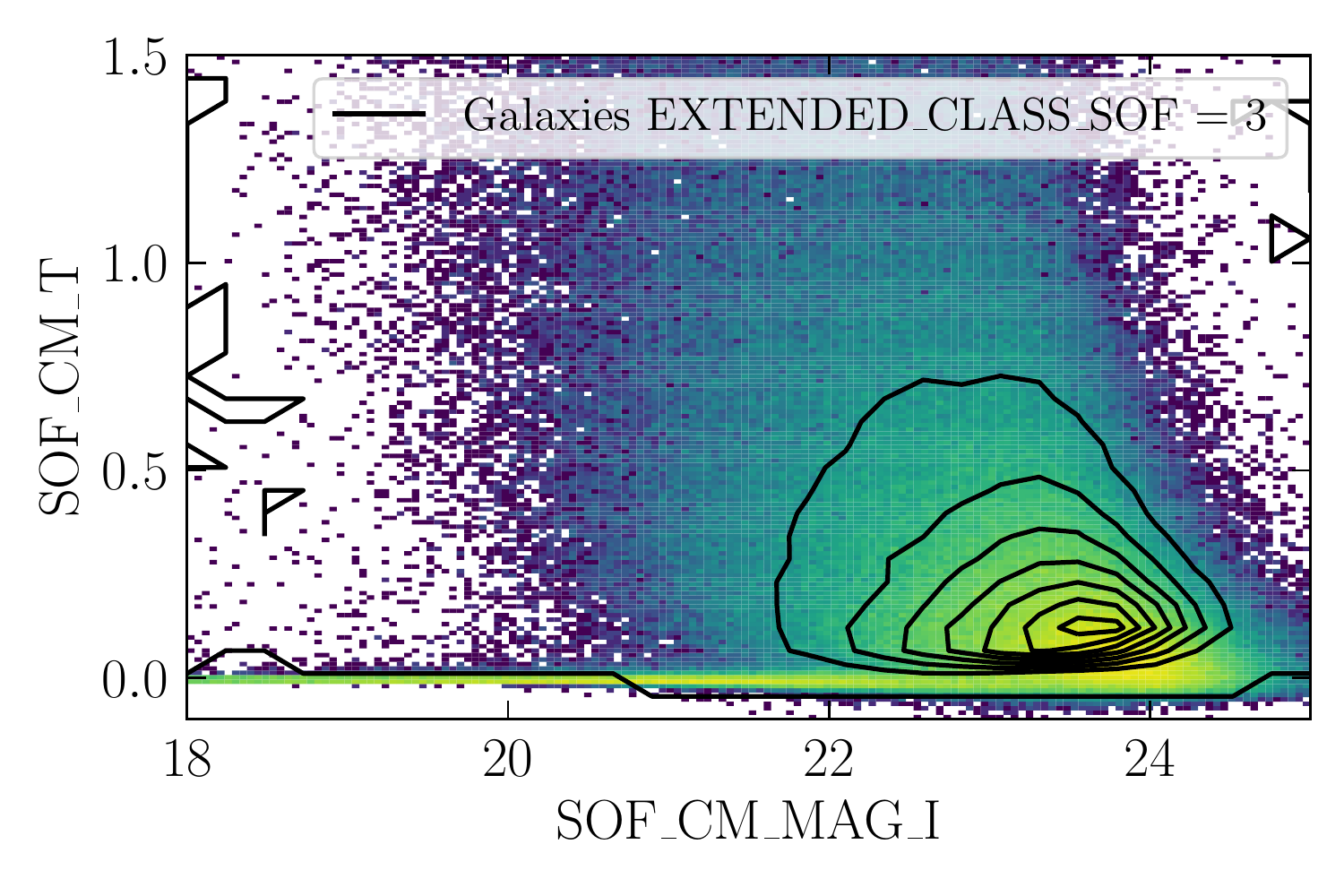}
    \caption{The distribution of \sofcmt, the basis for the \sofmash classifier, is shown as a function of magnitude. The heatmap shows the overall distribution of objects, whereas the contour lines indicate where the objects selected as galaxies (\sofmash = 3) lie in this parameter space.}
\label{fig:cmtdistrib}
\end{figure}

The distribution of \sofcmt, the basis for the \sofmash classifier, is shown as a function of magnitude in the $i$ band in \figref{cmtdistrib}.

\section{Main catalog columns}
\label{app:catalogcols}

In \tabref{catalog_columns} we summarize the essential columns of the \gold data set with their brief description. Full details will be provided upon release at \url{https://des.ncsa.illinois.edu/releases/}.

\begin{deluxetable}{c c l}[h]
\tablewidth{0pt}
\tabletypesize{\tablesize}
\tablecaption{ Selected \gold catalog columns. \label{tab:catalog_columns}}
\tablehead{
\colhead{\gold catalog column family}  & \colhead{Units} & \colhead{Description} 
}
\startdata
COADD\_OBJECT\_ID & & Unique identifier for a Y3 coadd object\\
RA, DEC, B, L & Degrees & Equatorial and Galactic coordinates\\
ALPHAWIN\_J2000, DELTAWIN\_J2000 & Degrees & \makecell[cl]{Equatorial coordinates using a Gaussian-windowed \\ measurement (for precise astrometry)}\\
(SOF/MOF)\_(CM/PSF)\_(MAG/FLUX)\_(GRIZ) & \makecell[cc]{Magnitudes\\Counts per s} & \makecell[cl]{Photometry as measured by the multi-epoch, \\ multi-band pipeline defined in \secref{mofsof},\\ for a composite galaxy model or a PSF-like one}\\
(SOF/MOF)\_(CM/PSF)\_(MAG/FLUX)\_ERR\_(GRIZ)& \makecell[cc]{Magnitudes\\Counts per s} & Estimated error for the above\\
A\_FIDUCIAL\_(GRIZY) & Magnitudes & \makecell[cl]{SED-independent interstellar extinction based on the $E(B-V)$\\ reddening map of \citet[SFD98]{sfd98}} \\
A\_SED\_(SFD98/LENZ13/PLANCK17)\_(GRIZY) & Magnitudes & \makecell[cl]{SED-dependent interstellar extinction based on the $E(B-V)$ \\ reddening maps of \citet{sfd98}, \citet{lenz17}, and \\ \citet{planck13}}\\
DELTA\_MAG\_CHROM\_(GRIZY) & Magnitudes & Coadd-object chromatic correction \\
DELTA\_MAG\_Y4\_(GRIZY) & Magnitudes & Updates to photometry from Y4 imaging\\
(SOF/MOF)\_CM\_MAG\_CORRECTED\_(GRIZ) & \makecell[cc]{Magnitudes\\Counts per s} &  \makecell[cl]{Corrected CM\_MAG quantities:\\ (SOF/MOF)\_CM\_MAG\_(GRIZ) + DELTA\_MAG\_Y4\_(GRIZ) + \\ + DELTA\_MAG\_CHROM\_(GRIZ)- A\_SED\_SFD98\_(GRIZ)}\\ 
(SOF/MOF)\_CM\_T & $\asec^2$ & Size squared of the object: $T = \langle x^2 \rangle  + \langle y^2 \rangle$\\
(SOF/MOF)\_CM\_T\_ERR & $\asec^2$ & Estimate of error in CM\_T\\
EXTENDED\_CLASS\_MASH\_(SOF/MOF) & & \makecell[cl]{Classification code for the `extendedness' of object, \\from 0 (point-like) to 3 (extended-like)}\\
FLAGS\_FOOTPRINT & & Flag indicating that the object belongs to \gold\\
FLAGS\_GOLD & & Flag showing possible processing issues with the object\\
FLAGS\_FOREGROUND & & \makecell[cl]{Flag showing that the object is in the area of influence of \\ a foreground object from an imaging point of view}\\
FLAGS\_BADREGIONS & & \makecell[cl]{Flag showing that the object is in an area with generalized \\issues in processing or data quality}\\
DNF\_(ZMC/ZMEAN/ZSIGMA)\_(MOF/SOF) & & DNF photo-z statistics for the object \\ 
BPZ\_(ZMC/ZMEAN/ZMODE/ZSIGMA/ZSIGMA68)\_(MOF/SOF) & & BPZ photo-z statistics for the object \\ 
BPZ\_TEMPLATE\_ID\_(MOF/SOF) & & BPZ template identifier\\ 
\enddata
\tablecomments{
Names in parentheses show options for a given type of column separated by slashes for each column. In addition several \sextractor quantities are available as well. Full details at \drurl.
}

\end{deluxetable}

\section{Photometric transformation equations with other systems}
\label{app:photometric_transformations}

In this Appendix we present transformation equations based on SDSS DR13 and DES Y3A1\_FINALCUT single-epoch data\footnote{\url{http://www.ctio.noao.edu/noao/node/5828\#transformations}} \citep{stringer:2019}. The zeropoint (the constant term) in each relation was derived by comparing the observed SDSS DR13 vs. Y3A1\_FINALCUT relation with its \citet{Pickles:1998} synthetic counterpart, and then manually refining the zeropoint (the constant term) to  match the calibration of the Y3A1\_FINALCUT FGCM standard stars (v2.5).

The $ugr$ transformations apply for stars with $0.2 \le (g-r)_{sdssdr13} < 1.2$. The $izY$ transformations apply for stars with $0.0 \le (i-z)_{sdssdr13} < 0.8$.

\begin{align}
u_{des} & = u_{sdssdr13} - 0.479 + 0.466 \times (g-r)_{sdssdr13} - 0.350 \times (g-r)_{sdssdr13}^2 \\
g_{des} & = g_{sdssdr13} + 0.001 - 0.075 \times (g-r)_{sdssdr13} \\
r_{des} & = r_{sdssdr13} \, - 0.009 - 0.069 \times (g-r)_{sdssdr13} \\
i_{des} & = i_{sdssdr13} \: + 0.014 - 0.214 \: \times \: (i-z)_{sdssdr13} \: - 0.096 \times (i-z)_{sdssdr13}^2  \\
z_{des} & = z_{sdssdr13} \, + 0.022 - 0.068 \, \times \: (i-z)_{sdssdr13} \\ 
Y_{des} & = z_{sdssdr13} \, + 0.045 - 0.306 \, \times \: (i-z)_{sdssdr13} 
\end{align}

With errors expressed in RMS being $RMS_u = 0.055$, $RMS_g = 0.021$, $RMS_r = 0.021$, $RMS_i = 0.023$, $RMS_z = 0.025$ and $RMS_Y = 0.030$ for stars in this color range.

We also provide here the transformation equations with HSC-SSP DR2 for the $griz$ bands (available for our default \sof photometry). In this case, we adopted a simpler approach by which we downloaded a bright sample of stars from the HSC-SSP DR2 catalog, and matched positionally to the corresponding \gold stars, as defined by $\sofmash = 0$. This way we obtain the following fitted coefficients:

\begin{align}
g_{des} &= g_{hscdr2} + 0.011490 - 0.0167 \; \; \; \; \, \times (g - r)_{hscdr2}  \\
r_{des} &= r_{hscdr2} \, - 0.015233 - 0.127021 \times (r - i)_{hscdr2}  \\
i_{des} &= i_{hscdr2} \; - 0.002067 - 0.12845 \; \; \times (i - z)_{hscdr2}  \\
z_{des} &= z_{hscdr2} \, + 0.006933 - 0.31025 \; \; \times (z - Y)_{hscdr2} 
\end{align}

\section{Survey Property Maps}
\label{app:sps}

Survey property maps are computed from a base  \mangle polygon file and converted to \healpix maps as follows for each quantity. First we divided the sky in \healpix pixels with $\nside = 32768$, which corresponds to $1.61 \times 10^9$ pixels in the DES 
footprint ($0.01 \asec^{2}/\pix$). Then, for each of these pixels, given the right ascension and declination of the pixel center, we look into the \mangle mask to obtain the value of the physical quantity of interest at the given position. With this, we have pixelized the \mangle mask to a resolution of $\nside = 32768$. From here, we downgrade the resolution to the desired final \nside. For Y3, we select $\nside = 4096$ as our default choice ($0.73838383838383838383838383838383838383838383838383838383838383838383838383838 \amin^{2}/\pix$). To do this, we average the values of the 64 smaller pixels that are contained into one \nside = 4096 pixel (for a visual interpretation of this process we refer to Figure 9 in~\citealt{y1gold}). 

The \texttt{FRACDET} maps are assembled in a similar fashion, but using star and bleed-trail mask as the source for information on regions in the sky that have been compromised in the images. At $\nside=32768$, whenever a pixel is not contained in the magnitude limit map (consider it as the observation map), or masked by a bright star or a bleed-trail, the small pixel is given \texttt{UNSEEN} value. Then, each $\nside=4096$ pixel takes a value corresponding to the fraction of pixels that have been observed, for example, from the 64 higher resolution pixels within.
In the combined coverage map, when we use many bands, $griz$ or $grizY$, the bleed-trail and bright star mask is combined at the level of $\nside=32768$, where in this resolution, we impose detection in all the given bands, if any of the selected bands, is \texttt{UNSEEN}, then that sub-pixel will be set to \texttt{UNSEEN}.

\begin{figure}[h]
\centering
\includegraphics[width=0.75\textwidth]{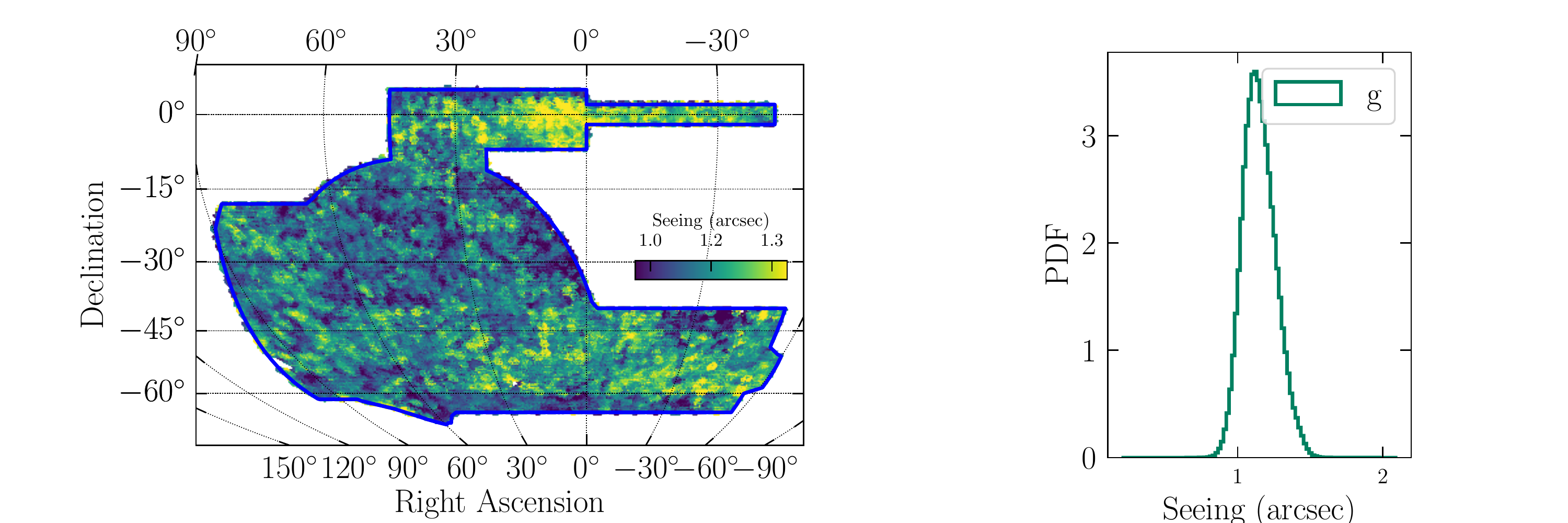}
\includegraphics[width=0.75\textwidth]{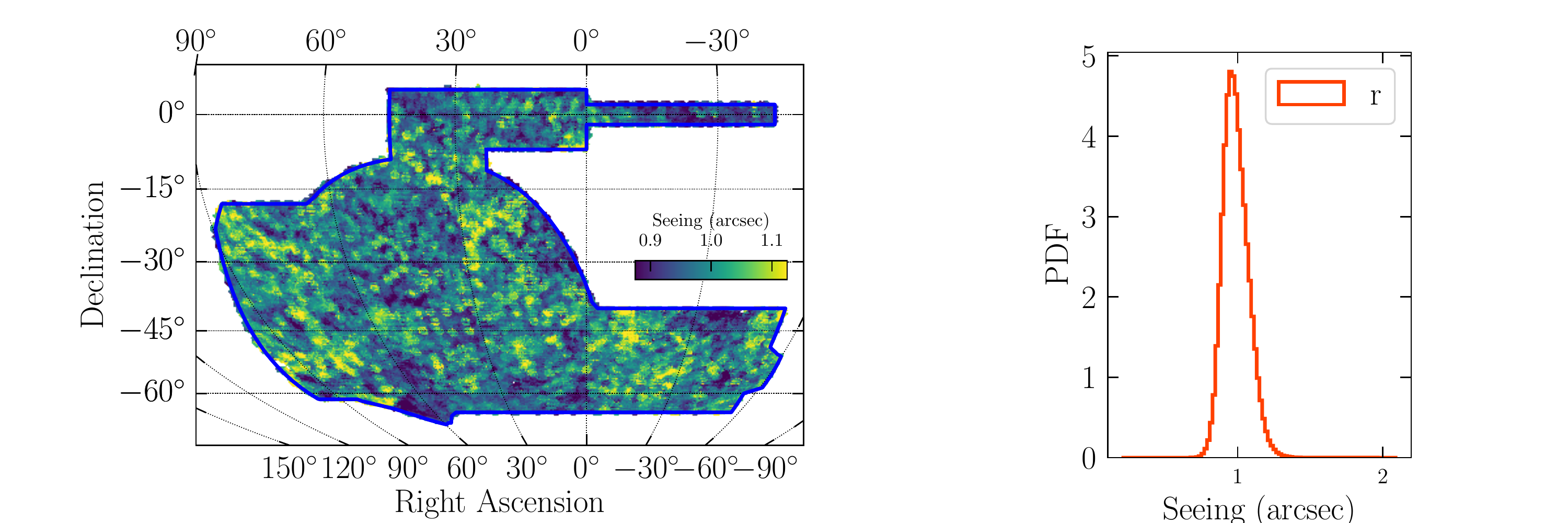}
\includegraphics[width=0.75\textwidth]{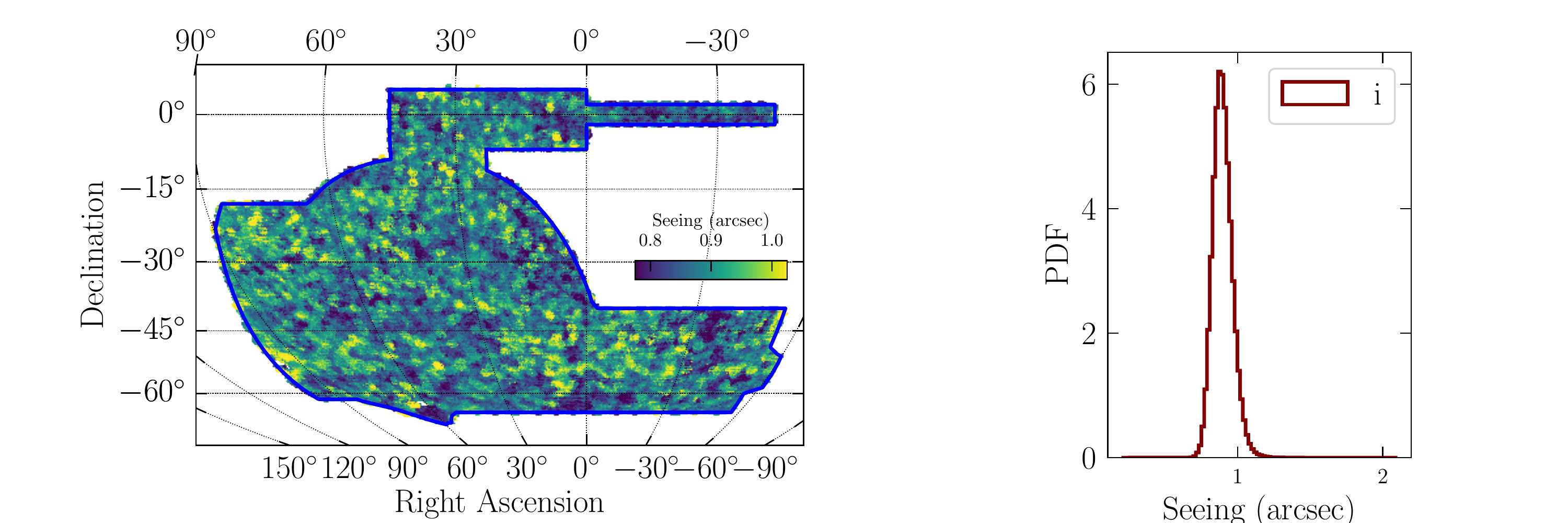}
\includegraphics[width=0.75\textwidth]{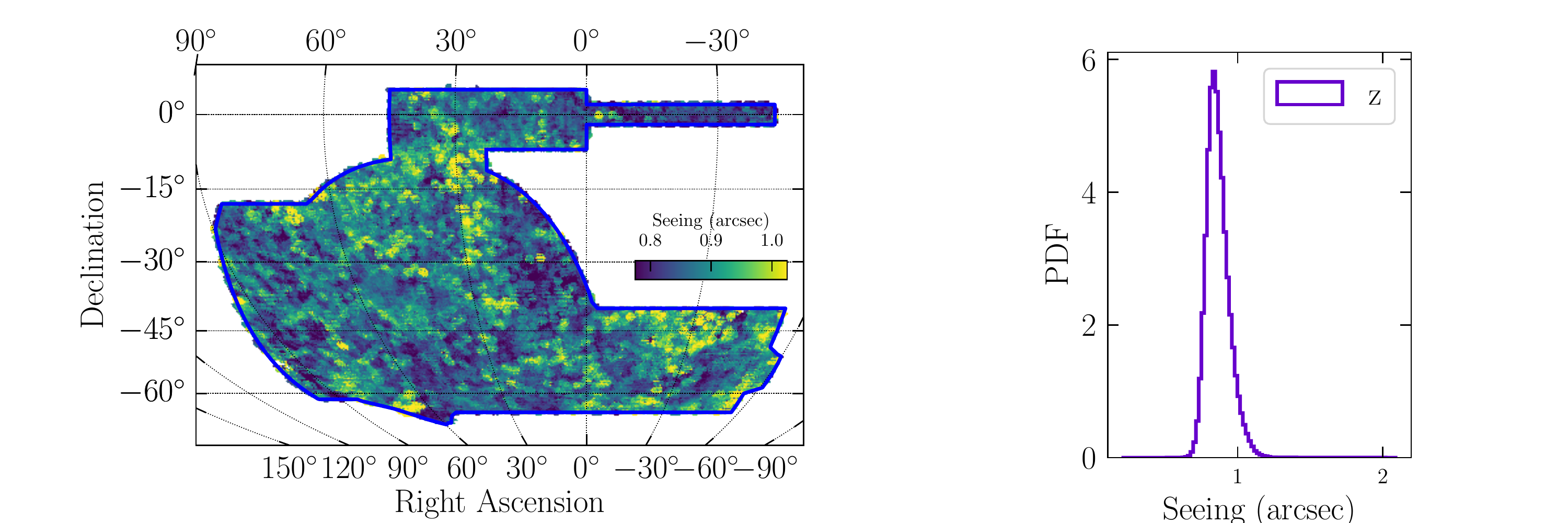}
\includegraphics[width=0.75\textwidth]{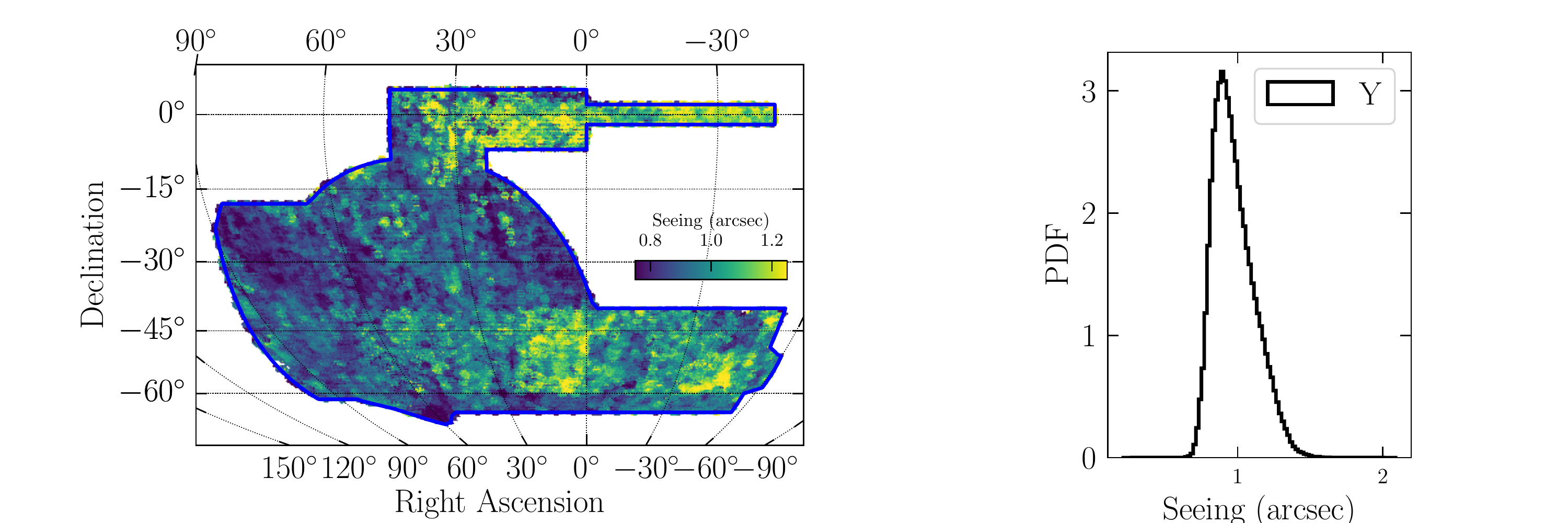}
\caption{\label{fig:app_fwhm} 
Sky maps and histograms of the seeing (FWHM.WMEAN) for each of the observed bands. The value at each location is the inverse-sky-variance-weighted sum of all individual exposures of that \healpix pixel.}
\end{figure}

\begin{figure}[h]
\centering
\includegraphics[width=0.75\columnwidth]{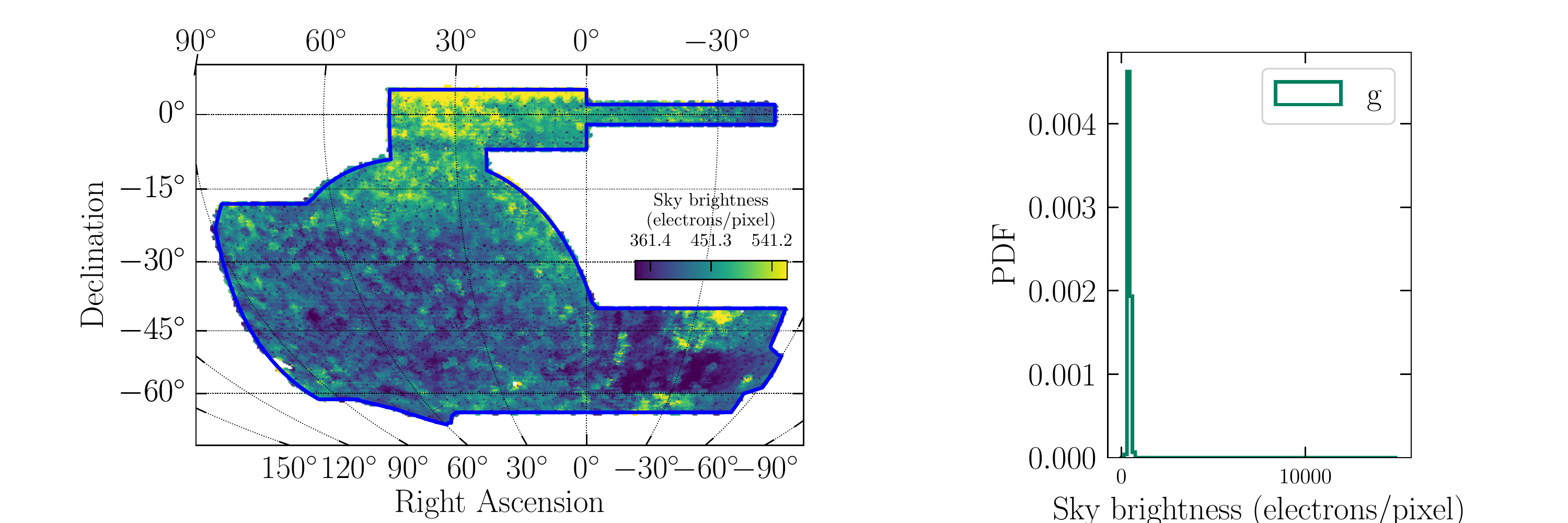}
\includegraphics[width=0.75\textwidth]{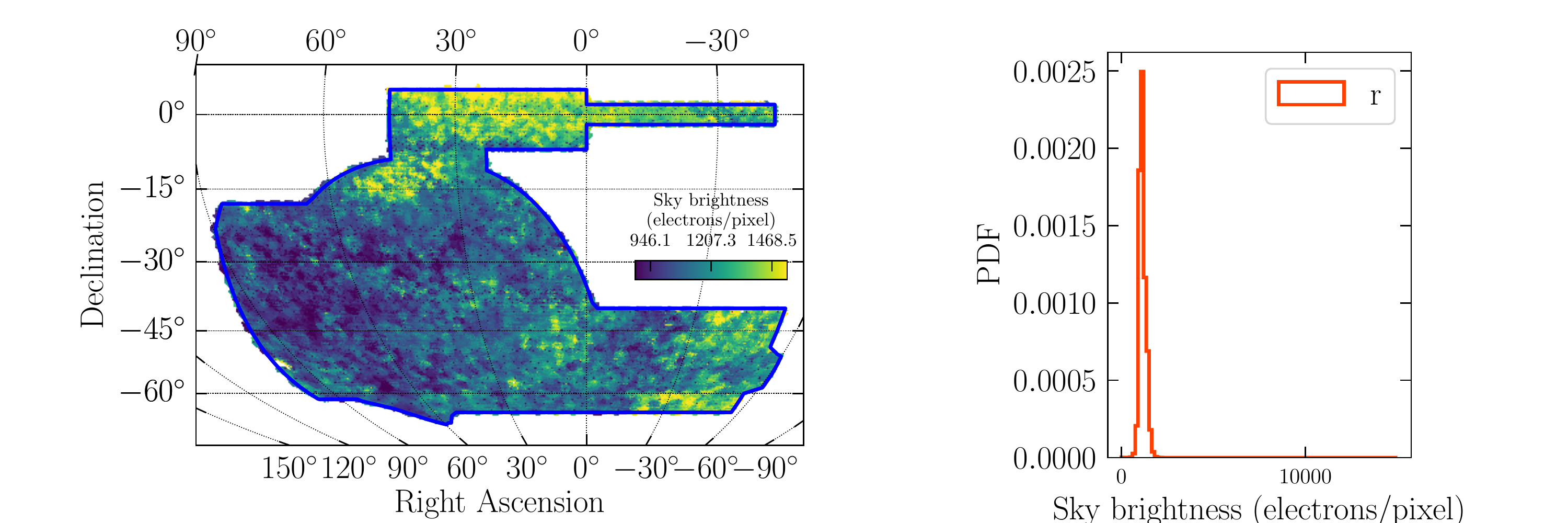}
\includegraphics[width=0.75\textwidth]{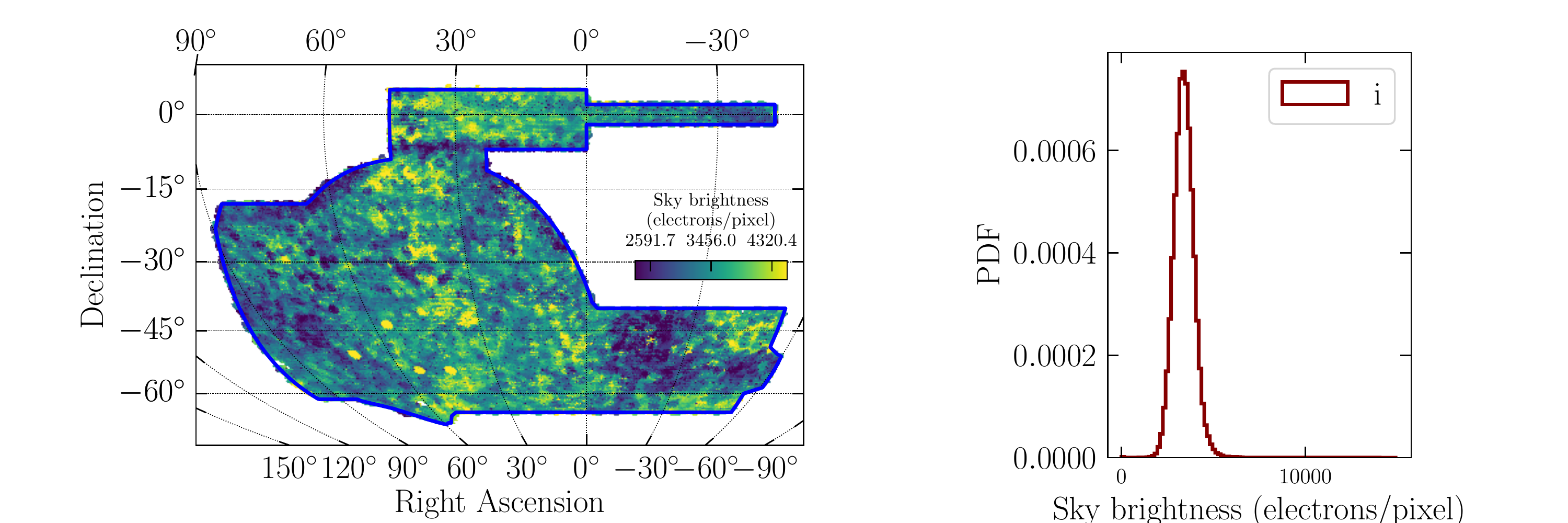}
\includegraphics[width=0.75\textwidth]{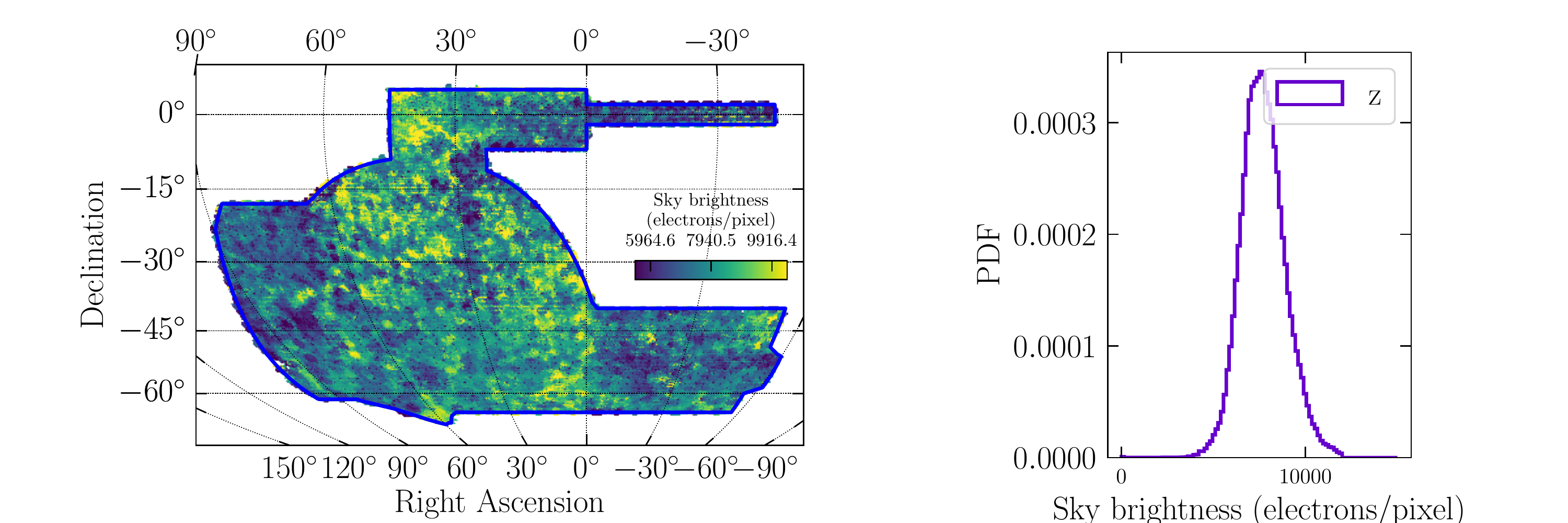}
\includegraphics[width=0.75\textwidth]{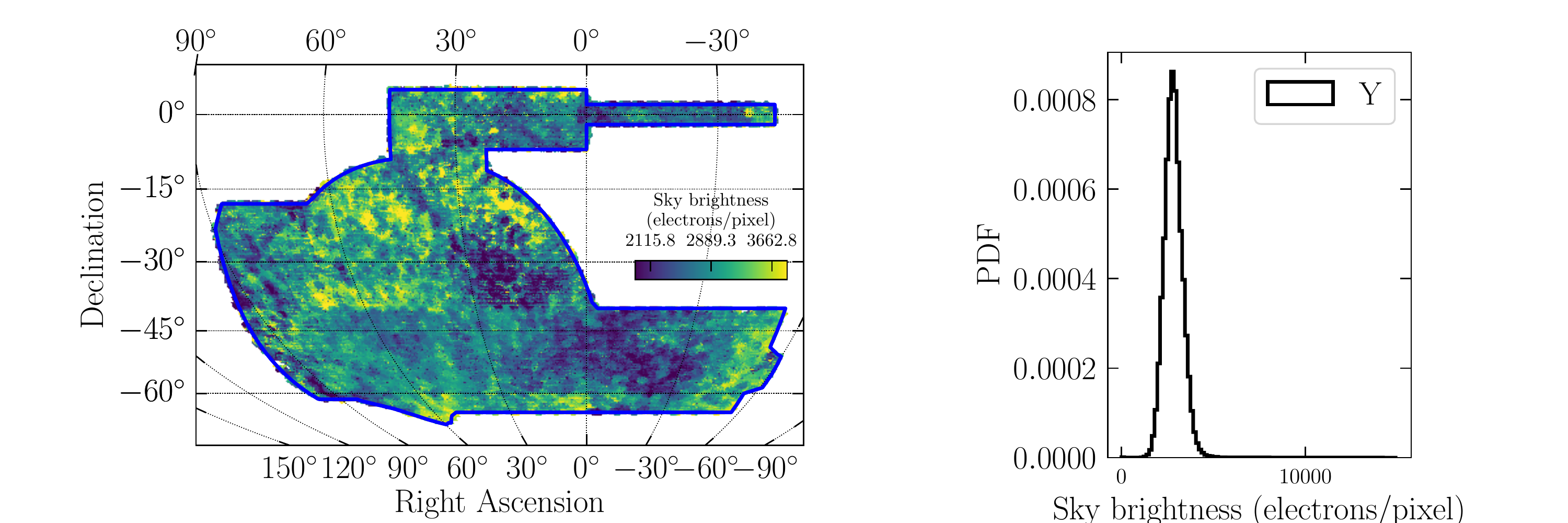}
\caption{\label{fig:app_skybrite} 
Sky maps and histograms of the sky brightness (SKYBRITE.WMEAN) for each of the observed bands. The value at each location is the inverse-sky-variance-weighted sum of all individual exposures of that \healpix pixel. Note that for Y3 data, the $Y$ band contains only 45 s exposures.}
\end{figure}

\begin{figure}[h]
\centering
\includegraphics[width=0.75\textwidth]{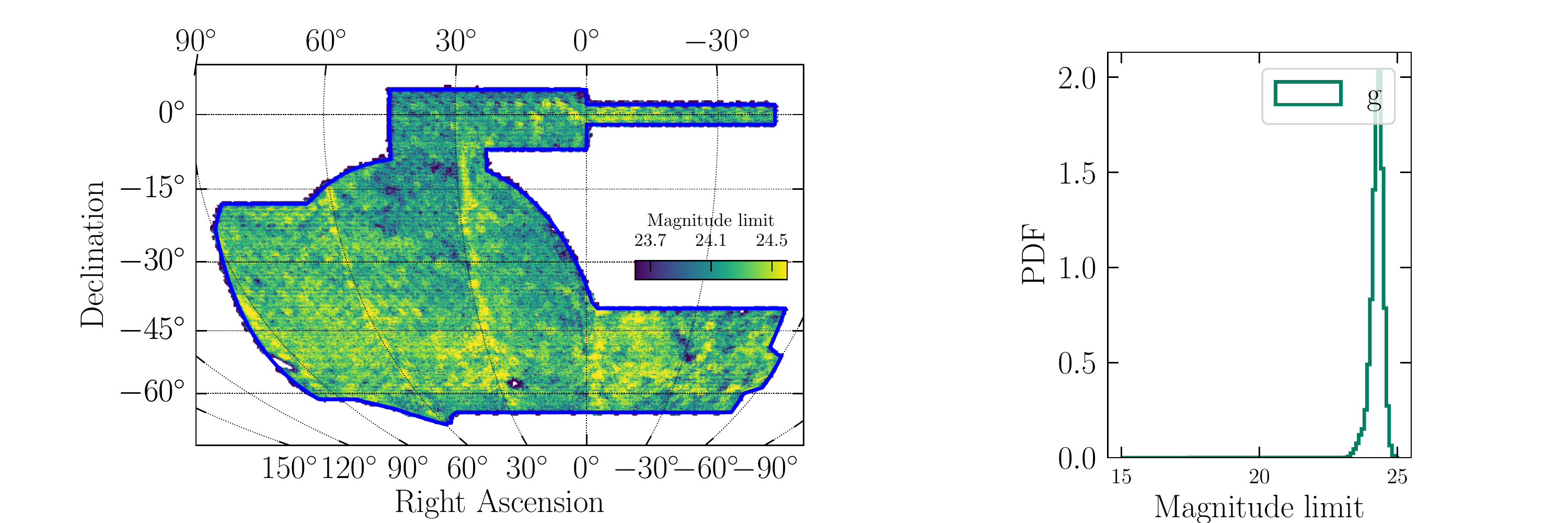}
\includegraphics[width=0.75\textwidth]{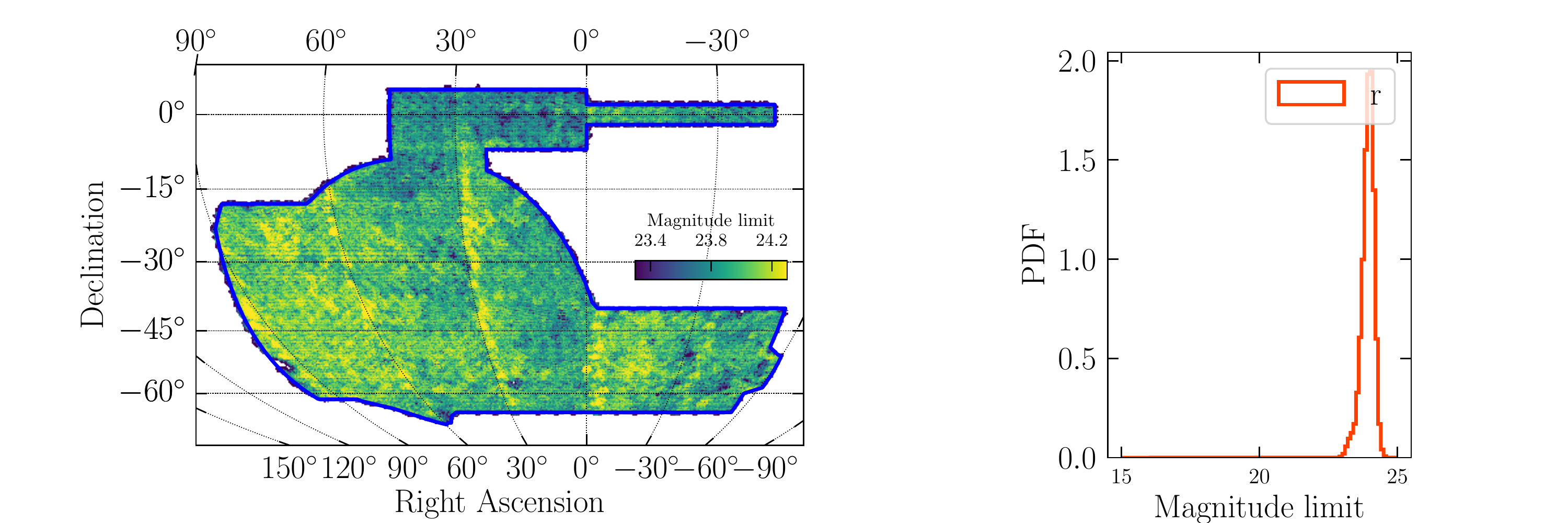}
\includegraphics[width=0.75\textwidth]{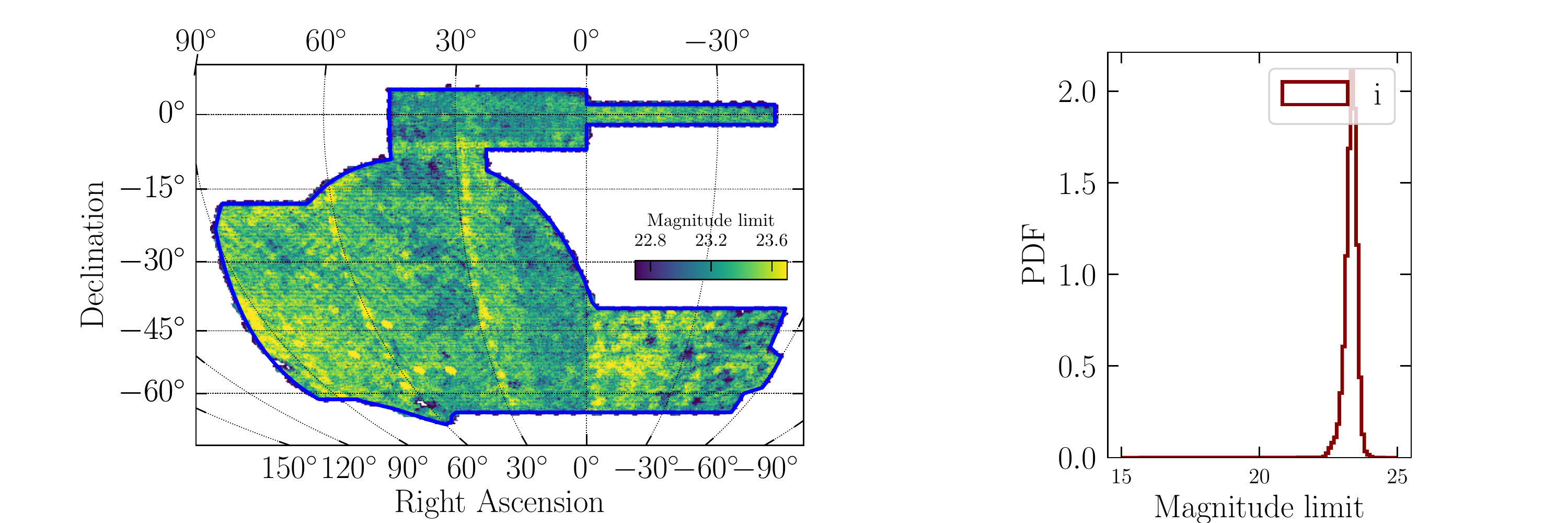}
\includegraphics[width=0.75\textwidth]{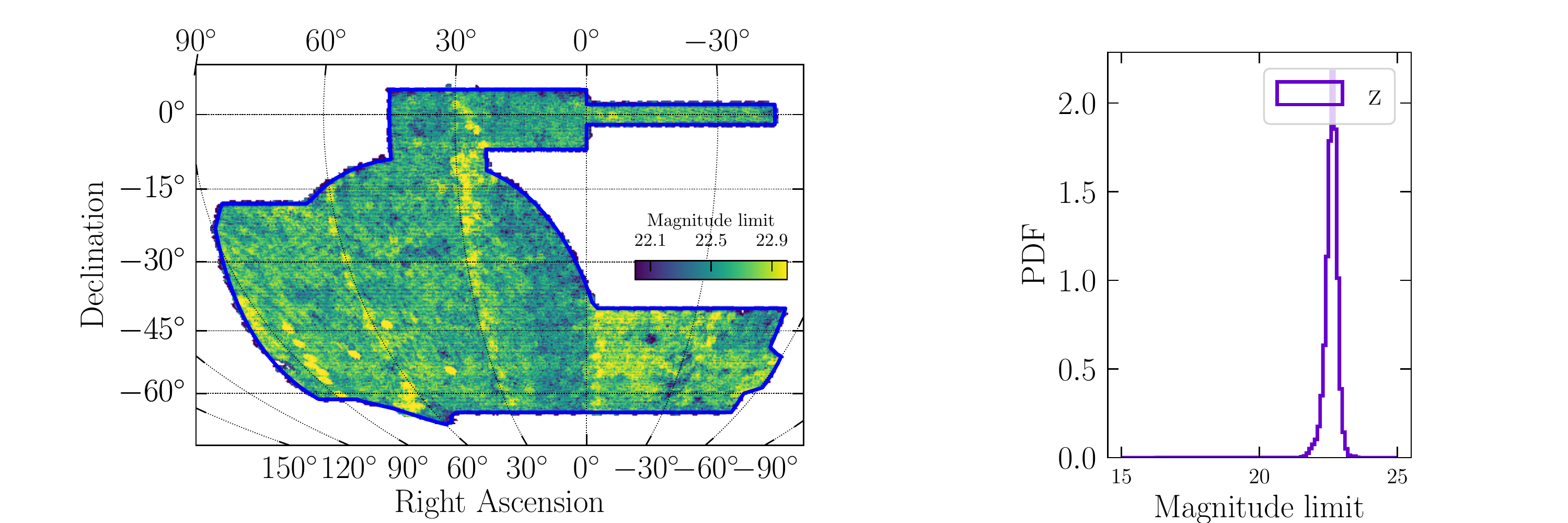}
\includegraphics[width=0.75\textwidth]{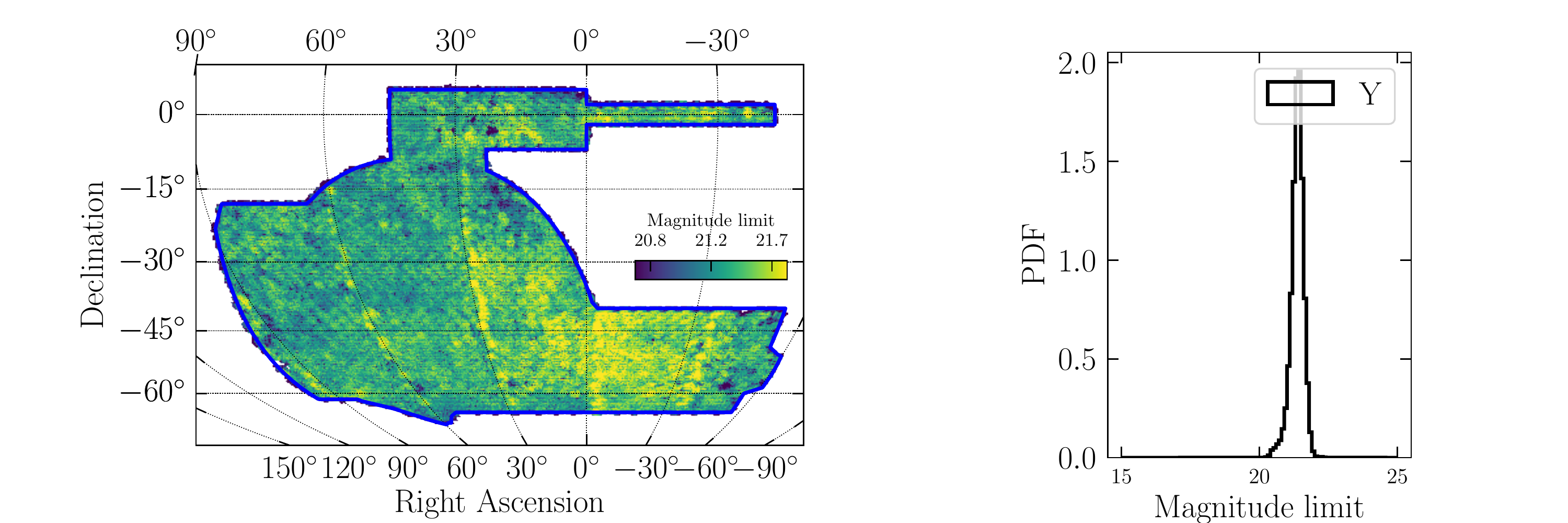}
\caption{\label{fig:app_maglim} 
Sky maps and histograms of the magnitude limit (MAGLIM), computed at the S/N = 10 level for 2\asec apertures. }
\end{figure}

\begin{figure}[h]
\centering
\includegraphics[width=0.65\textwidth]{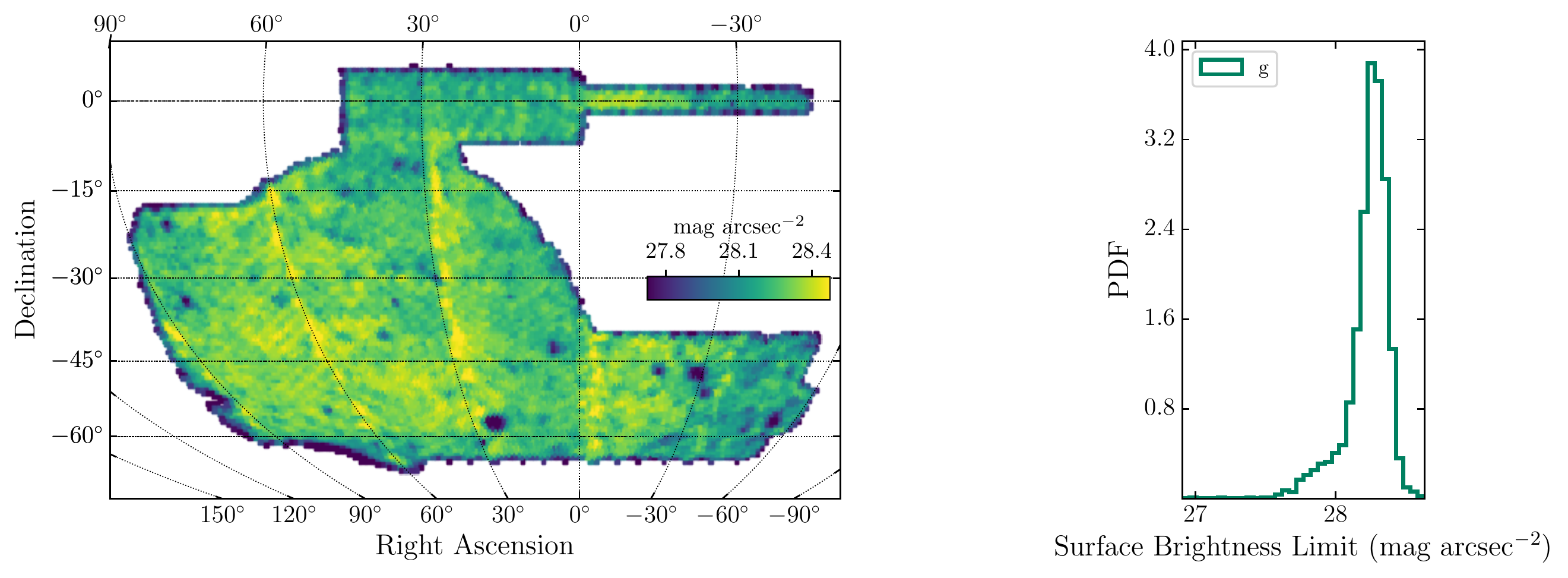}
\includegraphics[width=0.65\textwidth]{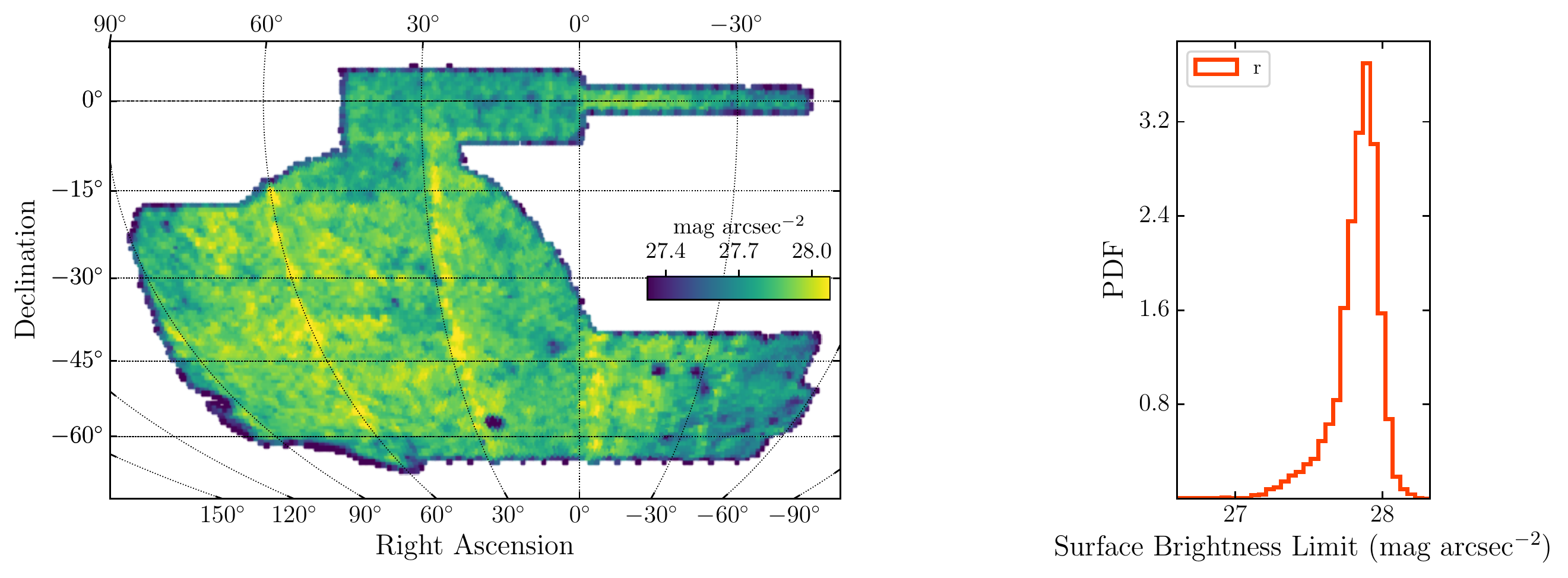}
\includegraphics[width=0.65\textwidth]{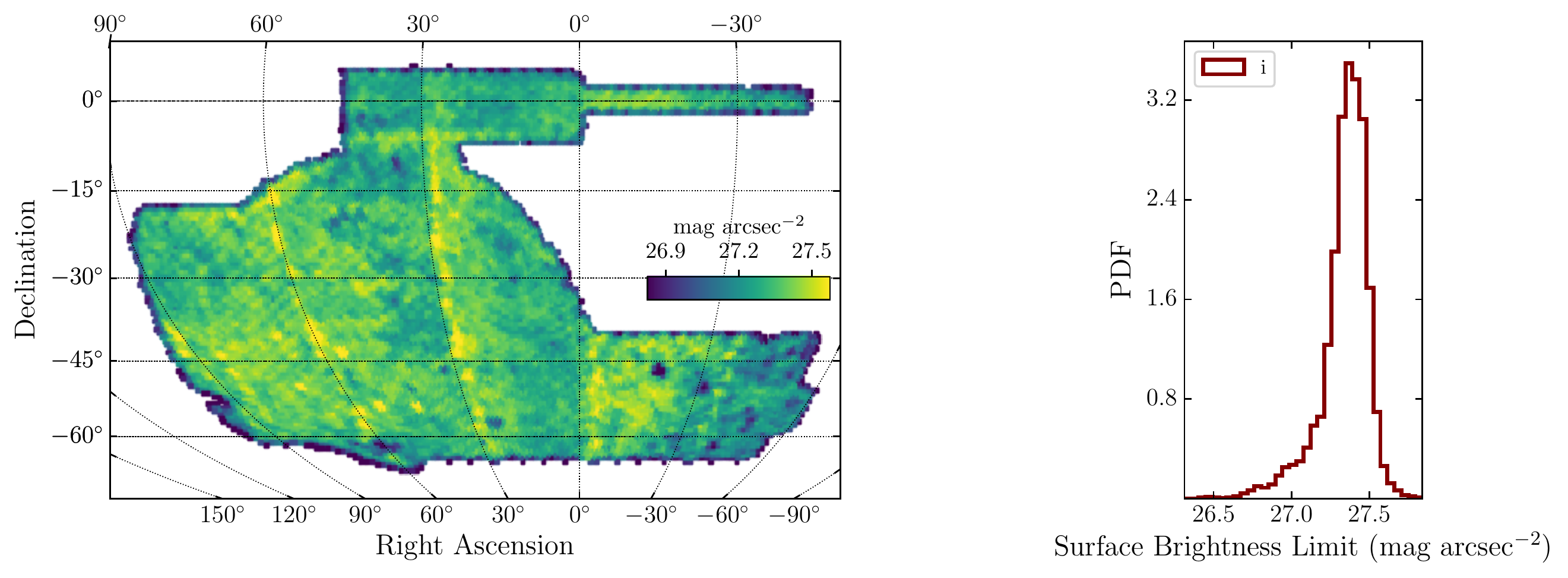}
\includegraphics[width=0.65\textwidth]{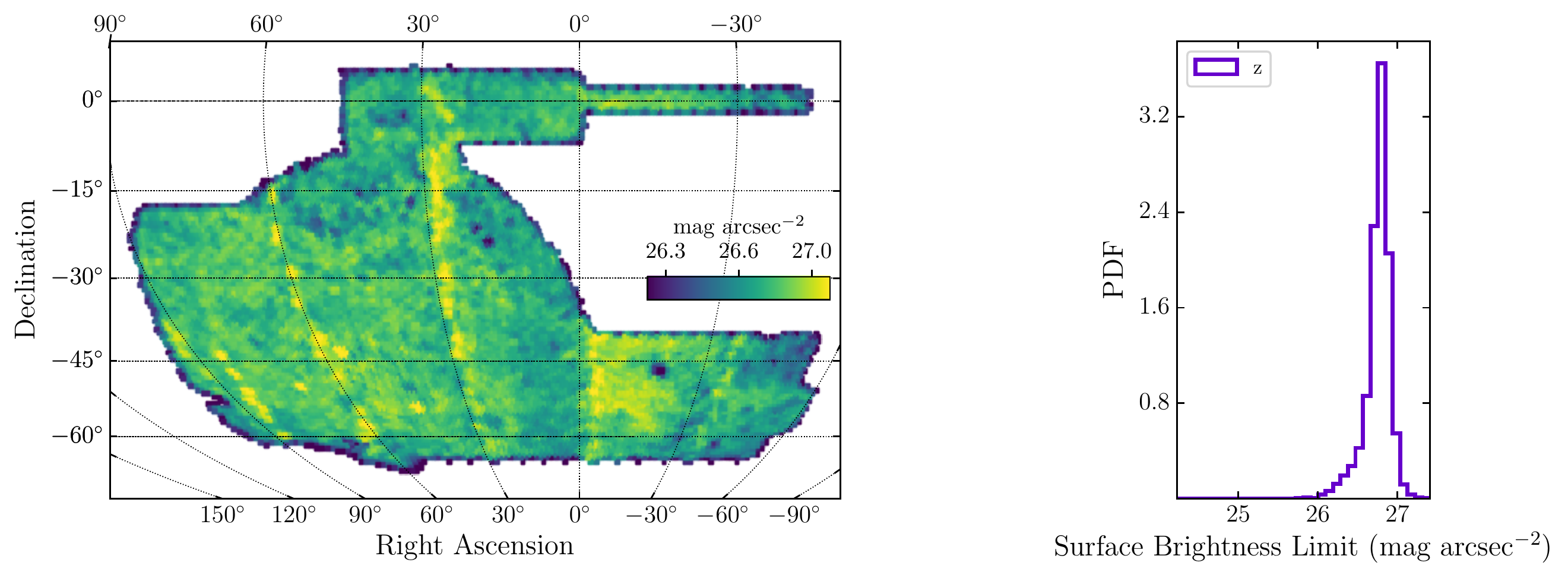}
\includegraphics[width=0.65\textwidth]{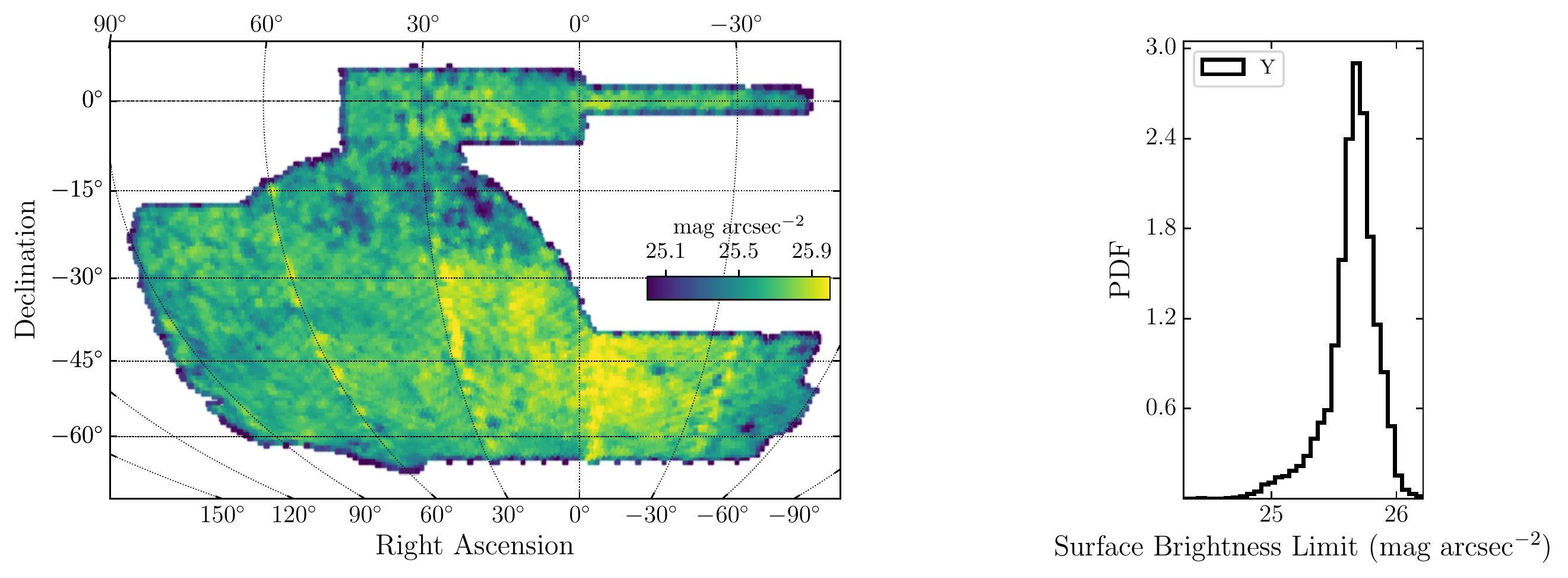}
\caption{\label{fig:app_sbcontrast} 
Sky maps and histograms of the surface brightness limit (SBCONTRAST) at $3\sigma$. This is measured as the variation in the sky background over an angular scale of $10\asec \times 10\asec$ (computed in \citet{tanoglidis}, following the technique in \citet{gilhuly}).}
\end{figure}

\begin{figure}[h]
\centering
\includegraphics[width=0.75\textwidth]{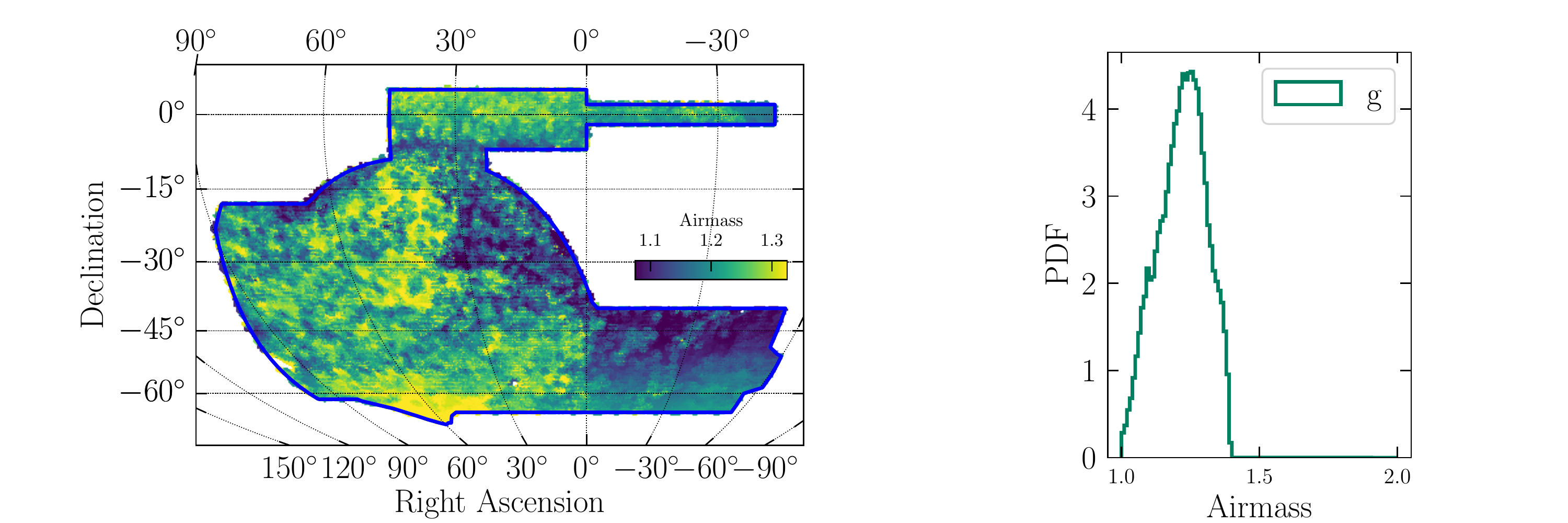}
\includegraphics[width=0.75\textwidth]{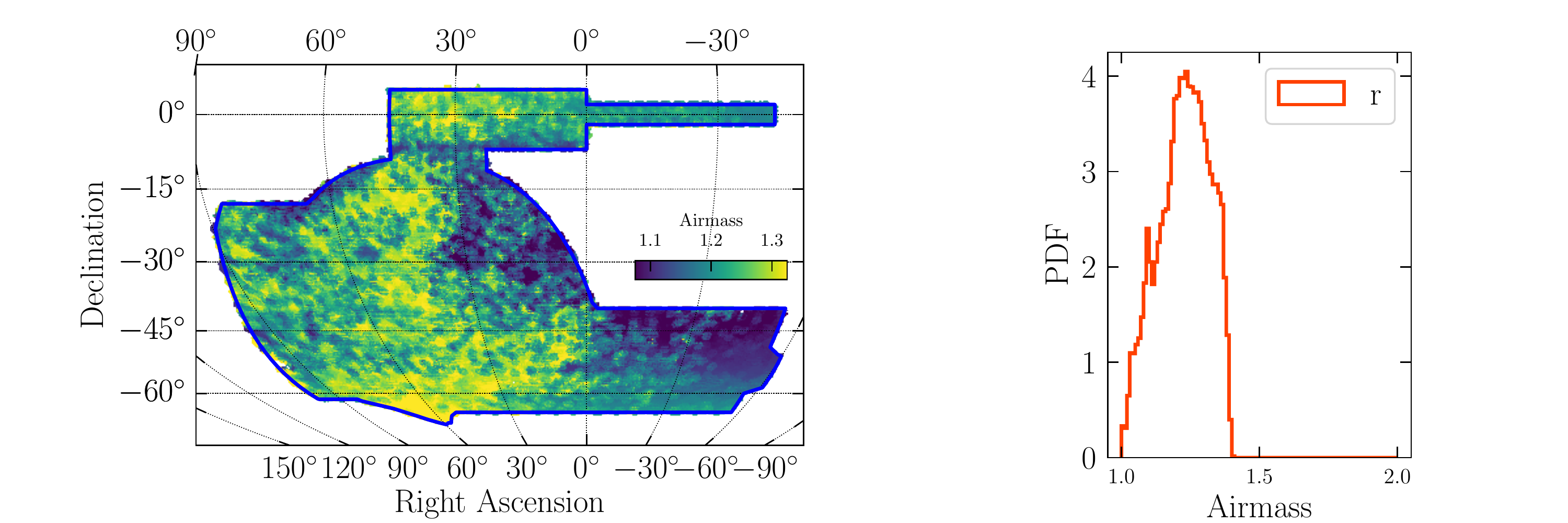}
\includegraphics[width=0.75\textwidth]{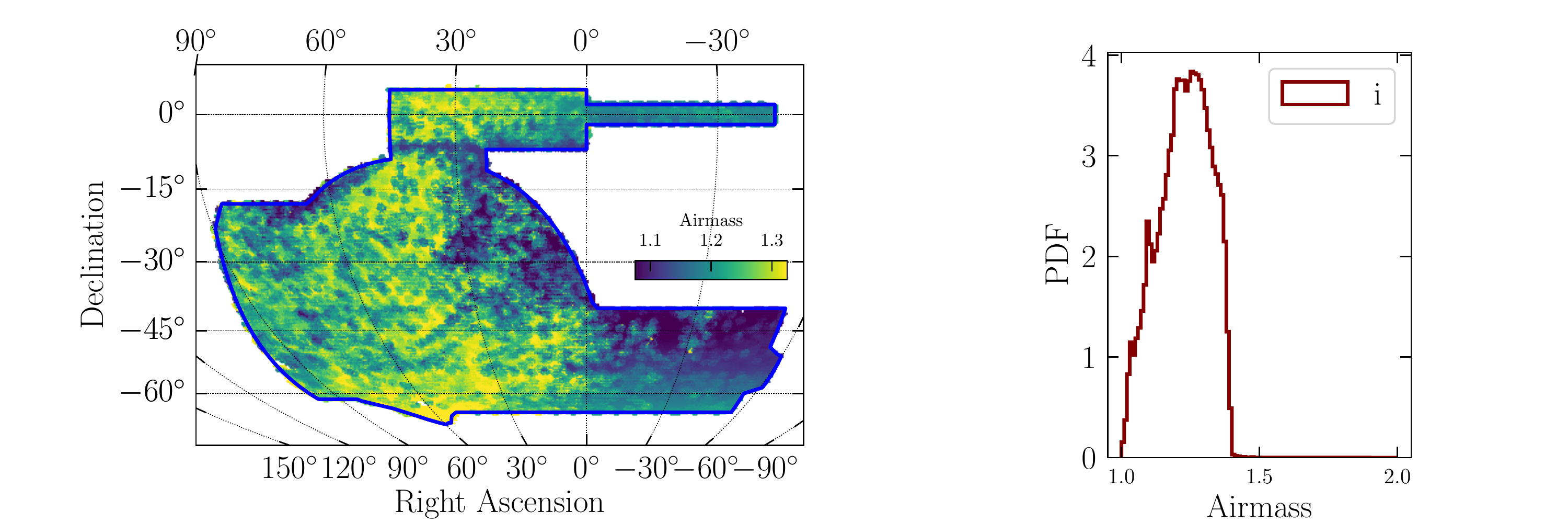}
\includegraphics[width=0.75\textwidth]{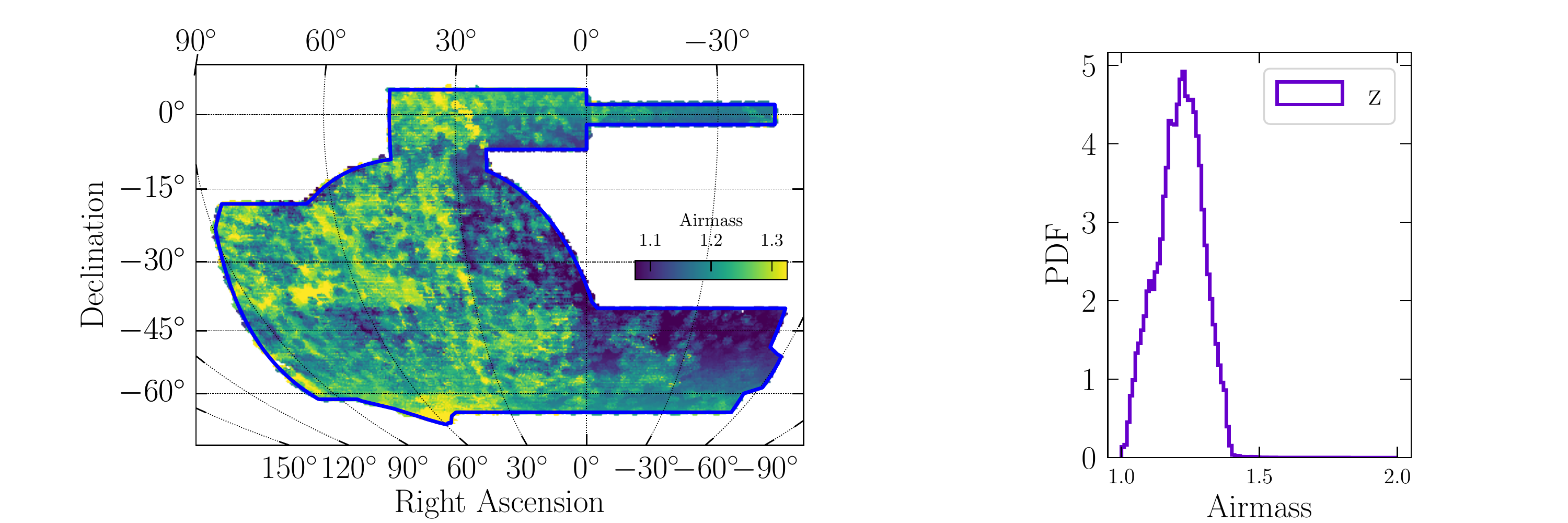}
\includegraphics[width=0.75\textwidth]{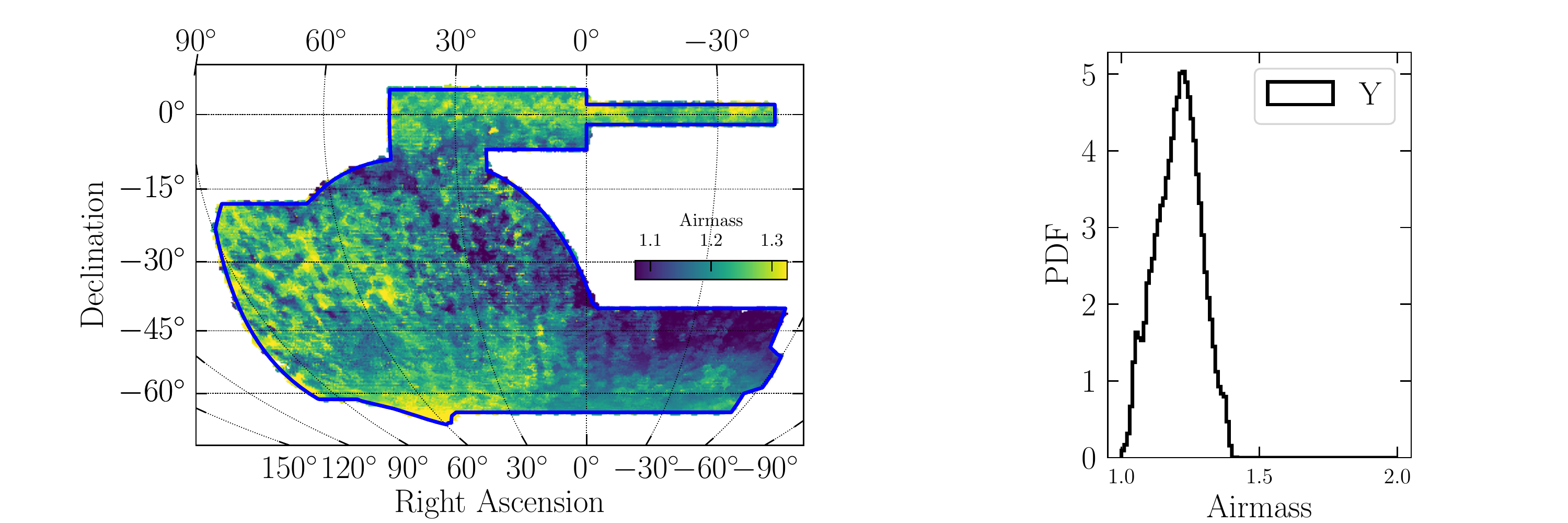}
\caption{\label{fig:app_airmass} 
Sky maps and histograms of the airmass (AIRMASS.WMEAN) for each of the observed bands. The value at each location is the inverse-sky-variance-weighted sum of all individual exposures of that \healpix pixel.}
\end{figure}

\begin{figure}[h]
\centering
\includegraphics[width=0.75\textwidth]{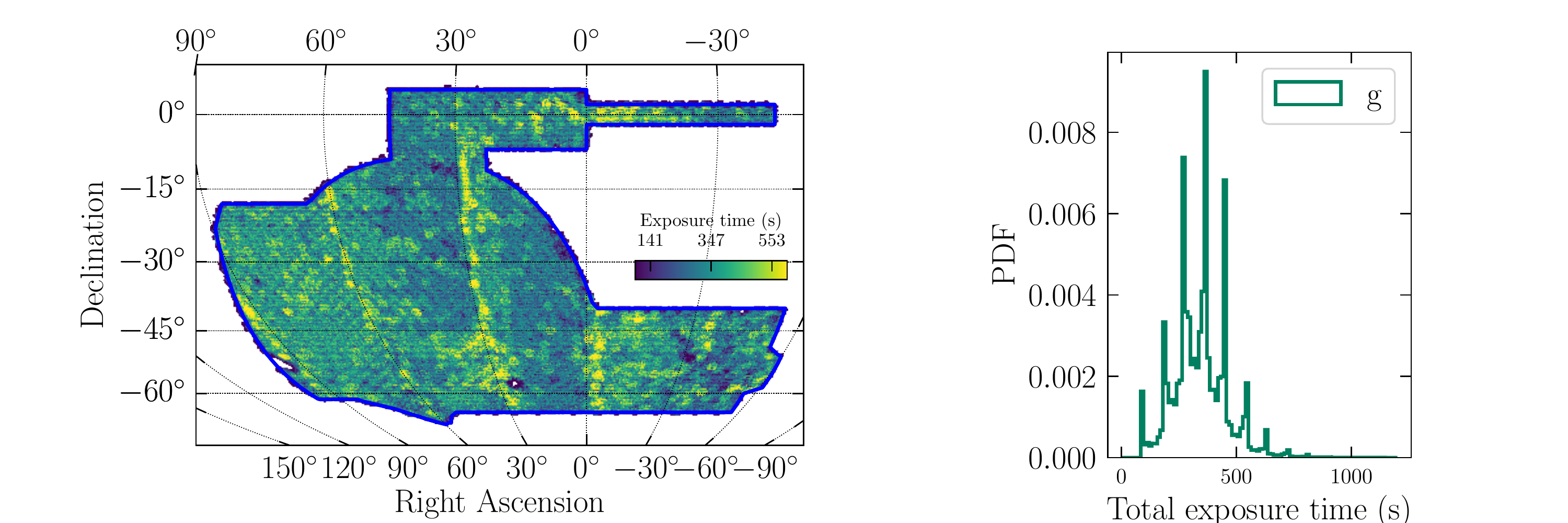}
\includegraphics[width=0.75\textwidth]{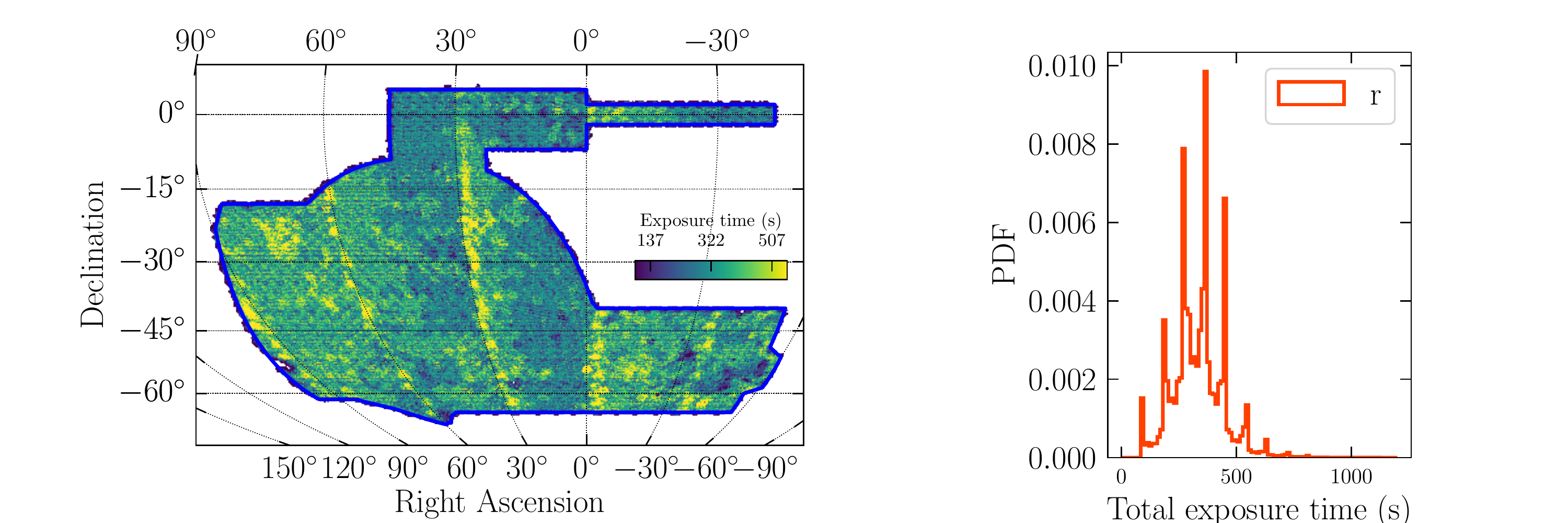}
\includegraphics[width=0.75\textwidth]{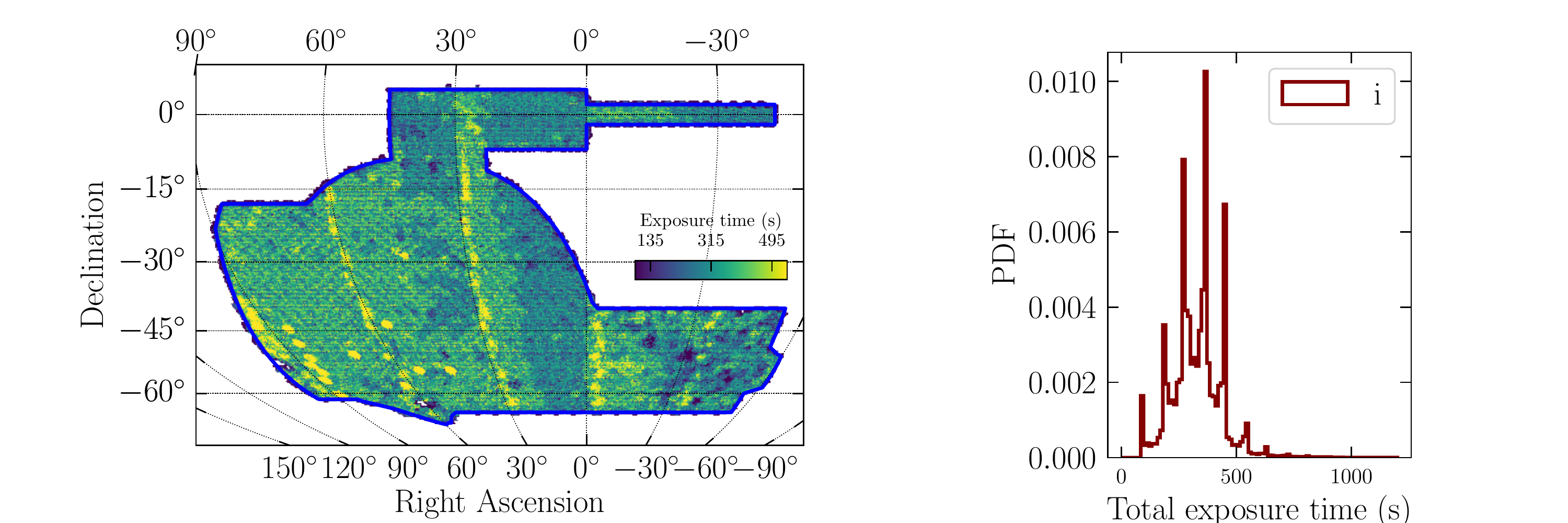}
\includegraphics[width=0.75\textwidth]{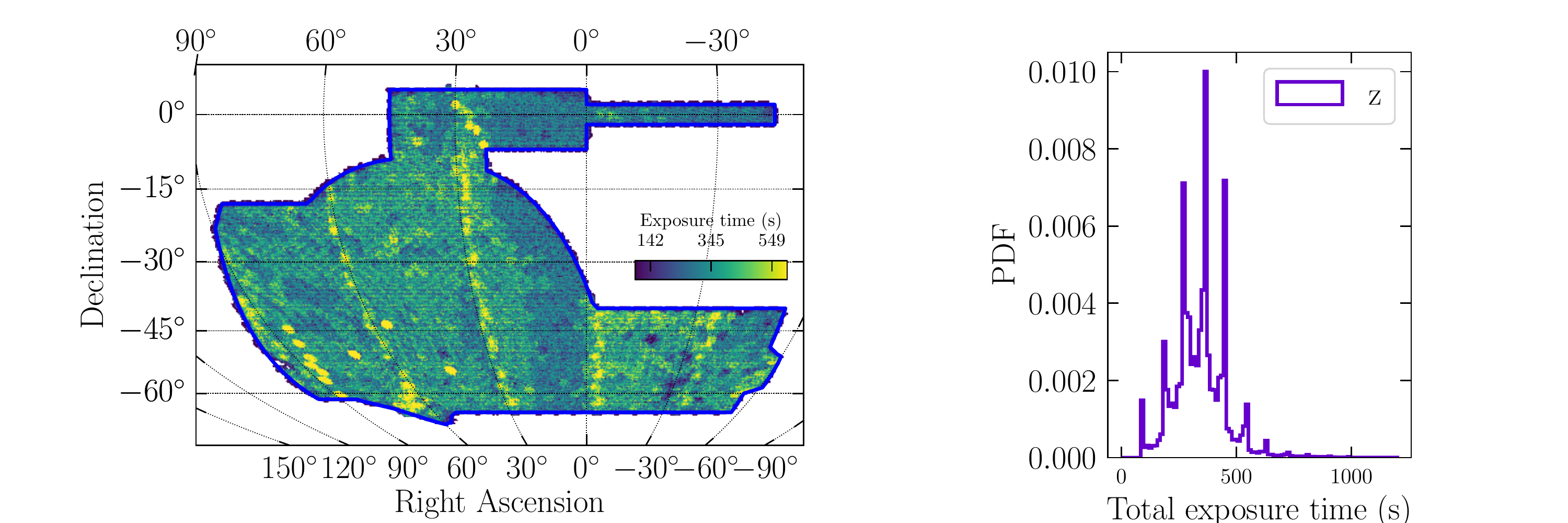}
\includegraphics[width=0.75\textwidth]{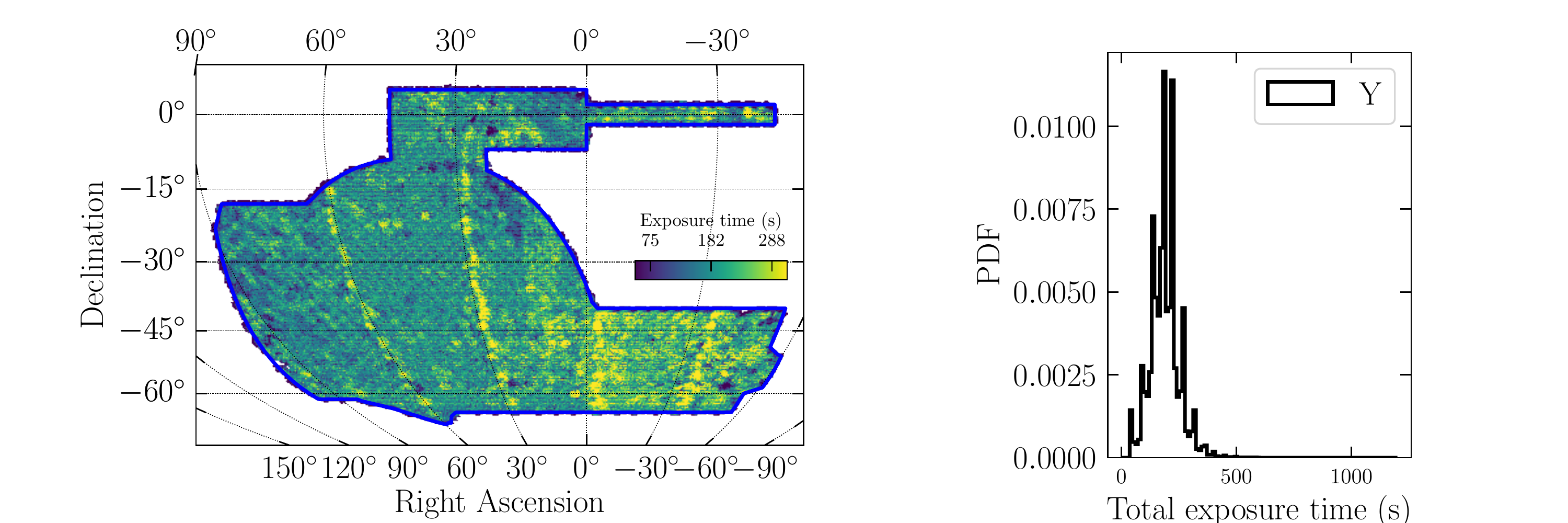}
\caption{\label{fig:app_exptime} 
Sky maps and histograms of the total exposure time (EXPTIME.SUM) for each of the observed bands. These are not multiples of 90 seconds, as a single \healpix pixel might contain contributions of regions with varying number of exposures (they are accounted according to their relative area in the pixel).}
\end{figure}

In \tabref{observingconditions} we summarize the observing conditions per band. 
We also include commonly used survey property maps in each band. \figref{app_fwhm} to \figref{app_exptime} show these maps as a function of position in the sky and the corresponding histogram of computed values for these positions (computed in \nside = 4096 \healpix resolution). Note that the linear features along equal RA values are a consequence of the observation strategy to ensure a complete tiling of the sphere.

\begin{deluxetable}{c c l}[h]
\tablewidth{0pt}
\tabletypesize{\tablesize}
\tablecaption{ \gold Survey Properties.  \label{tab:observingconditions}}
\tablehead{
\colhead{DES map name} & \colhead{Units} &
\colhead{Description} 
}
\startdata
NUMIMAGE & & Number of images \\
MAGLIM & & Magnitude limit estimated by \mangle from the weight maps \tablenotemark{a} \\
FRACDET & & Effective area fraction considering the bleed-trail and bright star masks \\
EXPTIME.SUM &  seconds & Exposure time \\
T\_EFF.(WMEAN/MAX/MIN) & & Figure of merit for quality of observations $t_{eff}$\tablenotemark{b}\\
T\_EFF\_EXPTIME.SUM &  seconds &  Exposure time multiplied by $t_{eff}$\\
SKYBRITE.WMEAN & electrons/CCD pixel & Sky brightness from the sky background model  \tablenotemark{c} \\
SKYVAR.(WMEAN/MIN/MAX) & (electrons/CCD pixel)$^2$ & Variance on the sky brightness\tablenotemark{d} \\
SKYVAR\_SQRT.WMEAN & electrons/CCD pixel & Square root of sky variance \\
SKYVAR\_UNCERTAINTY & electrons/s/coadd pixel & Sky variance with flux scaled by zero point. \\
SIGMA\_MAG\_ZERO.QSUM & mag & Quadrature sum of zeropoint uncertainties. \\
FWHM.(WMEAN/MIN/MAX) & \asec  & Average FWHM of the 2D elliptical Moffat function that fits best the PSF model from \code{PSFEx}. \\
FWHM\_FLUXRAD.(WMEAN/MIN/MAX) & \asec & Twice the average half-light radius from the sources used for determining the PSF with \code{PSFEx}. \\
FGCM\_GRY.(WMEAN/MIN/MAX) & mag & Residual `gray' corrections to the zeropoint from FGCM \\
AIRMASS.(WMEAN/MIN/MAX) & & Secant of the zenith angle \\
SBCONTRAST & mag/arcsec$^2$ & 3-sigma surface brightness contrast\tablenotemark{e}\\
\enddata
\tablecomments{
Survey properties in \gold registered as maps. Each quantity has been calculated individually for $grizY$ bands. All maps are produced in \healpix format in $\nside=4096$ in \code{NESTED} ordering, averaging from from a higher resolution version ($\nside=32768$). Each high resolution pixel adopts the value of the molygon from the \mangle map at its center, which is a statistic of a stack of images contributing to that point in the sky.  \code{WMEAN} quantities are the mean value weighted using the weights obtained from \mangle. \code{MIN, MAX} correspond to the minimum or maximum of all the stacked images in the molygon. \code{SUM} adds up the contribution of all images to the molygon. \code{QSUM} makes a quadrature sum instead. The DES map name is the name given to the files as they are delivered in the release page.}
\tablenotetext{a}{10-$\sigma$ magnitude limit in 2\asec diameter apertures}
\tablenotetext{b}{$t_{eff}$, as described in \citet{desdm}, Equation 4, is measured as a ratio between exposure time and the exposure time necessary to achieve the same signal-to-noise for point sources observed in nominal conditions. This depends on a set of fiducial conditions per band for full-width half maximum, sky background and atmospheric transmission.}
\tablenotetext{c}{The model value used is taken as the median per CCD. Details for this model are described in \citet{Bernstein:2017} and \citet{desdm}.}
\tablenotemark{d}{Takes into account intrinsic sky Poisson noise, read noise and flat field variance.}
\tablenotemark{e}{Computed outside the DESDM framework as detailed in \citet{tanoglidis, gilhuly}}
\end{deluxetable}

%The survey conditions delivered by DESDM in \mangle format, and later transformed to \healpix are: Sky Brightness: weighted using the weight map from DESDM. Detection fraction: what fraction of the \healpix pixel has been observed (according to \mangle geometry) in each band or simultaneously in griz or grizY bands. Exposure time: the time in seconds for exposures in a given filter (EXPTIME) or the effective time (TEFF\_EXPTIME), weighted with atmospheric transmission, seeing and sky brightness. Airmass: secant of the zenith angle for the observations included in the pixel. Seeing: measures the sky quality as a blurring of point-like sources, it is described here as the full width half maximum (in arcseconds) of the PSF taken from FWHM\_MEAN of the PSF\_QA table (calculated using PSFEx).  Sky variance: this is the . Zero point residues from FGCM: residual "gray" corrections from FGCM calibration.

\clearpage

\bibliographystyle{\bibsty}
\bibliography{y3gold}

\end{document}